\documentclass[superscriptaddress,nofootinbib]{revtex4-2}
\usepackage{graphicx}
\usepackage{dcolumn}
\usepackage{bm}
\usepackage{mathrsfs}
\usepackage{multirow}
\usepackage{amsmath}
\usepackage{amsfonts}
\usepackage[makeroom]{cancel}

\usepackage[colorlinks,
linkcolor={red!60!black!},
anchorcolor={green!80!black!},
urlcolor={blue!80!black!},
citecolor={blue!50!black!}]{hyperref}
\usepackage[table]{xcolor}
\usepackage{tipa}
\usepackage{physics}
\usepackage{caption}
\usepackage{subcaption}
\usepackage{bbold}
\usepackage{amssymb}
\usepackage{tikz}

\usepackage[left=2.5cm, right=2.5cm,bottom=3.5cm,top=3cm]{geometry}

\begin{document}

\title{Studying replica wormholes and the Page curve with simplicial quantum gravity}

\author{José Padua-Argüelles}
\email{jpaduaarguelles@pitp.ca}
\affiliation{Perimeter Institute, 31 Caroline Street North, Waterloo, ON, N2L 2Y5, Canada}
\affiliation{Department of Physics and  Astronomy, University of Waterloo, 200 University Avenue West, Waterloo, ON, N2L 3G1, Canada}

\begin{abstract}

\textbf{Abstract} 
The replica paradigm has emerged as a powerful tool for investigating the black hole information paradox, offering a semiclassical route to reproducing the Page curve and suggesting unitary evolution for evaporating black holes. However, existing analyses have relied on simplified models such as JT gravity, and mostly remain limited to the $n \to 1^+$ limit in Euclidean signature. This work develops a framework based on Quantum Regge Calculus (QRC) that provides a lattice-like approach to address these gaps. A triangulation scheme is introduced that accommodates both gravitational and radiation degrees of freedom, enabling explicit evaluation of the fundamental components of the Regge gravity and radiation actions in a spherically symmetric setting. The formulation naturally incorporates analytic continuation techniques to probe the role of complex saddles in Lorentzian signature. A proof-of-principle implementation is carried out within a controlled minisuperspace reduction, revealing semiclassical saddles in the $n \to 1^+$ limit that recover the Page transition.  While significant challenges remain (including the definition of the discrete configuration space, ambiguities in the gravitational measure, and the treatment of asymptotic boundaries), the framework developed here provides a promising foundation for further progress. The results suggest that sufficiently refined QRC calculations could extend the replica approach beyond existing models.
\end{abstract}

\maketitle

\tableofcontents
%{\let\clearpage\relax \tableofcontents}

\section{Introduction\label{sec:introduction}}

The past lustrum has witnessed a shift in perspectives regarding the black hole information paradox. It was long believed that preventing information loss in black hole evaporation required new UV physics. However, recent applications of the replica trick in gravity \cite{Penington:2019kki,Almheiri:2019qdq,Marolf:2020rpm,Chandrasekaran:2022asa} suggest that standard semiclassical path integral techniques may already encode some of the necessary conditions for unitarity, offering a promising perspective on the problem. More specifically, when entropies are computed using the replica trick, the Page curve emerges from the saddle point approximation of the gravitational path integral, thus recovering a key benchmark of unitarity. The essential mechanism behind this result is that the replica trick introduces a new saddle of non-trivial topology (assuming quantum gravity includes a sum over topologies). This saddle, associated with a so-called \emph{replica} wormhole, dominates at late times and accounts for the purification process. As well articulated in \cite{Marolf:2020rpm,Marolf:2021ghr}, while Hawking’s calculation correctly showed that the von Neumann entropy of radiation grows indefinitely under a semiclassical treatment, it has become clear that this does not necessarily imply a violation of unitarity. Instead, the quantity being computed must be reconsidered: rather than von Neumann entropy, an operationally well-defined alternative should be used. The so-called swap entropy provides one such alternative and justifies the use of the replica trick. Moreover, this perspective is naturally framed in real-time and thus provides a more direct and physically intuitive understanding of the underlying dynamics.

These arguments are based on standard path integral techniques, which lends them a certain robustness. Therefore, it is natural to ask whether they can be reproduced in other quantum gravity approaches, particularly those based on the path integral formalism, such as Quantum Regge Calculus (QRC) \cite{Rocek:1981ama,Hamber:2009mt} and spinfoams \cite{Baez:1997zt,Perez:2003vx}. If they can, these approaches may offer valuable insights into the physics of gravitational replicas from new perspectives, or potentially further reinforce the expected conclusions. As an example that justifies this hope, recent work \cite{Dittrich:2024awu} applied QRC to gain insights into continuum quantum gravity, studying a discrete analogue of the gravitational partition function that computes the quantum gravity Hilbert space dimension for a spatial ball with a positive cosmological constant and its connection to de Sitter horizon entropy.

This paper addresses these topics by presenting a computational framework for gravitational replica calculations within the context of discrete quantum gravity, specifically using quantum Regge calculus —an approach to defining lattice quantum gravity. The framework is designed to provide a computational handle to investigate open questions surrounding replica wormhole saddles, including their existence at finite replica number $n$ (a key issue from an operational standpoint), their role in the Lorentzian path integral (which is crucial given that they must be complex, \emph{e.g.} Euclidean \cite{Marolf:2020rpm,Louko:1995jw}), and corrections beyond the gravitational saddle point approximation around them. Moreover, the framework can hopefully serve as a bridge to the discrete quantum gravity community.

The structure of this paper is as follows: After contextualizing the discrete setup by first presenting the continuum, operational perspective (and presentation) of \cite{Marolf:2020rpm} in \S\ref{sec:continuum}, \S\ref{sec:discrete} then introduces the discrete quantum gravity scheme for computing swap entropies in spherically symmetric Hawking evaporation spacetimes. This section details the discretization process and the computations of the gravitational and matter actions, each covered in separate subsections. These actions are computed using complex Regge calculus, which facilitates discussions of contour deformations and complex saddles. The parametrization of the analytic continuation presented here is novel, providing an intermediate approach between the global generalized Wick rotation angle of \cite{Asante:2021phx} and the complexified edge lengths of \cite{Jia:2021xeh}. As this work aims to establish the foundation for a broader study of black hole evaporation with QRC, the actions are computed in a ‘modular’ way —meaning that the results provide essential building blocks applicable to any triangulation within this framework, regardless of its refinement.

All of these notions are instantiated in \S\ref{sec:application} through an application that captures the minimal number of matter degrees of freedom necessary for swap entropy calculations. The matter content considered is a minimally coupled, free, massless scalar field. This choice enables the derivation of analytic expressions for the matter sector, such as the matter effective action, and allows to eventually perform analytic continuations in the replica number $n$, which makes it possible to take the $n\to1^+$ limit. For more refined triangulations or different matter content, these methods will likely need to be respectively replaced by numerical approaches and appeals to Carlson’s theorem. The model is then restricted to a minisuperspace that is CPT-reduced, replica-symmetric, and designed to incorporate key features of evaporating black hole spacetimes. Within this minisuperspace, two saddle points are identified with the expected topologies from the literature, and their swap entropies exhibit the anticipated Page-transition-like behavior. Despite such restriction, this can be seen as a promising first result: It suggests that as long as the discretization is sufficiently fine to incorporate enough physics so that these spacetimes emerge as effective geometries, one would ultimately be able to recover the Page curve in more realistic scenarios. It also serves as a concrete example that clarifies how the framework is to be used in future applications.

Most of the explicit final results are exceptionally lengthy (reaching sizes of up to 450MB) and are therefore provided in a dedicated GitHub repository \cite{github}, which also contains the code used in their derivation. Some of these derivations rely on the Mathematica library for Regge calculus detailedly described in \cite{PaduaArguellesQRC:2025}. For ease of reference, Appendix \ref{app:entangling_integrals} complements the main discussion in \S\ref{sec:application} by presenting intermediary steps more faithfully to the computational implementation in the code.

As a preparation for the content of sections \S\ref{sec:discrete} and \S\ref{sec:application}, \S\ref{sec:overview} provides a non-technical summary of it. 

Finally, finishing remarks and an outlook are provided in \S\ref{sec:discussion}, such as lessons gathered for other Regge-like approaches to Quantum Gravity (\emph{e.g.} spinfoams), or the use of this framework in more refined setups.

\section{Page curve and swap entropy in the continuum\label{sec:continuum}}

One of the long-standing problems in the path of understanding the quantum nature of gravity is the black hole information paradox, quickly formulated by Hawking after realizing that, in the context of quantum field theory in curved backgrounds, black holes radiate according to asymptotic observers in (flat) future null infinity $\mathscr I^+$. Building up on \cite{Hawking:1975vcx}, in \cite{Hawking:1976ra} he noticed that due to such radiation, black holes formed from collapse should evaporate and reasoned that they would eventually disappear,\footnote{Another possibility is that the black hole shrinks to a minimum size, leaving behind a so called remnant. However these scenarios face alternative challenges, for example, regarding entropy bounds \cite{Susskind:1995da,Banks:2010zn}.\label{fnote:remnants}} leaving only their radiation behind. However, such radiation is thermal and its entropy increases with retarded time $u$ until evaporation ends. This, he argued, creates tension with the unitary evolution of quantum mechanics. A pure state $\ket{\varphi_-}$ prepared at past null infinity can evolve to form a black hole that eventually disappears, leaving a resulting final state that would be mixed, and thus has nonzero von Neumann entropy: In this way, information about the initial state is irremediably lost in the evaporation process. Therefore, his calculation showed that the unitarity principle of quantum mechanics is jeopardized by gravitational physics.

Since its appearance, the information paradox has fueled a wide spectrum of research concerning the interface between the quantum and gravitational realms. On one side of it lies the point of view that information is indeed lost, on the other the one that a detailed understanding of UV physics would show that unitarity is preserved —see \cite{Page:1993up,Mathur:2009hf} for reviews.

At first sight both extremes seem to require some radicality, be it by abandoning a foundational principle of physics, or by using some microscopic quantum gravitational input. This was the point of view for several decades. However, in tension with that expectation, the recent uses of the replica trick in gravity suggest that saddle point evaluations of the gravitational path integral are enough to reproduce the \emph{Page curve}.

\begin{figure}[h!]
    \centering
    \begin{subfigure}[c]{0.45\textwidth}
        \centering
        \includegraphics[width=\linewidth]{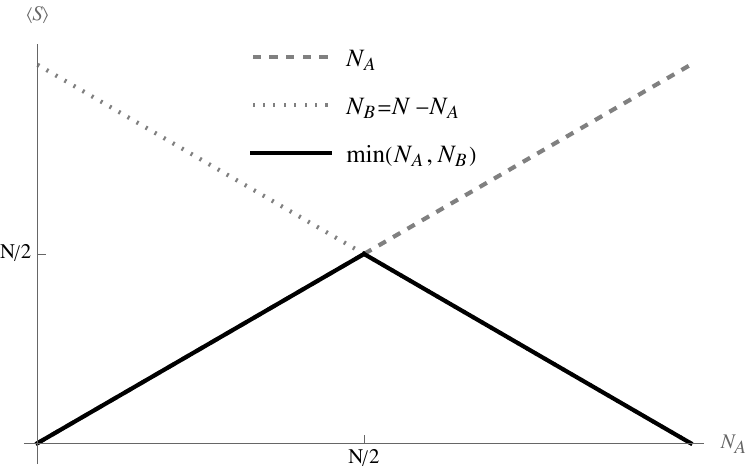}
        \caption{The average entropy of a two-part spin system, as evaluated in the limit of one subsystem being much smaller than the other, is given by the minimum of the subsystems' sizes.}
        \label{sfig:Page_spin}
    \end{subfigure}
    \hspace{1cm}
    \begin{subfigure}[c]{0.45\textwidth}
        \centering
        \includegraphics[width=\linewidth]{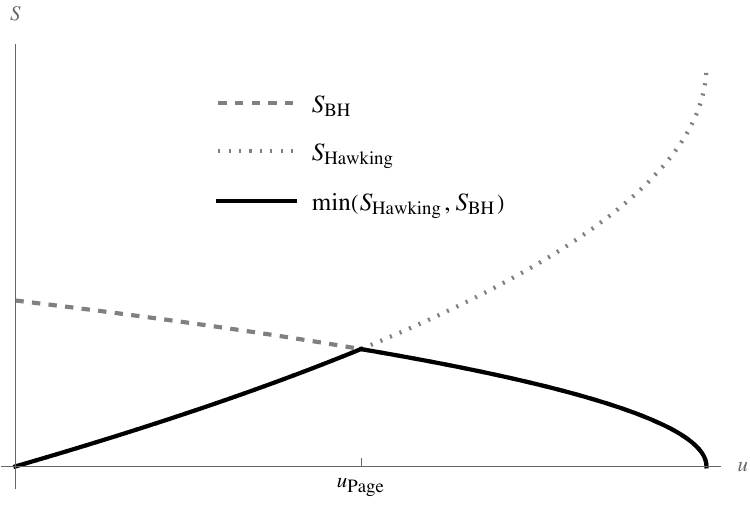}
        \caption{The entanglement entropy of black hole radiation is expected to follow the Page curve (a key signature of unitarity) rising as thermal radiation is emitted until reaching a maximum, at which the \emph{Page transition} marks the onset of purification. The curve initially rises, following Hawking’s result based on thermality, before decreasing in accordance with the Bekenstein entropy of the black hole.}
        \label{sfig:Page_BH}   
    \end{subfigure}
    \caption{Entropy (putative) behaviors for analogous systems: a two-part spin system with one subsystem shrinking and an evaporating black hole spacetime}
\end{figure}

Such curve was suggested by Page in \cite{Page:1993up} to be the one that the entropy of Hawking radiation should follow as a necessary condition for unitarity during the black hole evaporation process. He suggested that if the evaporation process was unitary, then the von Neumann entropy would initially increase, until the now called Page time marking the \emph{Page transitions}, after which the entropy would decrease until reaching zero once the black hole disappears, roughly following the behavior shown in \ref{sfig:Page_BH}. Indeed, as he observed, a similar thing happens to the average entropy of a system that could be considered analogous, a two-part spin system, as the size of one subsystem is reduced to zero \cite{Page:1993df}, because the average entropy scales as the entropy of the smaller sub-system, whenever one is much larger than the other —see \ref{sfig:Page_spin}. The Page curve suggests that an asymptotic observer might treat a black hole as a unitary quantum system with density of states $e^{S_\text{BH}}$, a property that has been called BH-unitarity in literature \cite{Marolf:2020rpm,Marolf:2021ghr}. Recent replica calculations show that the Page curve behavior can indeed apply to black holes.

Importantly, after its introduction to the gravitational context \cite{Penington:2019kki,Almheiri:2019qdq}, it was noted that the replica trick can be motivated by thinking in operational terms \cite{Marolf:2020rpm}. From this perspective, an actual observer’s measurements would not signal a unitarity problem: The work \cite{Marolf:2020rpm} suggests that measurement outcomes aiming at determining entropy will indicate that it indeed follows the \emph{Page curve}. This is the point of view taken in this work, so it will be useful to go over the key elements of its computational framework —see also \cite{Marolf:2021ghr} for a(nother) review.

Entropies cannot be determined by single systems, but ensembles of them. Thus an asymptotic experimentalist would need to prepare $n$ systems of black hole formation and evaporation, and then collect their respective radiation up to a retarded time $u$ in \emph{each}\footnote{Note that here an implicit assumption is being made: there are $n$ asymptotic regions that can be considered as disconnected.} asymptotic region $\mathscr I^+$ in order to determine the entropy of the radiation received up until $u$.

\begin{figure}[h!]
    \centering
    \includegraphics[width=0.5\textwidth]{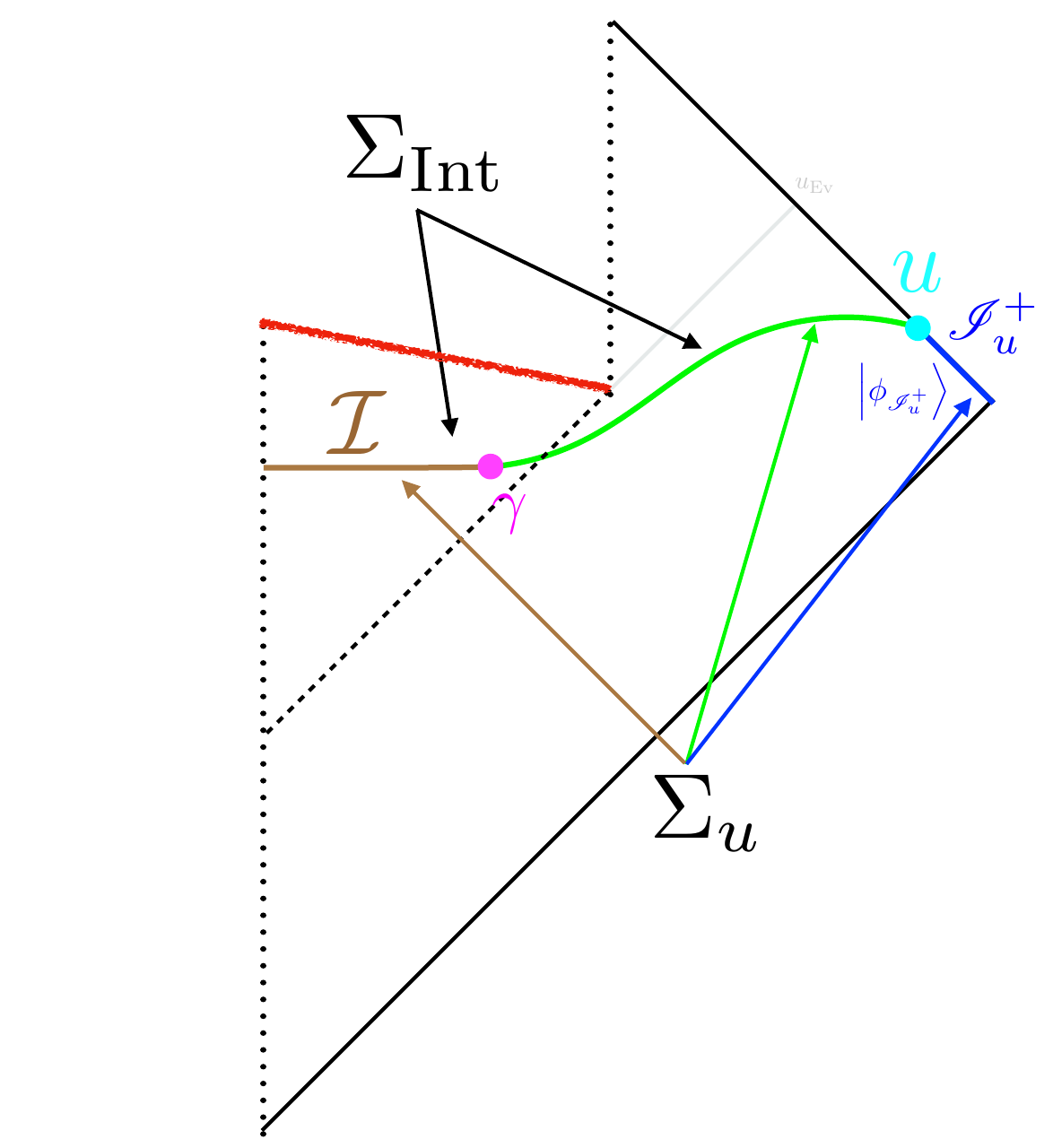}
    \caption{Penrose diagram representing a spacetime in which gravitational collapse and eventual black hole evaporation take place. The singularity (rough/red line) is not depicted horizontally as is customary, to emphasize the role of back-reaction. The figure also depicts all geometric subspaces and the observer’s Hilbert space states, as referenced in the text to describe the path integrals under study.}
    \label{fig:Hawking-Penrose}
\end{figure}

These data should be captured in terms of an observable and, as will be seen below, it can be the case for \emph{swap operators}\footnote{A technical observation is in place: Generically, swap operators are not hermitian. Nevertheless they can be measured, as they commute with their adjoint \cite{Marolf:2021ghr}.} $U\left(\sigma^{(n)}\right)$, which take the $n$-copy system and perform the cyclic permutation
\begin{equation}
    \sigma^{(n)}:i\overset{\sigma^{(n)}}{\rightarrow}\sigma^{(n)}_i=(i+1)\text{mod} n
\end{equation}
of the (different) $\mathscr I^+$ regions. More precisely, one can introduce a basis $\left\{\ket{\phi^1_{\mathscr I_u^+},\dots,\phi^n_{\mathscr I_u^+}}\right\}$ of states peaked on classical field configurations $\phi^i_{\mathscr I_u^+}$ defined on $\left(\mathscr I_u^+\right)^i$, the portions of the $i$-th $\mathscr I^+$ with retarded time not larger than $u$ (see FIG. \ref{fig:Hawking-Penrose})\footnote{Several criticisms could be raised regarding this diagram, as it implicitly assumes aspects beyond semiclassical control (\emph{e.g.} the neglect of the remnant scenario —see footnote \ref{fnote:remnants}). However, as will become clear later, the way copies of this spacetime are glued ensures that regions lacking semiclassical control do not contribute to the final calculation.}, and define the action of $U\left(\sigma^{(n)}\right)$ by 
\begin{equation}
   \bra{\phi^{1^*}_{\mathscr I_u^+},\dots,\phi^{n^*}_{\mathscr I_u^+}}U\left(\sigma^{(n)}\right)\ket{\phi^1_{\mathscr I_u^+},\dots,\phi^n_{\mathscr I_u^+}}=\prod_i \delta\left[\phi^{i^*}_{\mathscr I^+_u}-\phi^{\sigma^{(n)}_i}_{\mathscr I^+_u}\right].
    \label{eq:swap_operator}
\end{equation}

So see why $U\left(\sigma^{(n)}\right)$ is of interest, one can analyze its expectation value under an initial state $\ket{\varphi_-}$, according to an asymptotic observer that has access to retarded time up to $u$.

The `standard’ method (to be generalized below) for computing the expectation value of \emph{any} asymptotic observable $\mathcal{O}_{\mathscr{I}^+_u}^{(n)}$ in the $n$-copy system proceeds as follows: Each portion of $\mathscr I^+_u$ must be completed into a Cauchy slice $\Sigma_u$ by attaching an \emph{Interior} spacelike surface $\Sigma_\text{Int}$. This allows for the computation of the reduced density matrix $\hat\rho^{(n)}$ of the $n$-copy system by partially tracing $\ket{\varphi_-}\bra{\varphi_-}^{\otimes n}$ over inaccessible states, such as the peaked states $\ket{\phi^i_{\Sigma_\text{Int}}}=\ket{\phi^1_{\Sigma_\text{Int}}, \dots, \phi^n_{\Sigma_\text{Int}}}$ defined on $\Sigma_\text{Int}$. (Refer to FIG. \ref{fig:Hawking-Penrose} for reference). The `standard' computation can therefore be performed by means of functional integrals \emph{via}\footnote{The following trace operations are technically bold, as local von Neumann algebras in non-trivial, non-gravitational QFTs are of type III$_1$ \cite{10.1143/PTP.32.956}. Nonetheless, even in such cases, formal path integral manipulations often yield correct results and are therefore widely used. \cite{Witten:2018zxz}}
\begin{gather}
    \expval{\mathcal O^{(n)}_{\mathscr I^+_u}}_u:=\text{tr}_{\mathscr I_u^+}\left(\hat\rho^{(n)}(u)\mathcal O^{(n)}_{\mathscr I^+_u}\right
    )=\text{tr}_{\mathscr I_u^+}\left(\tr_{\Sigma_\text{Int}}\left(\ket{\varphi_-}\bra{\varphi_-}^{\otimes n}\right)\mathcal O^{(n)}_{\mathscr I^+_u}\right)\nonumber\\
    =\nonumber\\
    \int\mathrm D^n\Big[\phi_{\mathscr I_u^+}\Big]\mathrm D^n\Big[\phi^*_{\mathscr I_u^+}\Big]\mathrm D^n\Big[\phi_{\Sigma_\text{Int}}\Big]\prod_i\braket{\phi^i_{\mathscr I_u^+},\phi^i_{\Sigma_\text{Int}}}{\varphi_-}\braket{\varphi_-}{\phi^{i^*}_{\mathscr I_u^+},\phi^i_{\Sigma_\text{Int}}}\bra{\phi^{1^*}_{\mathscr I_u^+},\dots,\phi^{n^*}_{\mathscr I_u^+}}\mathcal O^{(n)}_{\mathscr I^+_u}\ket{\phi^1_{\mathscr I_u^+},\dots,\phi^n_{\mathscr I_u^+}}\nonumber\\
    =\nonumber\\
    \int\mathrm D^n\Big[\phi_-\Big]\mathrm D^n\Big[\phi^*_-\Big]\times\nonumber\\
    \times\prod_i \braket{\phi^i_-}{\varphi_-}\braket{\varphi_-}{\phi^{i^*}_-}\times\nonumber\\
    \times\int\mathrm D^n\Big[\phi_{\mathscr I_u^+}\Big]\mathrm D^n\Big[\phi^*_{\mathscr I_u^+}\Big]\bra{\phi^{i^*}_{\mathscr I_u^+}}\mathcal O^{(n)}_{\mathscr I^+_u}\ket{\phi^i_{\mathscr I_u^+}}\left(\int\mathrm D^n\Big[\phi_{\Sigma_\text{Int}}\Big]\prod_i\braket{\phi^{i^*}_-}{\phi^{ i^*}_{\mathscr I_u^+},\phi^i_{\Sigma_\text{Int}}}\braket{\phi^i_{\mathscr I_u^+},\phi^i_{\Sigma_\text{Int}}}{\phi^i_-}\right),
    \label{eq:expectation_value_fixed_topo}
\end{gather}
Where in the last line resolutions of the identity using states $\ket{\phi_-}$ peaked in field configuration on every $\mathscr I^-$ were introduced.

\begin{figure}[h!]
    \centering
    \includegraphics[width=0.5\textwidth]{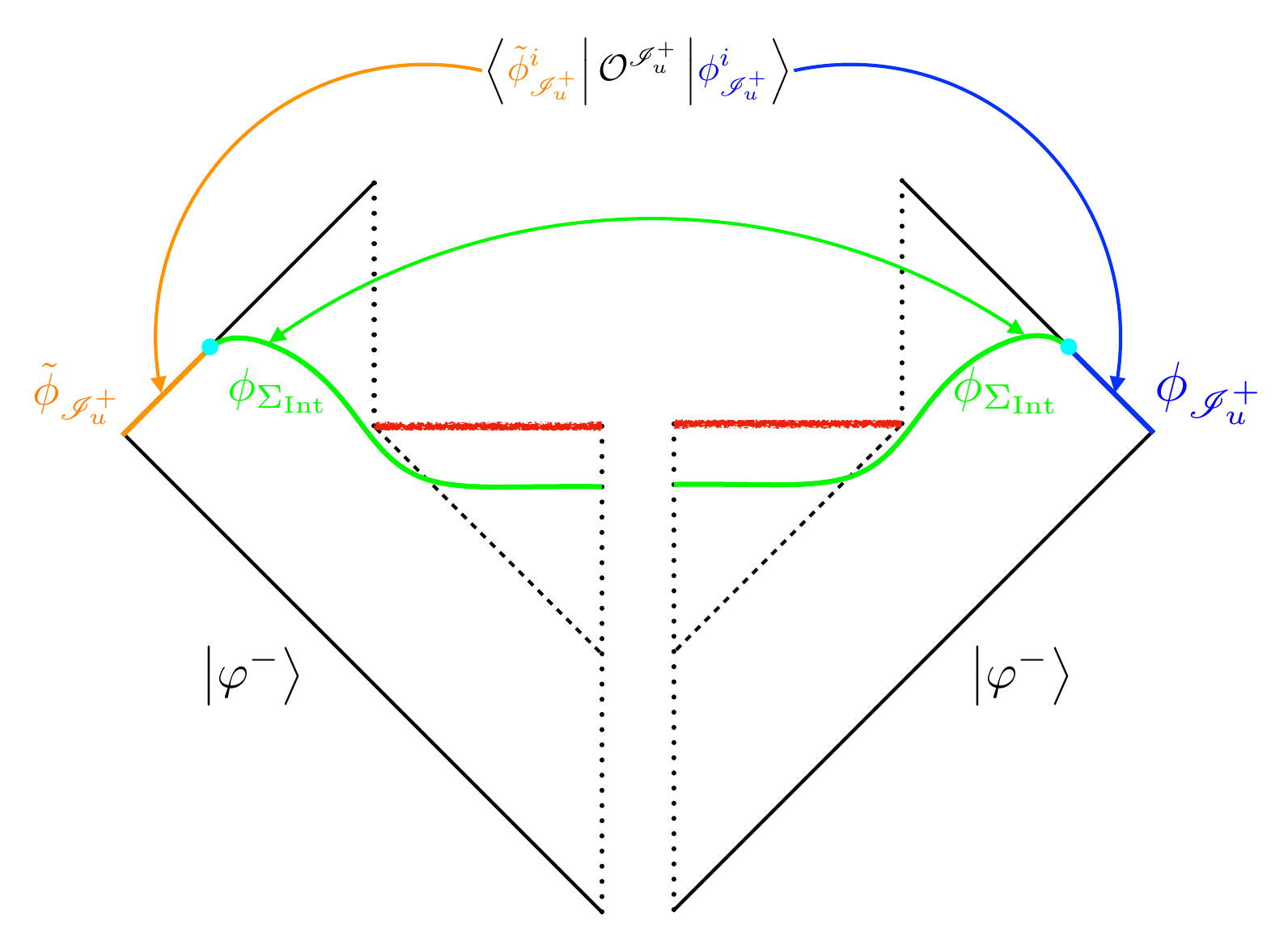}
    \caption{Depiction of the path integral computing the expectation value of an observable $\mathcal O^{(n)}_{\mathscr I^+_u}$, with respect of the initial state $\ket{\varphi^-}$, for the case $n=1$. If $\mathcal O^{(n)}_{\mathscr I^+_u}=\mathbb 1_{\mathscr I^+_u}$, the path integral computes the normalization constant $\text{Tr}\left(\rho^{(1)}(u)\right)$.}
    \label{fig:normalization}
\end{figure}

The last expression can be read in a way that will be useful later, and illustrated in FIG. \ref{fig:normalization} for the simplest case $n=1$, namely:
\begin{itemize}
    \item The middle line takes care of preparing peaked states $\ket{\phi_-^i}$ on each $\mathscr I^-$ from the initial states $\ket{\varphi_-}$ and similarly for dual states. Note that relevant to reproduce Hawking's results is the case in which $\ket{\phi_-}$ is the vacuum state in Minkowski spacetime, then $\braket{\phi_-^i}{\varphi_-}$ is the vacuum wave-functional and can be computed with a path integral on half of Euclid's space \cite{Weinberg_1995}.
    
    \item The lower line, read from right to left consists of the probability amplitude for the peaked state $\ket{\phi^i_-}$ to propagate to the peaked state $\ket{\phi^i_{\mathscr I_u^+},\phi^i_{\Sigma_\text{Int}}}$ and for $\ket{\phi^{i^*}_-}$ to propagate to $\ket{\phi^{i^*}_{\mathscr I_u^+},\phi^i_{\Sigma_\text{Int}}}$ backwards in time. The former can naturally be computed, up to a normalization constant, with a path integral of $e^{\imath S\left[\phi^i_-;\phi^i_{\mathscr I_u^+},\phi^i_{\Sigma_\text{Int}}\right]}$ over bulk configurations and similarly for the later but using $e^{-\imath S^*\Big[\phi^{i^*}_-;\phi^{i^*}_{\mathscr I_u^+},\phi^i_{\Sigma_\text{Int}}\Big]}$, with $S^*$ the CPT conjugate of the action. Since configurations are identified at the interior portion of the partial slices and there is summation over them, one can equivalently consider just one path integral over one larger spacetime glued along the interior sub-slice. This then computes the transition amplitude of the boundary configurations at portions of the two $\mathscr I^+$. The product then gives the total amplitude $\mathcal W\left[\left\{\phi_{\mathscr I_u^+}^i\right\},\left\{\phi_{\mathscr I_u^+}^{ i^*}\right\}\right]$ over the copies. A simple rearrangement shows that $\mathcal W$ can be understood providing the matrix elements of a reduced density matrix, \emph{viz.}%\footnote{Observe that taken together, these last two points indicate that the formulation corresponds to a path integral on a Schwinger-Keldysh contour.}
    \begin{equation}
        \mathcal W\left[\left\{\phi_{\mathscr I_u^+}^i\right\},\left\{\phi_{\mathscr I_u^+}^{i^*}\right\}\right]=\bra{\phi_{\mathscr I_u^+}^1,\dots,\phi_{\mathscr I_u^+}^n}\text{tr}_{\Sigma_\text{Int}}\left(\bigotimes_{i=1}^n\ket{\phi_-^i}\bra{\phi_-^i}\right)\ket{\phi_{\mathscr I_u^+}^{ 1^*},\dots,\phi_{\mathscr I_u^+}^{n^*}}.
        \label{eq:reduced_density_decoupled}
    \end{equation}
    Finally one integrates these amplitudes, weighed by the operator's matrix elements, over all such boundary configurations. 

    \item And as a last step, one integrates over all field configurations in the $\mathscr I^-$'s —upper line.
\end{itemize} 

Returning to the case $\mathcal O^{(n)}_{\mathscr I^+_u}=U\left(\sigma^{(n)}\right)$, the delta functional implements a cyclic gluing of the different asymptotic regions. When $n=1$, this amount to gluing the right and left $\mathscr I^+_u$'s in FIG. \ref{fig:normalization} directly, and since for $n=1$, $U\left(\sigma^{(n)}\right)=\mathbb{1}_{\mathscr I^+_u}$, the final result should therefore be the trace of the reduced density matrix $\hat\rho^{(1)}(u)$, \emph{i.e.} 1. Note that to obtain this through path integral calculations one would need to keep track of all normalizations. As it is easier to instead normalize final results, this will be the approach taken from now on: then the path integral computation just discussed actually computes the normalization constant $r$ of the operator
\begin{equation}
    \rho^{(1)}(u)=:r\hat\rho^{(1)}(u).
\end{equation}
The above discussion regarding $n=1$ can also be phrased slightly differently. One can now consider
$$\mathcal O^{(n)}_{\mathscr I^+_u}=\ket{\phi''_{\mathscr I^+_u}}\bra{\phi'_{\mathscr I^+_u}},$$
then \eqref{eq:expectation_value_fixed_topo} computes the matrix element $\bra{\phi'_{\mathscr I^+_u}} \rho^{(1)}(u)\ket{\phi''_{\mathscr I^+_u}}$. The integrals over field configurations in the $\mathscr I^+_u$'s are killed by the deltas of the matrix elements of $\mathcal O^{(n)}_{\mathscr I^+_u}$, which also fix the boundary conditions of the fields to the primed and double primed configurations. Thus, gluing these asymptotic regions in the $\expval{U\left(\sigma^{(n)}\right)}_u$ calculation amounts to setting left and right indices of $\rho^{(1)}(u)$ to be equal and summing over them, so one is indeed performing the trace of $\rho^{(1)}(u)$.

\begin{figure}[h!]
    \centering
    \begin{subfigure}[c]{0.4\textwidth}
        \centering
        \includegraphics[width=\linewidth]{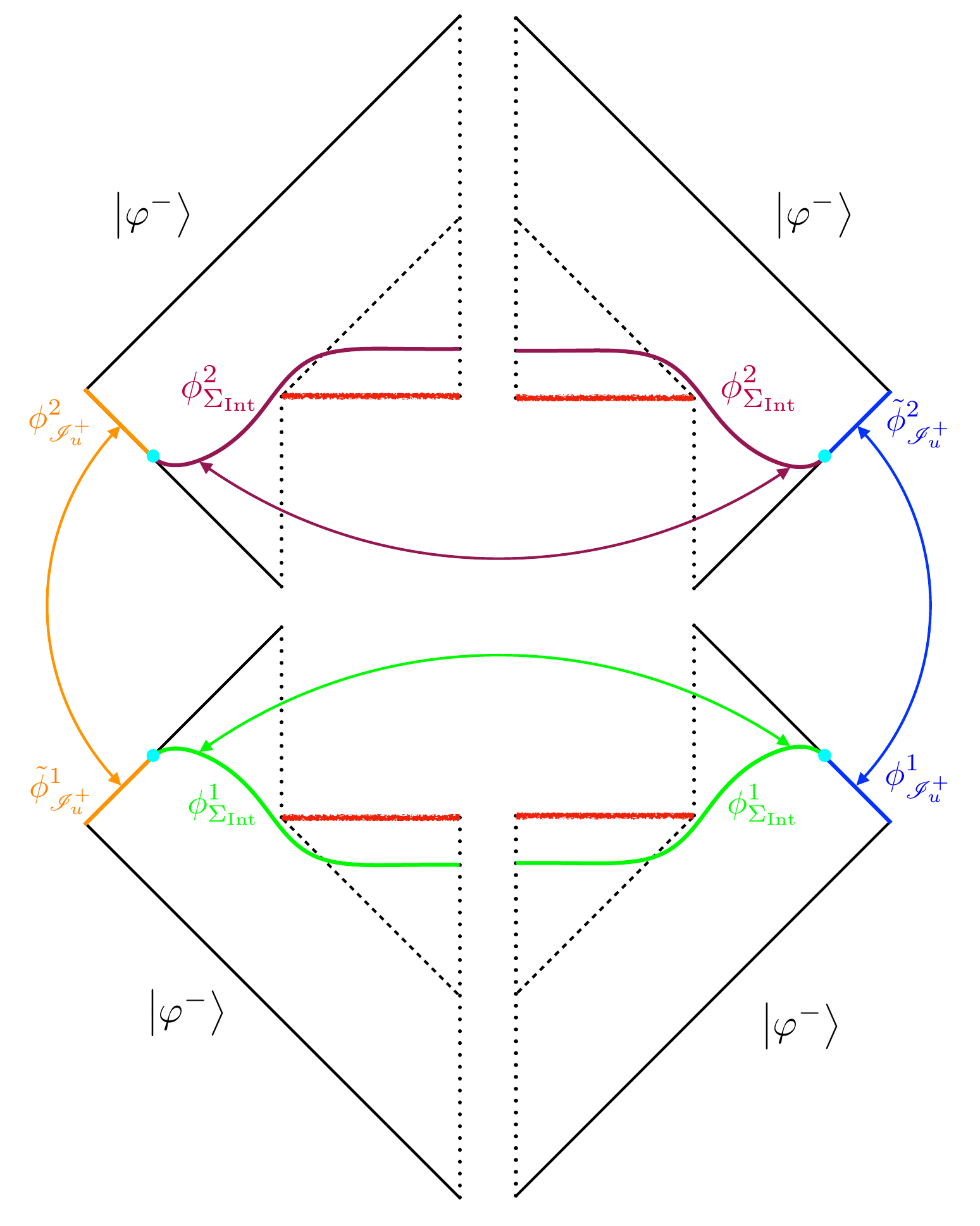}
        \caption{Contribution of the Hawking topology to the path integral $\text{Tr}\left(\rho^{(n=2)}(u)\right)$. If this were the only contribution, the swap entropy would coincide with the $(n=2)-$Rényi entropy of the Hawking radiation collected by the asymptotic observer.}
        \label{sfig:Hawking2}
    \end{subfigure}
    \begin{subfigure}[c]{0.54\textwidth}
        \centering
        \includegraphics[width=\linewidth]{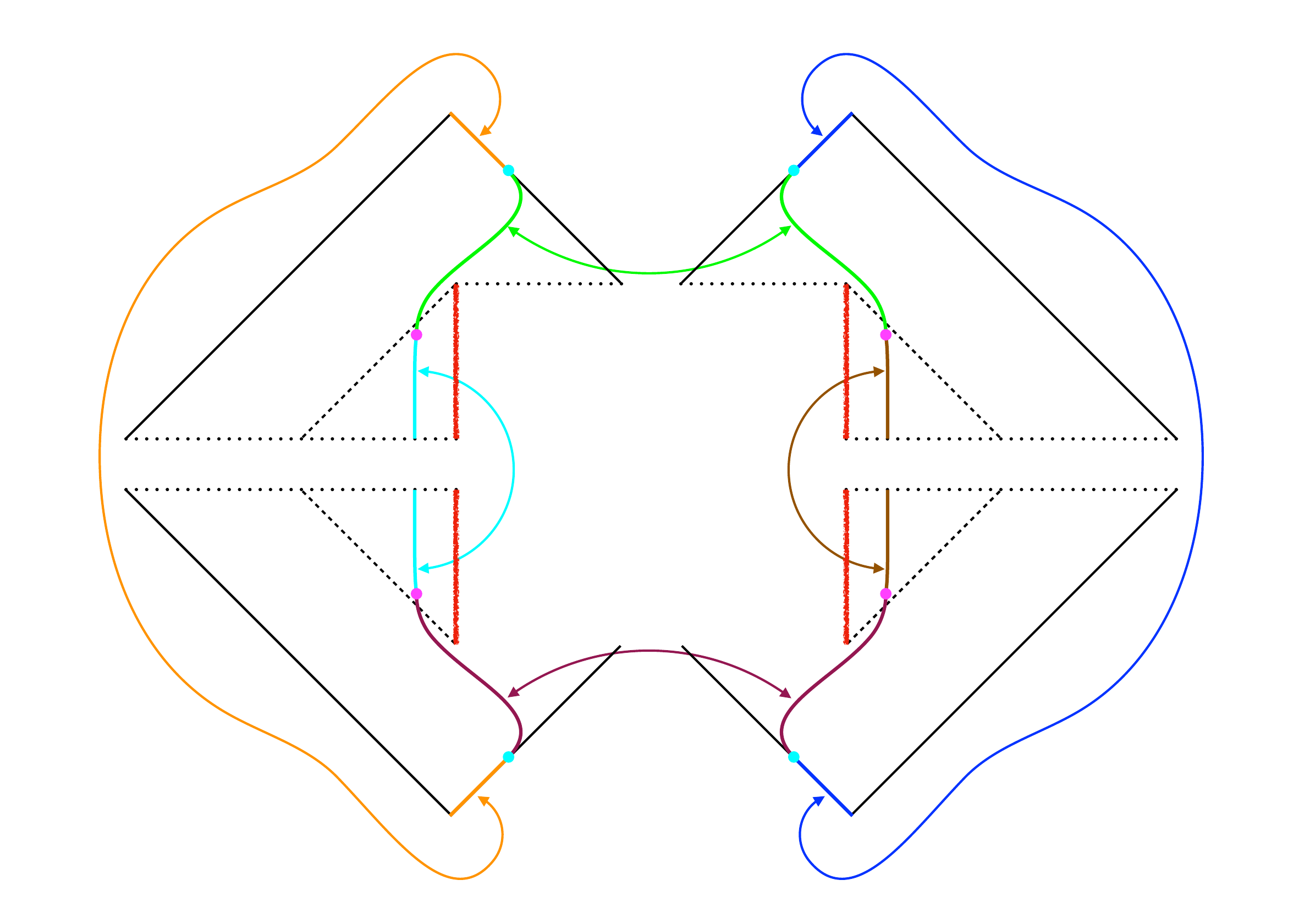}
        \caption{Contribution of the wormhole topology to the path integral $\text{Tr}\left(\rho^{(n=2)}(u)\right)$. This is the natural topology for the computation of the $(n=2)-$Rényi entropy of the bulk region outside the island.}
        \label{sfig:swap2}   
    \end{subfigure}
    \label{fig:swap2}
    \caption{Different fixed-topology path integrals appearing in the calculation of the swap entropy.}
\end{figure}

This new perspective can be now used to interpret the $n=2$ calculation. In such case $\sigma^{(n)}$ is non-trivial and \emph{a} generalization of the $n=1$ computation is now illustrated by FIG. \ref{sfig:Hawking2}, so it follows from the above discussion that the lower and upper parts compute matrix elements of $\rho^{(1)}$ and the side gluing perform a matrix multiplication and trace, so
\begin{equation}
    \expval{U\left(\sigma^{(n=2)}\right)}_u=\text{Tr}\left(\left(\hat\rho^{(1)}\right)^{n=2}\right)=\frac1{r^{n=2}}\text{Tr}\left(\left(\rho^{(1)}\right)^{n=2}\right).
    \label{eq:sigma_expec_H2}
\end{equation}
After this discussion, the claim above regarding how the expectation values $U\left(\sigma^{(n)}\right)$ \emph{may} encode the entropy can be verified. Indeed, consider the \emph{swap} entropy, defined by
\begin{equation}
    S^{(n)}_\text{Swap}\left[\rho^{(n)}(u)\right]:=-\frac1{n-1}\log\langle U\left(\sigma^{(n)}\right)\rangle_u.
    \label{eq:swap_entropy}
\end{equation}
%
%It is important to note that the path integral used to compute the right hand side of eq. \eqref{eq:swap_entropy}, namely the one depicted in FIG. \ref{sfig:Hawking2}, needs to be regularized since it diverges with the area of the celestial sphere at retarded time $u$,\footnote{\color{red} Other explanation} as will become clear later. As such, the divergence is not dependent on the physics of interest and in fact is also there for Minkowski spacetime so nothing relevant is lost when regularizing. The regularization can be done, \emph{e.g.} by subtracting the Minkowski divergence.

Equation \eqref{eq:sigma_expec_H2} implies that $S^{(2)}_\text{Swap}$ is the $n=2$ Rényi entropy. Notably, it is expressed in terms of the expectation value of an observable, giving it a direct operational definition.

In general, using the computation schemes defined above one finds that $S^{(n)}_\text{Swap}$ is the $n$-Rényi-entropy:
\begin{equation}
   S^{(n)}_\text{Rényi}[\rho]:=-\frac1{n-1}\log\text{Tr}\left(\hat\rho^n\right)=-\frac1{n-1}\log\frac{\text{Tr}\left(\rho^n\right)}{(\text{Tr}\rho^{(1)})^n}.
   \label{eq:Rényi}
\end{equation}
%
%As mentioned above, the path integral computing the right hand side of \eqref{eq:Rényi} needs to be normalized to provide the proper result. This is done by a factor of $\frac 1{r^n}=\frac1{(\text{Tr}\rho)^n}$ inside the trace, with $r$ computed as instructed by FIG. \ref{fig:normalization} —\emph{cf.} discussion around it.
 
$S^{(n)}_\text{Swap}$ can then be formally analytically continued\footnote{In general, there may be several ways to perform the extension. However, under some assumptions uniqueness can be guaranteed —for example those of \emph{Carlson's theorem}.} to non integer values of $n$ and its $n\to 1^+$ limit may be taken. And as is well known, this limit of Rényi entropies produces that of von Neumann. The physical interpretation of such limit is that it describes what would be deduced by the experimentalist to be the entropy of a single spacetime from the replicated experiment. If nothing from the discussion above were modified, then the asymptotic observer would deduce the von Neumann entropy for the single black hole and thus be in agreement, in the gravitational saddle point approximation, with Hawking's result, violating unitarity.

However, in quantum gravity, the discussion above may need to be generalized, specifically \eqref{eq:reduced_density_decoupled} requires revision. As will be seen, this generalization would lead to correlations between the $n$ copies and thus to a departure from the $n$-Rényi entropy.

As mentioned above, $\mathcal W$ can be computed using a Feynman path integral over the radiation fields, but also over gravity, \emph{i.e.} geometry. If the gravitational path integral is to include a sum over topologies, consistent with the rationale of summing over geometries, which are not just metrics, but (manifold,metric) pairs, then there may be gravitational saddles which do not correspond to a decoupled density matrix such as $\bigotimes_{i=1}^n\ket{\phi_-^i}\bra{\phi_-^i}$. Indeed, one such possibility would be to have a saddle in a spacetime topology such as the one shown in FIG. \ref{sfig:swap2}. In other words, the calculation laid out above was fixing a particular topology of the spacetime in the path integral computing $W$ and there may be other topologies that have saddle points with the same boundary data on the partial future null infinities. 

Expressed differently, expectations of a quantum gravity path integral suggest that what one should consider instead of eq. \eqref{eq:expectation_value_fixed_topo} is
\begin{gather}
        \expval{\mathcal O^{(n)}_{\mathscr I^+_u}}_{\hat\rho^{(n)}}= \int\mathrm D^n\Big[\phi_-\Big]\mathrm D^n\Big[\phi^*_-\Big]\times\nonumber\\
        \times\prod_i \braket{\phi^i_-}{\varphi_-}\braket{\varphi_-}{\phi^{i^*}_-}\times\nonumber\\
    \times\int\mathrm D^n\Big[\phi_{\mathscr I_u^+}\Big]\mathrm D^n\Big[\phi^*_{\mathscr I_u^+}\Big]\bra{\phi^{1^*}_{\mathscr I_u^+},\dots,\phi^{n^*}_{\mathscr I_u^+}}\mathcal O^{(n)}_{\mathscr I^+_u}\ket{\phi^1_{\mathscr I_u^+},\dots,\phi^n_{\mathscr I_u^+}}\mathcal W\left[\left\{\phi_{\mathscr I_u^+}^i\right\},\left\{\phi_{\mathscr I_u^+}^{ i^*}\right\}\right],
    \label{eq:expectation_value_var_topo}
\end{gather}
with
\begin{equation}
    \mathcal W\left[\left\{\phi_{\mathscr I_u^+}^{i^*}\right\},\left\{\phi_{\mathscr I_u^+}^i\right\}\right]=\sum_\mathcal M\int\mathrm D[\phi_\text{Bulk}]e^{\imath S_\text{Matter}+\imath S_\text{Gravity}},
    \label{eq:reduced_density_var_topo}
\end{equation}
where the sum is over manifolds and the integral over matter fields as well as the metric, and the appropriate CPT conjugation has taken place. This expression is naturally just a ``statement of intent'' \cite{Ambjorn:2024pyv}, and making sense of it is the goal of approaches like QRC or spinfoams. It is only used in replica calculations of this section in that it motivates considering contributions from gravitational saddle point evaluations on non-trivial topologies like that of \ref{sfig:swap2}. 

For the calculation of the swap entropy, it is therefore of interest to consider \emph{replica wormhole} geometries like those of \ref{sfig:swap2}. The first thing to notice is that such a non-trivial topology is incompatible with an everywhere smooth metric of Lorentz signature. Indeed, note that the \emph{bifurcation} or \emph{entangling} surface connecting all spacetime branches (bulk/pink dot) has a singular causal structure: attached to it there are four light cones, instead of two. And for any $n$ one finds $2n$ light cones. As a result, the Ricci curvature is distributional, which, as discussed in\footnote{See also \cite{Lewkowycz:2013nqa} and \cite{Sorkin:2019llw} for Euclidean and discrete counterparts, respectively.} \cite{Neiman:2013ap,Bodendorfer:2013hla,Neiman:2013lxa,Marolf:2020rpm,Colin-Ellerin:2020mva}, gives an imaginary contribution to the Einstein-Hilbert action\footnote{In principle, this expression has a sign ambiguity, corresponding to a branch cut in the gravitational action. The branch chosen here not only yields a reasonable result but also aligns with the “fluctuation convergence criterion” discussed in recent and earlier works \cite{HalliwellContours,Louko:1995jw,Kontsevich:2021dmb,Witten:2021nzp,Lehners:2021mah,Jonas:2022uqb,Lehners:2022xds}. For a more detailed discussion, see \cite{Dittrich:2024awu} and references therein.}
\begin{equation}
    \text{Im}S^{(n)}_\text{EH}=(n-1)\frac A{4G},
    \label{eq:imaginary_contribution}
\end{equation}
with $A$ the area of this (light-)conical singularity.

In part motivated by the replica calculations under discussion, but also by circumventing the conformal factor problem in gravitational thermodynamics \cite{Marolf:2022ybi}, as well as their natural appearance in discrete quantum gravity \cite{Asante:2021phx,DittrichConfig:2025} or in calculations of the dimension of quantum gravitational Hilbert spaces \cite{Dittrich:2024awu}; this mild, finite action\footnote{Note that the singularities are potentially problematic, because when the number of light cones is less than two, they give exponential enhancement to their respective histories. So they could, in principle, yield an uncontrollably divergent integral. However, the works \cite{Marolf:2022ybi,Dittrich:2024awu} have explored mechanisms in which the enhancement is not fatal, and even sensible.} type of conical singularities has been conjectured to be relevant when treating Lorentzian gravitational path integrals —see also the two-dimensional work of \cite{Louko:1995jw} dealing with real-time topology change. Their appearance is indeed striking, but maybe just as much as histories with backward-in-time or superluminal segments in the free relativistic particle path integral \cite{987181}, or the dominant role of continuous yet nowhere smooth paths in non-relativistic path integrals, such as that for the harmonic oscillator \cite{gall2016brownian}.

Therefore, for any given wormhole topology, once all matter degrees of freedom are integrated, one has the contribution to the swap operator expectation value
\begin{equation}
    \frac1{r^n}\int\mathrm D\left[g^{(n)}\right]e^{\imath S^{(n)}_\text{Gravity}\left[g^{(n)}\right]+W^{(n)}_\text{Eff}\left[g^{(n)}\right]},
    \label{eq:semi_effective_patt_Integral}
\end{equation}
where the matter effective action $W_\text{Eff}$ is defined as the logarithm of the matter path integral for a given replica wormhole metric $g$, which includes also an integral over the vacuum wave-functional's argument. Note that the normalization constant has the same path integral form, \emph{viz.}
\begin{equation}
    r=\int\mathrm D\left[g^{(1)}\right]e^{\imath S^{(1)}_\text{Gravity}\left[g^{(1)}\right]+W^{(1)}_\text{Eff}\left[g^{(1)}\right]},
\end{equation}
except that the topology is that of FIG. \ref{fig:normalization}.

These are the integrals to which the saddle point approximation will be employed. (Note however, that the matter path integral —which may be considered to include gravitons— is performed exactly.)

One therefore has that in a semiclassical limit, the contribution of these \emph{assumed} saddles to $\expval{U\left(\sigma^{(n)}\right)}_u$ would be 
\begin{equation}
    \frac1{r_*^n}e^{\imath S^{(n)}_\text{EH}\left[g^{(n)}_*\right]+W^{(n)}_\text{Eff}\left[g^{(n)}_*\right]},
    \label{eq:wormhole_contribution}
\end{equation}
with
\begin{equation}
    r_*\approx e^{\imath S^{(1)}_\text{EH}\left[g^{(1)}_*\right]+W^{(1)}_\text{Eff}\left[g^{(1)}_*\right]}.
    \label{eq:semiclassical_norm}
\end{equation}
\emph{If} these putative saddles $g^{(n)}_*$ are replica symmetric, meaning the corresponding geometries are symmetric under permutations of the $n$ copies and therefore are captured by the geometry of one copy, which will be considered as captured by $g^{(1|n)}$. Then, one has (see \cite{Marolf:2020rpm,Colin-Ellerin:2020mva} —and also \cite{Lewkowycz:2013nqa}— for the continuum result, but in \S\ref{ssec:Regge_action} this will be explicitly shown in the discrete setup)
\begin{equation}
    S^{(n)}_\text{EH}[g^{(n)}_*]=n S^{(1)}_\text{EH}\left[g^{(1|n)}_*\right]+\imath (n-1)\frac {A[i_\gamma]}{4G},
    \label{eq:S_EH_additivity}
\end{equation}
with $i_\gamma$ the metric of the splitting surface $\gamma$. 

One can write $W^{(n)}_\text{Eff}[g^{(n|1)}]$ in similar terms by the defining $\Delta W_\text{Matter}$ via
\begin{equation}
    W^{(n)}_\text{Eff}\left[g^{(n)}_*\right]=:n W^{(1)}_\text{Eff}\left[g^{(1|n)}_*\right]-(n-1)\Delta W_\text{Matter}[i_\gamma].
    \label{eq:W_Eff_additivity}
\end{equation}
Physically, $\Delta W_\text{Matter}$ can be understood as the matter $n$-Rényi entropy that could be captured in $\Sigma_\text{Int}\setminus\mathcal I$, where $\mathcal I$ (referred to as an $\mathcal I$sland in other contexts) is the region in $\Sigma_\text{Int}$ inside the splitting surface. This follows from a similar reasoning to the one above: For example, in the case $n=2$ (FIG. \ref{sfig:swap2}) the path integral (with a fixed metric) can be understood as the purity of the un-normalized reduced state in the Hilbert space of states in $\Sigma_\text{Int}\setminus \mathcal I$ and upon normalization the first term in the effective action cancels exactly —note that for this cancellation the assumption of having a \emph{fixed and replica symmetric} spacetime is necessary.

%Note that since this relies on the cancelation upon normalization, the replica symmetry below is of greater importance.

The reason $\Delta W_\text{Matter}$ only depends on $i_\gamma$ comes from the fact that diffeomorphism invariance of the path integral ensures that the outcome is independent of the choice of $\Sigma_u$ and depends only on the geometries of the surfaces singled out in the calculation: that of the splitting surface $\gamma$ and of the celestial sphere at $u$, $S^2_u$.

Thus, $g^{(n)}_*$, being a saddle, is such that
\begin{equation}
    0=n\frac{\delta}{\delta g^{(1|n)}_*}\left(\imath S^{(1)}_\text{EH}+W^{(1)}_\text{Eff}\right)-(n-1)\frac{\delta S_\text{Gen}}{\delta i_\gamma},
\end{equation}
where $\Delta W$, motivated by its physical meaning, was grouped with the area term into the generalized entropy $S_\text{Gen}=\frac A{4G}+\Delta W_\text{Eff}$ .

This expression provides a sufficient condition for $g^{(n)}$ to be a replica symmetric saddle, namely that its `reduced' part $g^{(1|n)}$ is a one copy saddle $g^{(1)}_*$ \emph{and} importantly that this one-copy spacetime has a so-called \emph{Quantum Extremal Surface} (QES), meaning that the variations of the generalized entropy with respect to $\gamma$ vanish. 

\emph{If} these \emph{replica saddles} exist and are of this form, then it is a matter of simplifying \eqref{eq:wormhole_contribution} using \eqref{eq:semiclassical_norm}, \eqref{eq:S_EH_additivity} and \eqref{eq:W_Eff_additivity} to see that they contribute semiclassically to $\expval{U\left(\sigma^{(n)}\right)}_u$ with simply $e^{-(n-1)S_\text{Gen}}$.

Other than the replica saddles, there may also be the Hawking saddles, which as argued above produce Rényi entropies. So \emph{if}\footnote{This assumption will be maintained throughout this work. However, note that more complex topologies appear to be relatively suppressed according to equation \eqref{eq:imaginary_contribution}, making this assumption reasonable. That said, this may require revisiting, as the core idea of the replica mechanism is that the replica topology contribution can be comparable to that of the Hawking topology.} only these topologies give dominant contributions, what has been argued is that in a semiclassical limit
\begin{equation}
    \expval{U\left(\sigma^{(n)}\right)}_u\approx e^{-(n-1)S^{(n)}_\text{Rényi}}+e^{-(n-1)S_\text{Gen}}\approx e^{-(n-1)\min\left\{S^{(n)}_\text{Rényi},S_\text{Gen}\right\}}.
\end{equation}
And therefore (\emph{cf.} eq. \eqref{eq:swap_entropy}),
\begin{equation}
    S_\text{Swap}^{(n)}(u)=\min\left\{S^{(n)}_\text{Rényi}(u),S_\text{Gen}(u)\right\}
    \label{eq:swap_n_Page}
\end{equation}
Upon taking the $n\to 1^+$ limit, this becomes
\begin{equation}
    S_\text{Swap}:=\lim_{n\to 1^+}S_\text{Swap}^{(n)}(u)=\min\left\{S_\text{vN}(u),S_\text{Gen}(u)\right\},
    \label{eq:swap_Page}
\end{equation}
with $S_\text{vN}$ the von Neumann entropy of $\mathscr I^+_u$.

As argued above, if only the Hawking topology were to contribute, then one would have $S_\text{Swap}(u)=S_\text{vN}(u)$ for all $u$, \emph{i.e.} the picture described by Hawking in which the entropy grows indefinitely. However, eq. \eqref{eq:swap_n_Page} shows that the swap entropy generalizes Rényi entropies and may not give the $n$-Rényi entropy for a state $\rho^{(n)}\neq\rho^{\otimes n}$.

Now, as $S_\text{vN}$ starts begin zero, it should dominate in \eqref{eq:swap_Page} when $u$ is small enough. However, as the black hole evaporates $S_\text{Gen}=\frac A{4G}+\Delta W_\text{Matter}$ is expected to go to zero, because:
\begin{enumerate}
    \item For large enough retarded times, the QES is expected to lie close to the horizon and therefore $A(u)$ approaches zero as $u$ goes to the Ev\emph{aporation} time $u_\text{Ev}$ \cite{Penington:2019npb,Almheiri:2019psf}.

    \item As noted above $\Delta W_\text{Matter}$ can be interpreted as the $n$-Rényi entropy of radiation that could be captured in $\Sigma_\text{Int}\setminus\mathcal I$. Considering the previous point this means that in the limit $u\to u_\text{Ev}$, this matches that of the radiation that has not been captured in $\mathscr I^+_u$. But as $u$ approaches $u_\text{Ev}$, the black hole is assumed to totally evaporate and therefore it is reasonable to expect that all radiation has been captured by this time, thus this contribution to the swap entropy should also approach zero.\footnote{Interestingly, as detailedly pointed out in \cite{Marolf:2020rpm}, in the $u\to u_\text{Ev}$ limit the replica saddles extrapolate to the ``Polchinski-Strominger saddles'' discussed in \cite{Polchinski:1994zs}, which provide a path integral geometric interpretation of why this contribution vanishes: In this limit the path integrals effectively `un-swap' and exactly cancel with the normalization path integrals. In fact, the resolution to the information paradox suggested in \cite{Polchinski:1994zs}, closely resembles the one provided by the replica paradigm, but is nevertheless unsatisfactory under closer look on fronts such as needing to go beyond semiclassical approximations, or hinting at a persistence of unitarity violation —see \cite{Polchinski:1994zs,Marolf:2020rpm} for this analysis.}
   
\end{enumerate}
Therefore, one expects that $S_\text{vN}$ dominates for small retarded times as it increases from zero, and that $S_\text{Gen}$ dominates for large retarded times when it decreases to zero as the black hole evaporates.

In fact, for late times, the Bekenstein-Hawking entropy is expected to dominate in the generalized entropy, so that the Page curve is indeed followed. \cite{Marolf:2021ghr}

In summary, the swap entropy follows the Page curve in the limit $n \to 1^+$. And it is a quantity that depends on an observable of the asymptotic experimentalist conducting $n$ experiments and collecting data before retarded time $u$. This is the operational conclusion of \cite{Marolf:2020rpm} advertised earlier.

Given that $S^{(n)}_\text{Rényi}(u)$ is expected to behave as $S_\text{vN}(u)$, then the same story is anticipated for $S^{(n)}_\text{Swap}$. However, note that several assumptions were made above, in particular, it was assumed that there were, for the one copy spacetimes, saddle geometries that posses a QES. Results supporting their existence for finite $n$, are significantly scarce (but existent: \emph{e.g.} \cite{Mirbabayi:2020fyk} for $n=2$), although the situation is better for $n\approx 1$ (\emph{e.g.} the seminal works \cite{Penington:2019kki,Almheiri:2019qdq}, or \cite{Chandrasekaran:2022asa}). The situation is similar for more general saddles $g^{(n)}_*$. And in any case, their relevance for a Lorentzian setup such as the one considered here does not seem to have been addressed in literature except perhaps for the related discussion \cite{Held:2024qcl}. This is a particularly intriguing question, given that replica saddles cannot be Lorentzian, a consequence of the action being complex \cite{Sorkin:2019llw}, and thus correspond to complex geometries that do not lie in the original real-time integration, so an appropriate deformation needs to be justified —as an example the saddles discussed in \cite{Penington:2019kki,Chandrasekaran:2022asa,Almheiri:2019psf} are \emph{Euclidean}. Additionally most of the work remain detached from 4-dimensional Einstein gravity, remaining in more controllable setups sutch as JT gravity. These issues were a partial motivation for the work herein presented, which it is turn to discuss after having exposed the continuum computational framework.

\section{Overview of the proposal and the first results \label{sec:overview}}

This section gives a non-technical overview of the discrete quantum gravity program to study black hole evaporation in the replica paradigm, and the initial application and results that have been obtained. It thus provides a general summary of the following sections.

Aiming to explore open questions in said context, the proposal is to employ Regge calculus —a lattice-like formulation of general relativity— as a tool for performing \emph{explicit} path integral calculations that are otherwise intractable in the continuum, except in highly simplified cases, such as JT gravity. In practice, this replaces the path integrals from the previous section with finite-dimensional integrals, making computations feasible and allowing for lattice-based evaluations, particularly for the swap entropy. Key questions include whether replica saddles exist for finite $n$ and whether they, being complex, actually contribute to the relevant \emph{real-time} gravitational path integrals. A first step before taking the continuum limit (fine-grained limit of the lattice), would be to evaluate these issues for several triangulations and see if the results are generic or not. Later, one could in principle approach the difficult problems of evaluating corrections to the semiclassical gravitational evaluations and eventually take the continuum limit. This paper begins to address the first step.

In Regge calculus, a continuous spacetime is triangulated by gluing piecewise linear polytopes, such as simplices or hypercubes living on Minkowski spacetime. The geometry of these polytopes, captured by a finite number of degrees of freedom, is then understood to discretize the metric and a path integral over metrics can thus be performed in the discrete as a finite dimensional integral over these geometric parameters.

Thus, the first step in this program is to design an efficient triangulation scheme. This is what \S\ref{ssec:discretization} deals with. The scheme proposed there has the virtues of being applicable to generic Penrose diagrams, but more pertinent to the present discussion, it facilitates contour deformations in discrete path integrals and allows for boundary conditions of mixed signature.

The use of Regge calculus is justified by the fact that to its discrete geometries one can associate a so-called Regge action, which converges to the Einstein-Hilbert action in the continuum limit. \S\ref{ssec:Regge_action} details its definition and computation in full generality, enabling the calculation of the Regge action for any instance of the triangulation scheme using `fundamental modules’ —elemental computational blocks that combine to yield the full action. While these expressions are too long to be included here, they are provided in the GitHub repository \cite{github} together with their computation, for completeness and availability.

To calculate these modular elements, a discrete version of spherical symmetry was imposed, which eliminates the possibility of discrete gravitons running the evaporation process. This motivates the introduction of matter degrees of freedom to the framework and correspondingly a discussion on how matter is treated fully quantum mechanically in Regge calculus, aligning with the swap entropy calculations in the previous section. Particularly \S\ref{ssec:matter} focuses on a minimally coupled, massless, and non-interacting quantum scalar field in this framework. The final expressions relevant to this, and their derivation, are also provided in \cite{github}.

Finally, \S\ref{sec:application} presents a concrete example applying this framework. The calculation assumes metric fluctuations only in regions near the would-be singularity, a physics informed choice that also highlights challenges in handling asymptotic regions. The computation then proceeds in two main steps:
\begin{itemize}
    \item First, matter is integrated out analytically for both the Hawking and wormhole topologies, assuming replica and CPT symmetry. This yields explicit expressions for the matter effective actions, which can be analytically continued in the replica number $n$.  It is the sheer size of these expressions that motivated presenting these and other results (and their explicit computations) in the GitHub repository. The Mathematica output for each effective action is approximately 450MB. Despite this, numerical evaluations are remarkably efficient, typically requiring less than 0.03s on a personal computer.

    \item The second step of the application is a search of semiclassical gravitational saddles on both Hawking and wormhole topologies. This search was performed on a minisuperspace that was replica and CPT-symmetry and further reduced so that in the end only one independent geometric variable remains. This variable corresponds to the area of the splitting surface for the wormhole topology and its analogue for the Hawking topology. This minisuperspace was primarily employed for practical reasons, though it was designed to capture certain anticipated characteristics of evaporating black hole spacetimes that might emerge as effective geometries resulting from partial gravitational integrations.
\end{itemize}
In this simplified scenario, which assumes Euclidean geometry, saddle points were identified in the limit $n\to1^+$ that replicate the Page transition: Specifically, the saddle point relevant to the Hawking topology exhibits a swap entropy component that grows as discrete retarded time progresses. In contrast, the saddle in the wormhole topology gives a decreasing component. Ultimately, these saddles exchange dominance of their contributions to the swap entropy.

The triangulation employed in this study is fairly coarse —although it nevertheless incorporates all essential elements necessary for swap entropy calculations. Considering this, along with the minisuperspace restriction, the result should be regarded primarily as an initial validation step, or optimistically as an insight into the generality of replica saddles. Regardless, these findings justify further advancing this framework to ultimately tackle the driving questions, focusing on enhancing computational infrastructure for more thorough numerical investigations. To this end, this exercise has provided a practical demonstration of how to manage the technicalities inherent in the proposed line of research.  A broader evaluation of the progress made, necessary improvements, and natural extensions are provided in the final section.

\section{Discrete framework for evaporating black holes\label{sec:discrete}}

\subsection{Discretization scheme \label{ssec:discretization}}

Now a discretization scheme of the evaporating black holes and replica wormhole spacetimes described in \S\ref{sec:continuum} will be presented. This has the goal of setting the grounds for a discrete framework that allows one to explore open questions with regards to the replica trick in the context of black hole evaporation and its implications for the Page curve. And also to serve as a stepping stone to bring this paradigm in contact with quantum gravity approaches which rely on discretizations such as spinfoams or quantum Regge calculus. The scheme will be applied to a minimal construction in the next section. As a promising first result, one finds that the corresponding swap entropy displays a Page-like transition. 

Although the presentation will be done in four-dimensional spacetimes, it can easily be modified to work in any.

\begin{figure}
    \centering
    \includegraphics[width=0.5\textwidth]{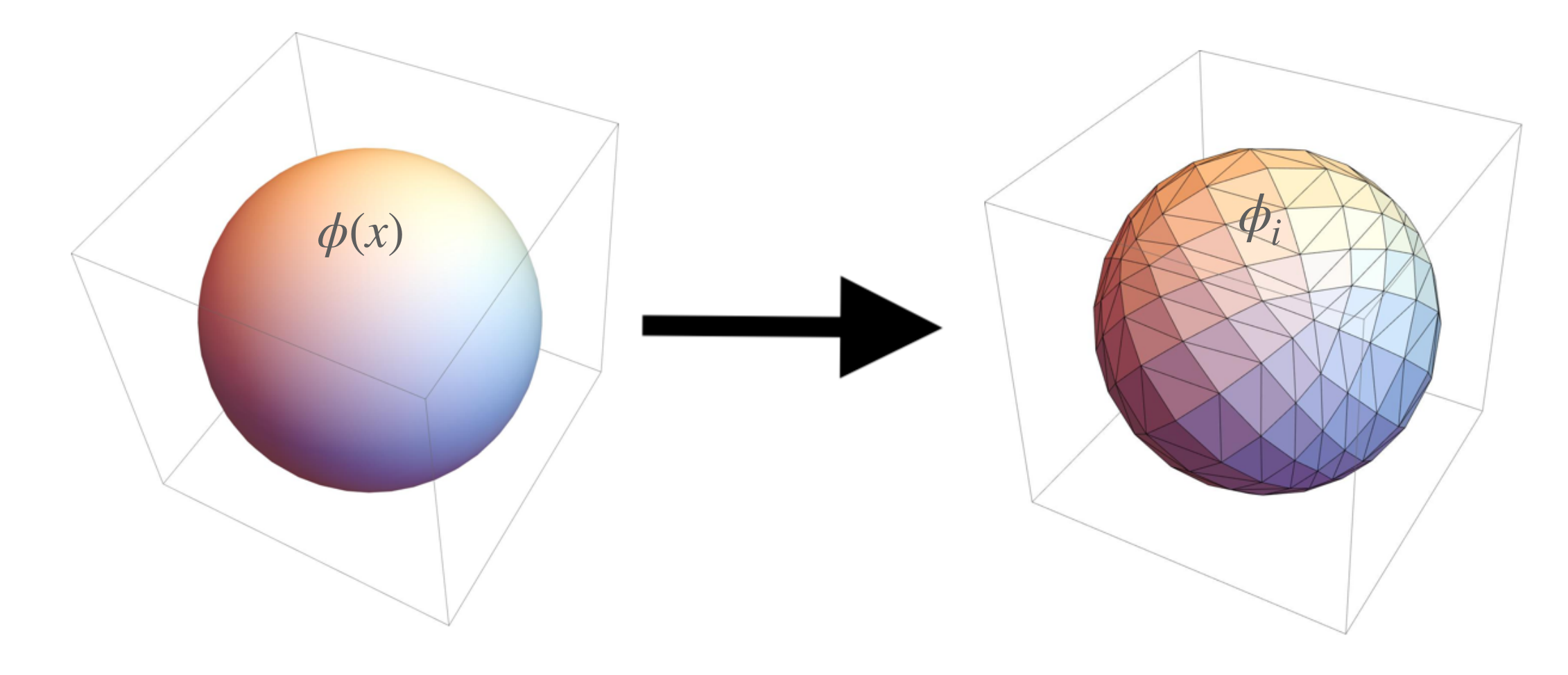}
    \caption{The discretization process exemplified by a scalar field theory on a sphere. The continuum manifold of the sphere is replaced by a triangulation. Upon triangulating, the continuous field $\phi$ labeled by spacetime points $x$ becomes a discrete field labeled by vertex labels $i$. Similarly the metric is replaced by the triangle's lengths.}
    \label{fig:Triangulate}
\end{figure}

The model is discrete in a lattice-like sense: Instead of considering a smooth spacetime and fields on it, it is based on a piecewise linear manifold and (discrete) fields on it. Although it need not be one (not even a simplicial complex), these piecewise linear manifolds are commonly referred to as `triangulations'. As an effect of the discretization, one truncates the infinite degrees of freedom of the starting field theory to a finite number, and thus, either an appropriate continuum limit needs to be addressed, or the discretization should be fine enough so that it captures the physics of interest. An exemplification of how a 2-dimensional theory could be triangulated is presented in FIG. \ref{fig:Triangulate}.

The way fields are defined on a triangulation, and the dynamical features of gravity and matter are discussed in the following two subsections, respectively. But before, the `triangulations' that make up the herein proposed scheme ought to be introduced.

\begin{figure}
    \centering
    \includegraphics[width=0.3\textwidth]{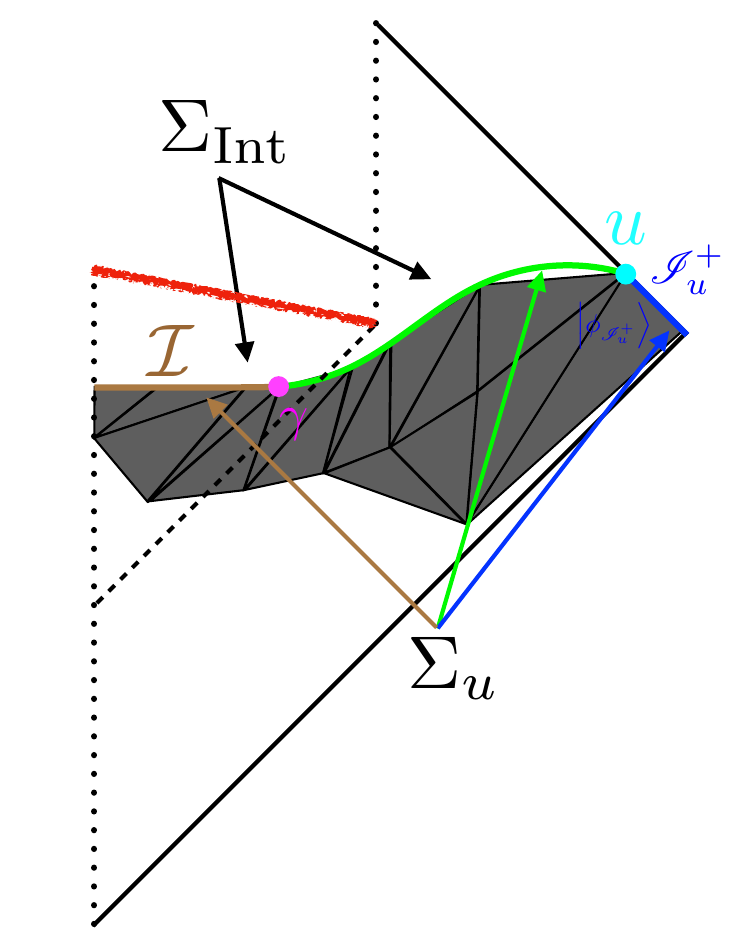}
    \caption{The discretization scheme employs piecewise linear polytopes whose topology corresponds to the triangles in the Hawking-Penrose diagram. While most points in the diagram represent topological spheres, some, such as those in the vertical dotted lines, correspond to actual points. Nevertheless, all cases can be covered with the (class of) polytope(s) discussed in the text. Future work will explore how to deal with the asymptotic regions of spacetime.}
    \label{fig:scheme}
\end{figure}

\begin{figure}
    \centering
    \begin{subfigure}[c]{0.45\textwidth}
        \centering
        \includegraphics[width=\linewidth]{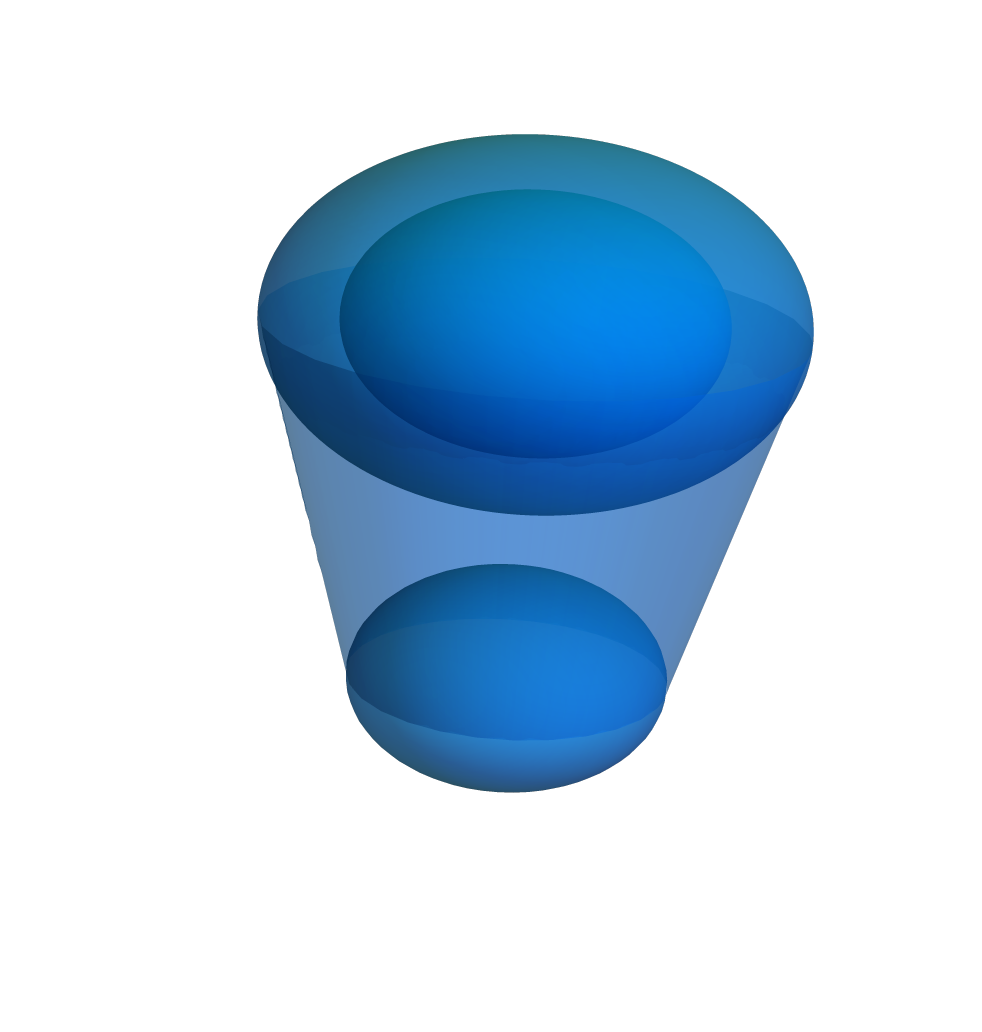}
        \caption{Topology corresponding to triangles in a (Hawking-)Penrose diagram, formed by three (2-spherical) annuli glued cyclically.}
        \label{sfig:continuous_diagram_triangle}
    \end{subfigure}
    \hspace{1.5cm}
    \begin{subfigure}[c]{0.4\textwidth}
        \centering
        \includegraphics[width=\linewidth]{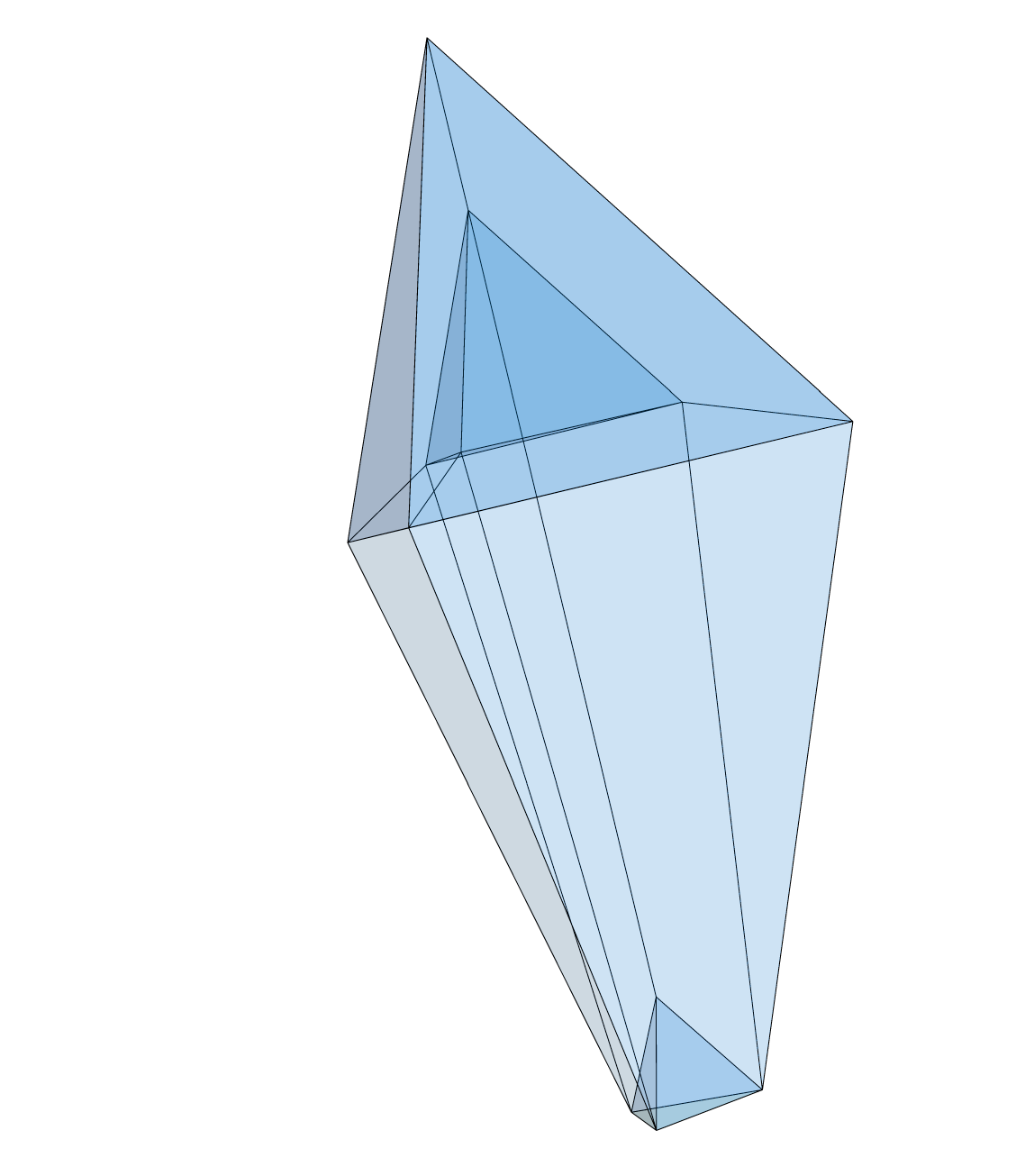}
        \caption{Discrete version of the continuous topology corresponding to a triangle in the (Hawking-)Penrose diagram, formed by three tetrahedral shell annuli glued cyclically. This structure consists of four polytopes, each made up of three triangular frusta (truncated triangular pyramids) glued cyclically. It is these latter fundamental building blocks form the basis of the discretization scheme.}
        \label{sfig:discrete_diagram_triangle}   
    \end{subfigure}
    \label{fig:diagram_triangle}
    \caption{Continuous and discrete depictions of the topology of a triangle in a (Hawking-)Penrose diagram. The cases of diagram triangles with degenerate spheres can be covered by shrinking a sphere (left) or tetrahedral shell (right) to points.}
\end{figure}

The basic idea is to first triangulate Hawking's diagram of an evaporating black hole spacetime, or rather the topological space $\mathcal M$ on which this process can occur, as illustrated by the triangles in FIG. \ref{fig:scheme}. The number of triangles is a flexible aspect of the scheme and can be adjusted based on the specific requirements of the application and similarly for the number of triangles per vertex —aspects of this flexibility and beyond are discussed (further) in \S\ref{sec:discussion}. Now, since the figure depicts a Penrose diagram, it is necessary to interpret what these triangles represent in four dimensions. Neglecting the boundary of the diagram, the vertices correspond to 2-spheres, meaning that lines connecting them represent annuli. Consequently, the triangles correspond to regions bounded by three annuli, with these annuli connecting three 2-spheres in a cyclic manner. This structure is illustrated in FIG. \ref{sfig:continuous_diagram_triangle}, where the annuli are the regions between the upper spheres, between the lower and the upper-inner sphere, and between the lower and upper-outer spheres.   By flipping the orientation of these regions, they can be glued according to the instructions provided by the triangulated Penrose diagram. To also consider triangles touching the boundary one might need to shrink some of these spheres to points —Remember, for example, that points in the vertical dotted lines are not spheres but correspond to vanishing coordinate radius.

\begin{figure}
    \centering
    \begin{subfigure}[c]{0.3\textwidth}
        \centering
        \includegraphics[width=\linewidth]{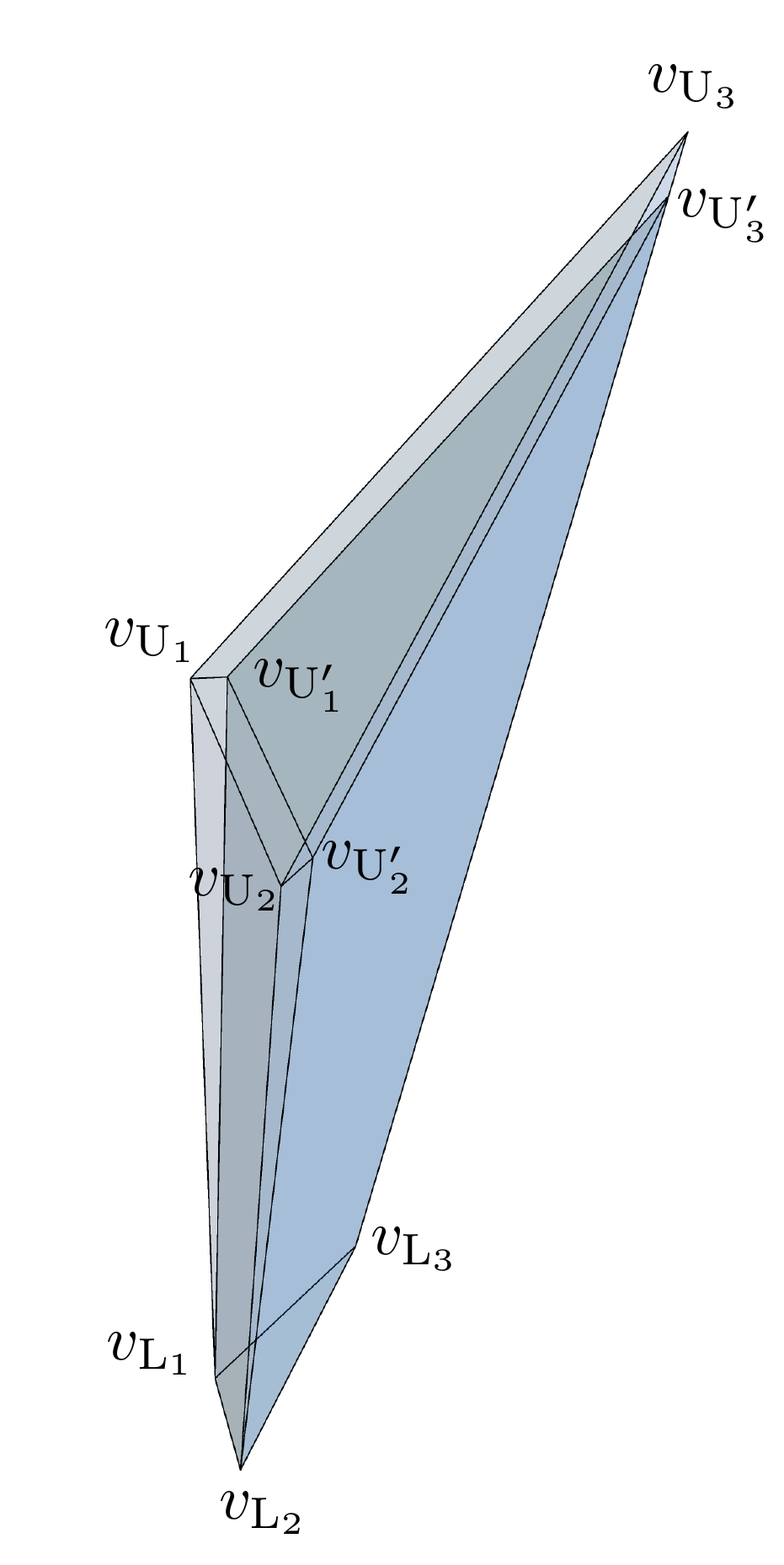}
        \caption{Cartoon of the fundamental polytope of the (four-dimensional) discretization scheme, showing the vertex nomenclature and vertex connectivity.}
        \label{sfig:4D_polytope}
    \end{subfigure}
    \hspace{2.5cm}
    \begin{subfigure}[c]{0.3\textwidth}
        \centering
        \includegraphics[width=\linewidth]{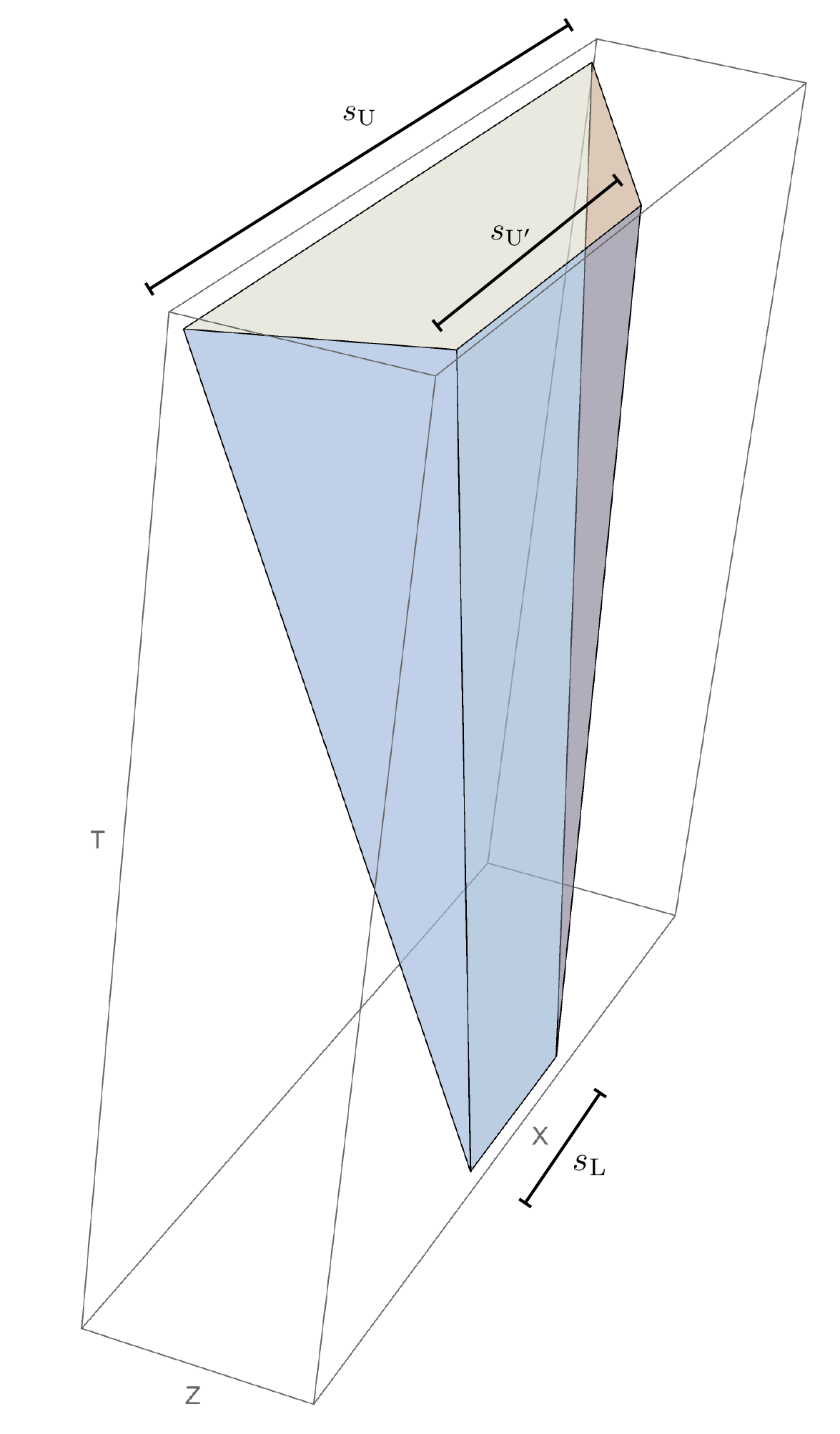}
        \caption{Illustration of the fundamental building block in a three-dimensional spacetime, including geometric nomenclature and a depiction of its embedding into flat spacetime.}
        \label{sfig:3D_polytope}   
    \end{subfigure}
    \label{fig:polytope}
    \caption{Depictions of the fundamental polytopes in four (left) and three (right) dimensions. Gluing three and four of them, resp., would result in the discrete analogue of triangle diagrams.}
\end{figure}

\begin{figure}
    \centering
    \includegraphics[width=0.75\linewidth]{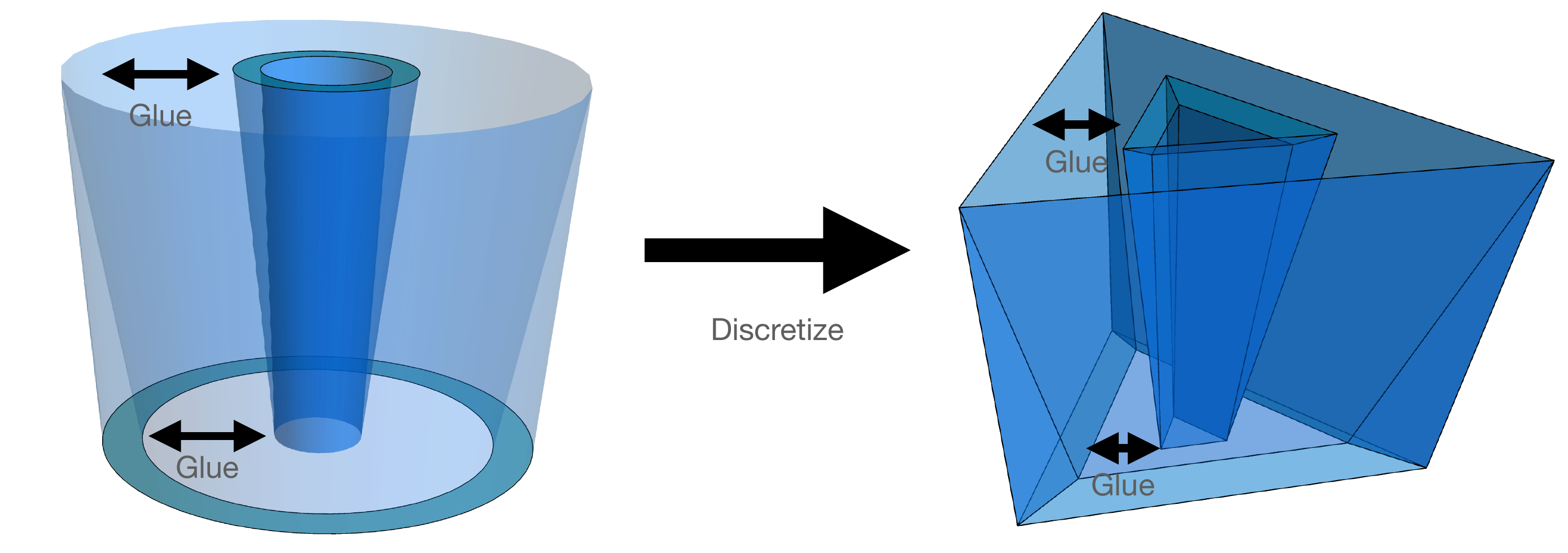}
    \caption{Three dimensional depiction of how the polytopes corresponding to diagram triangles are to be glued, and the continuum counterpart.}
    \label{fig:3D_quadrilateral}
\end{figure}

\begin{figure}
    \centering
    \includegraphics[width=0.75\textwidth]{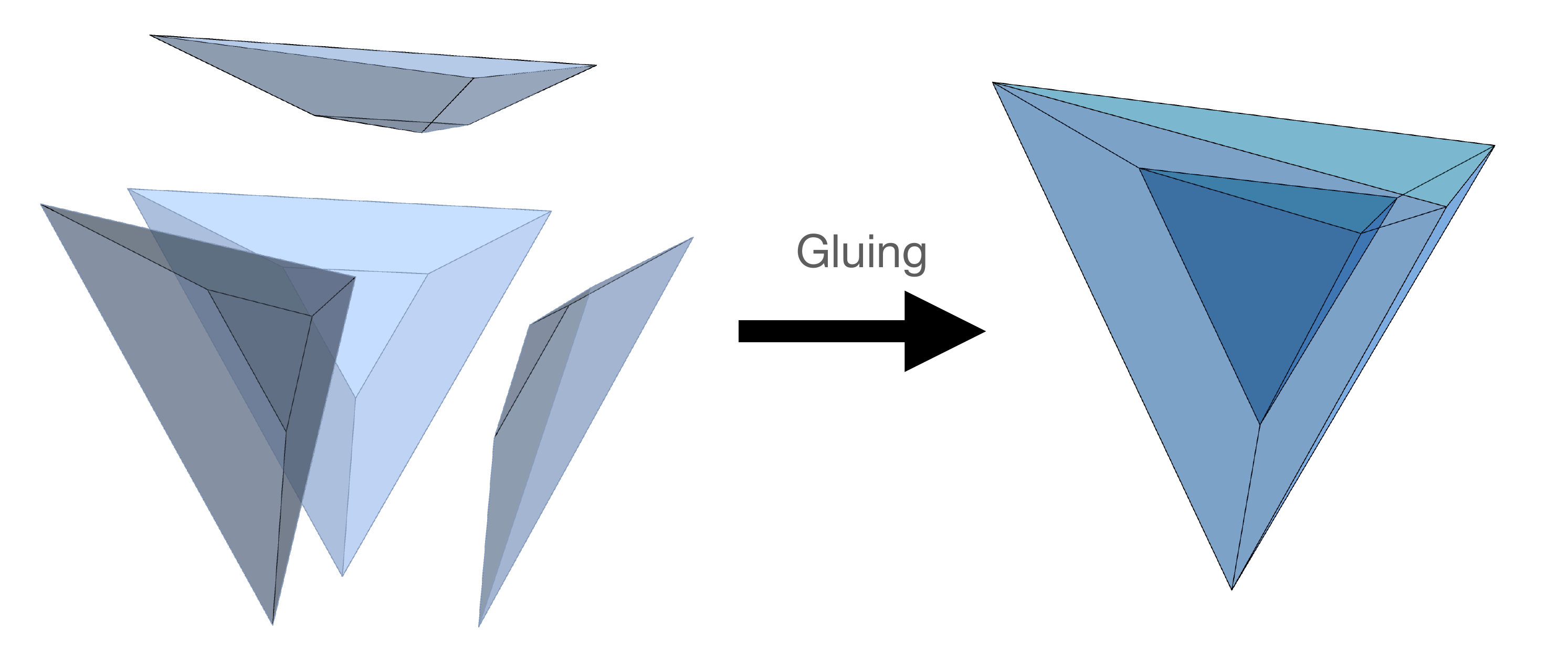}
    \caption{Illustration of how gluing triangular frusta results in a tetrahedral shell annulus.}
    \label{fig:gluing_frustra}
\end{figure}

Having identified the topological structures represented by the diagram's triangles, it becomes possible to define the polytopes —or building blocks— used to construct the piecewise linear geometries discretizing spacetime. The polytopes are defined by triangulating the 2-spheres using the boundaries of tetrahedra (to be generalized below) and discretizing the annuli by connecting the vertices of these tetrahedra in a one-to-one fashion. This matching must satisfy a transitive-like condition: if vertex $v_A$ of tetrahedron $A$ is connected to vertex $v_B$ of tetrahedron $B$, and $v_C$ of $C$ is connected to $v_A$, then the vertex of $C$ connected to $B$ must be $v_C$. The resulting discretization, shown in FIG. \ref{sfig:discrete_diagram_triangle}, reveals a ‘triangulation’ composed of four polytopes with identical topology glued together. These polytopes are constructed by connecting three triangles at their vertices in a one-to-one manner, consistent with the transitive property described above, as illustrated in FIG. \ref{sfig:4D_polytope} —or in FIG. \ref{sfig:3D_polytope} which for better visualization shows a three dimensional version. For additional clarity, FIG. \ref{fig:3D_quadrilateral} illustrates the continuous/discrete correspondence in the 3D case, where tetrahedra are replaced by triangles. Figure \ref{fig:3D_quadrilateral} also shows the gluing procedure of annuli. It might be helpful to analyze figure \ref{fig:gluing_frustra} as well, displaying how the annuli are discretized by a gluing of triangular \emph{frusta} (truncated pyramids).

In summary, the polytopes of FIG. \ref{sfig:4D_polytope} are glued together in groups of four to form the counterparts of the triangles in FIG. \ref{fig:scheme}, making them the fundamental building blocks of the triangulations used to discretize the evaporating black hole spacetimes of interest. At this point it should be clear that these fundamental cells can be glued in groups larger than four as well, in a way that allows triangulating spheres in a more refined way than with tetrahedral shells. However, for concreteness and clarity the rest of the discussion will refer to tetrahedral shells, as the generalization to finer shells (\emph{e.g.} icosahedral shells) should be straightforward. Needless to say that these triangulations should be regular if one keeps the restriction of spherical symmetry.

The strategy for triangulating spacetimes is then either to adopt a sufficiently fine discretization that captures the relevant physics or to develop a scheme for taking the continuum limit —see \cite{Dittrich:2012jq} for an example. Deferring the latter approach to future work, the focus from this point onward will be on the former.

Apart from the continuum limit, it is also important to carefully address how the discrete framework extends to asymptotic infinity. One approach is to perform calculations for a given triangulation and then take the limit as the boundary geometry extends to infinity in some appropriate sense (\emph{e.g.}, by taking certain edge lengths to be infinite). However, this may effectively coarse-grain an infinite portion of spacetime, potentially interfering with a subsequent refinement limit or with microscopic physics at infinity. Alternatively, a hybrid limit involving both refinement and large boundary lengths in the asymptotic region could be considered. Yet another perspective arises from noticing that the most relevant quantum gravitational effects are expected to occur near the classical singularity, suggesting that the gravitational path integral, and back reaction, may only need to be considered within a finite region, while classical gravity suffices to describe the dynamics away from this region. However, the precise delineation of the region where quantum gravitational effects become significant remains unclear. The proposed discretization scheme is flexible enough to explore these possibilities, and such an investigation is essential for drawing conclusive statements about the replica paradigm within quantum Regge calculus. The application below might give some intuition of how some of these procedures could be taken, but a detailed analysis is left for future work. 

Once the spacetime $\mathcal M$ has been discretized to produce a triangulation $\triangle(\mathcal M)$, one can construct the wormhole or Hawking spacetimes for any $n$, by gluing $2n$ copies of these triangulation as instructed by the path integral —see \S\ref{sec:application} for an explicit example. 

So far, only the topology of the building blocks has been specified, but naturally, their geometry must also be defined. As mentioned above, the triangulation is piecewise linear, meaning the polytopes are considered embeddable into a \emph{linear} spacetime with some metric $\eta$. Their geometry is then determined by the different ways they can be embedded, up to rigid transformations. This can be specified through a set of geometric invariants that typically include the lengths of the polytope edges. (For some polytopes, this information alone is not exhaustive —consider, for example, a quadrilateral. In such cases, a common procedure is to subdivide the polytope into some that do have this property, \emph{e.g.} simplices). These geometric invariants must satisfy compatibility conditions, typically a set of inequalities generalizing the triangle inequalities, ensuring that a consistent embedding into the linear spacetime actually exists.

With this geometric structure in place, it is now possible to specialize the discussion to spherically symmetric spacetimes. While the general framework allows for broader applications, this restriction provides a concrete setting in which to explore the discretization scheme and its implications.

Given that the path integral of interest corresponds to one over Lorentzian geometries, it is natural to choose $\eta$ to be the Minkowski metric. Then, the way the spherically symmetric restriction is implemented in the discrete, at the level of the polytopes, is by requiring the triangles to be equilateral and that the edges connecting a pair of triangles have the same length. One can then embed the polytope using equation \eqref{eq:embedding}. (\emph{Cf.} FIG. \ref{sfig:4D_polytope} —note the nomenclature L for Lower and U for Upper. See also the 3D version in FIG. \ref{sfig:3D_polytope} for easier visualization of the coordinate variables' meaning.)
\begin{gather}
    v_{\text{L}_1}=\left(t_{\text{L}},0,\frac{s_{\text{L}}}{\sqrt 3},z_{\text{L}}\right),\quad
    v_{\text{L}_2}=\left(t_{\text{L}},-\frac{s_{\text{L}}}{2},-\frac{s_{\text{L}}}{2\sqrt 3},z_{\text{L}}\right)\quad
    v_{\text{L}_3}=\left(t_{\text{L}},\frac{s_{\text{L}}}{2},-\frac{s_{\text{L}}}{2\sqrt 3},z_{\text{L}}\right),\nonumber\\
    v_{\text{U}_1}=\left(t_{\text{U}},0,\frac{s_{\text{U}}}{\sqrt 3},z_{\text{U}}\right),\quad
    v_{\text{U}_2}=\left(t_{\text{U}},-\frac{s_{\text{U}}}{2},-\frac{s_{\text{U}}}{2\sqrt 3},z_{\text{U}}\right)\quad
    v_{\text{U}_3}=\left(t_{\text{U}},\frac{s_{\text{U}}}{2},-\frac{s_{\text{U}}}{2\sqrt 3},z_{\text{U}}\right)\nonumber\\
    v_{\text{U}'_1}=\left(t_{\text{U}'},0,\frac{s_{\text{U}'}}{\sqrt 3},z_{\text{U}'}\right),\quad
    v_{\text{U}'_2}=\left(t_{\text{U}'},-\frac{s_{\text{U}'}}{2},-\frac{s_{\text{U}'}}{2\sqrt 3},z_{\text{U}'}\right)\quad
    v_{\text{U}'_3}=\left(t_{\text{U}'},\frac{s_{\text{U}'}}{2},-\frac{s_{\text{U}'}}{2\sqrt 3},z_{\text{U}'}\right).
    \label{eq:embedding}
\end{gather}
These coordinates are such that the triangles are located at time coordinates $t_I$, $Z$ coordinate $z_I$, and have edge lengths $s_I$. In this way, the geometry of each polytope is parametrized by $3 \times 3 = 9$ variables. Now, for this class of polytopes, the edge lengths fully determine their geometry, and there are only $3+3=6$ independent edge lengths. This implies that the coordinate-based description is redundant. However, when gluing two polytopes together, their shared frustum's geometry must match, and this matching is enforced at the level of coordinates. Doing so introduces a number of constraints greater than the number one would have using geometric invariants. As a result, for a full triangulation, the redundancy in the coordinate description is ultimately reduced to just three degrees of freedom. These residual degrees of freedom can be linked to the diagram boundary and may not be freely specifiable —particularly if they correspond to the vertical lines in the Hawking-Penrose diagram, which represent the spatial origin $r=0$.  

A clearer understanding of this redundancy should emerge in the concrete example discussed in the next section. For now, one can gain some intuition by considering the following: the ambiguity in defining the geometry of a single polytope is at most three-dimensional, since one can always set two of the $z_I$ variables and one of the $t_I$ variables to zero using Poincaré transformations —To visualize this and what follows, it may be helpful to refer to the three-dimensional example provided in \ref{sfig:3D_polytope}. Now, any two neighboring polytopes can always be embedded and glued in flat spacetime, so once one polytope is embedded, it is always possible to attach to it one of its neighboring polytopes. In doing so, a frustum is identified, so one is introducing only three new geometric degrees of freedom, corresponding directly to the three additional coordinate variables. This guarantees that no extra ambiguity is introduced at each step. 

By continuing this gluing process polytope by polytope, the geometric configuration of the full triangulation is determined, and the only remaining redundancy originates from the initial polytope —potentially a boundary polytope. It is important to clarify that in general, following this construction the full configuration cannot be embedded into flat spacetime without breaking connectivity of some of the polytopes, or some of them overlapping, as the configuration may correspond to a curved geometry. This point will be further detailed in the next section.  

As mentioned earlier, replica saddles, if they exist, cannot be Lorentzian and must correspond to complex (possibly Euclidean) geometries. To find their discrete counterpart, it is necessary to consider complex geometries for the polytopes. This is achieved by allowing the coordinates to become complex as well. If complex replica saddles exist in a given setup, the key question is whether the real-time contour, defined by real coordinate variables, can be deformed to pass through this saddle in a way such that its contribution will dominate in the semiclassical limit —see \cite{Dittrich:2024awu,Asante:2021phx} for examples of QRC addressing these questions. 

While any of these coordinates could, in principle, be complexified, a natural starting point (motivated by generalizing the Wick rotation procedure) is to consider only the $t_I$'s as complex. More specifically, given that the $t_I$'s are treated as time variables that can be Wick rotated to produce Euclidean geometries, it is of interest to consider —as a reminder U (L) stands for Upper (Lower)—
\begin{equation}
    t_{\text{U}}=t_{\text L}+r_{t_{\text{U}}}e^{\imath\phi_{\text{U}}},\quad t_{\text{U}'}=t_{\text L}+r_{t_{\text{U}'}}e^{\imath\phi_{\text {U}'}}.
\end{equation}
Note that by gluing polytopes together, the bulk time variables $t_\text{L}$ also naturally acquire a complex-polar decomposition.

This way, when all time coordinates of a polytope are imaginary, the geometry is Euclidean, and one can interpolate between Lorentzian and Euclidean geometries via intermediate complex geometries. Crucially, this interpolation can be performed locally within the triangulation, which is essential if one want to ensure that the boundary geometry remains Lorentzian —for instance, light-like in the case of $\mathscr I_u^+$.

Naturally, an analogous procedure can be used to complexify the $s_I$ and $z_I$ variables. This may be of interest given the fact that inside horizons the causal nature of spacetime coordinates might flip, so such generalization might be required whenever non-Euclidean complex saddles are expected.

In summary, with the choices made above, the geometry of the triangulation is fully determined by the variables $s_I$, $t_I$ (or $r_{t_I}$ and $\phi_I$) , and $z_I$. So gravitational dynamics must be discretized as dynamics for them. The way this will be done is by using the so-called \emph{quantum Regge calculus} \cite{Rocek:1981ama,Hamber:2009mt}.

\subsection{Dynamics for the discretized geometries\label{ssec:Regge_action}}
\subsubsection{Regge action(s)}

Quantum Regge calculus attempts to make sense of quantum gravity by considering a triangulated spacetime and discretizing the gravitational path integral as 
\begin{equation}
        \int\mathrm D[g]e^{W_\text{EH}}\longrightarrow \int\mathrm\mathrm dl_i\mu(l_i)e^{W_\text{R}},
        \label{eq:SOH_discretization}
\end{equation}
where $l_i$'s are the lengths of the triangulation's segments (and/or any other needed geometric parameters), $\mu$ is a measure factor, and $W_\text{R}$ ($W_\text{EH}$) is the –possibly complex– \emph{Regge} (Einstein-Hilbert) exponent, which is introduced below. That is, motivated by the fact that a triangulation's geometry is fully determined by the $l_i$'s, they are considered to discretize the metric. The path integral exponents correspond to minus the Euclidean action if dealing with Euclidean `quantum' gravity, or $\imath$ times the Lorentzian action, when working in that signature. Importantly, the integral over length assignments is restricted to those satisfying the generalized triangle inequalities (in the appropriate signature).\footnote{Notably, this restriction is loosened in other simplicial quantum gravity frameworks, such as spinfoams —see \cite{Dittrich:2023ava} (and references in there) for a discussion on this.}

The main justification for \eqref{eq:SOH_discretization} comes from the fact that as the triangulation is refined, the Regge action approximates increasingly better the Einstein-Hilbert action \cite{Cheeger:1983vq,Feinberg:1984he} and, the solution to the discrete equations of motion reproduce the continuum ones in the continuum limit \cite{Rocek:1981ama,Rocek:1982tj,barrett1988convergence,Barrett:1988wd}. (For reviews on classical Regge calculus see \cite{Misner1973,Williams:1991pj}.)

It is therefore remarkable that the Regge exponent takes the simple form\footnote{In this expression, it is assumed that the cosmological constant vanishes. The generalization to $\Lambda \neq 0$ simply adds a term equal to minus the sum of the volumes of the polytopes that make up the triangulation.}
\begin{equation}
    W_\text{R}=\sum_{b\in\text{Bones}} A(b)\epsilon(b),
    \label{eq:Regge_exponent}
\end{equation}
in units where $8\pi G=1$, which will be adopted from now on. Here the sum is over the codimension-two faces ($b$ones) of the triangulation in question (\emph{e.g.} the trapeziums in the frusta of the triangulation above), and $A(\square)$ outputs their $A$rea. $\epsilon$ is a notion of curvature in the triangulation localized at each bone,\footnote{Given this information, compare \eqref{eq:Regge_exponent} with $-\int\sqrt{g} R$.} the \emph{deficit angle}, defined as
\begin{equation}
    \epsilon=m_b\pi-\sum_{c\supset b}\theta_{b\subset c},
    \label{eq:deficit_angle_definition}
\end{equation}
where, the sum runs over all top-dimensional building blocks ($c$ells) $c$ that contain the bone $b$, and what is being summed are the dihedral angles at $b$ in each $c$. That is, the internal angle between the unique pair of codimension-one cells in $c$ that meet at $b$.  In the bulk, bones have $m_b = 2$, whereas for boundary bones, $m_b$ is typically chosen as $1$. However, a more refined choice is to take $m_b$ as the multiple of $\frac{1}{4}$ so that $m_b\pi$ best approximates the expected real part of the dihedral angle at that bone. This ensures that, in Lorentzian signature, the typical contribution of a boundary bone to the action remains real, since real parts of dihedral angles in such cases always take values that are multiples of $\frac{\pi}{4}$, leading to a cancellation for typical configurations. 

For visualization, consider the special case where the top dimension is three, so that a tetrahedron is a valid cell. In this case, the dihedral angles are associated with the edges, with each edge angle corresponding to the internal angle (within the tetrahedron) between the two triangles that share it.

Mathematically, the dihedral angle can be computed via the inner product of the normals (one outward- and one inward-pointing) to the codimension-one cells meeting at $b$ in $c$, using
\begin{equation}
    \theta(n_1,n_2)=\imath\log\left(\frac{n_1\cdot n_2+\sqrt{(n_1\cdot n_2)^2-(n_1\cdot n_1)(n_2\cdot n_2)}}{\sqrt{n_1\cdot n_1}\sqrt{n_2\cdot n_2}}\right)
    \label{eq:dihedral_angle_definition},
\end{equation}
where the inner product is defined with $\eta$  \cite{Jia:2021xeh,Asante:2021phx}. The square root and logarithm take the same values as their principal branches, and are extended to the whole complex plane by setting $\sqrt{-1}=\imath$, and $\log(-1)=\imath\pi$.\footnote{The reason this needs to be specified is that for Euclidean geometries, the square root in the numerator is negative due to the Cauchy-Schwarz inequality, placing it on the branch cut of the principal logarithm, so a prescription is required. A similar thing happens in the Lorentzian regime with the denominator whenever one of the normals is timelike. } This formula can be understood to follow from the familiar expression $\vec a\cdot_\text{E}\vec b=\sqrt{\vec a\cdot_\text{E} \vec a}\sqrt{\vec b\cdot_\text{E}\vec b}\cos\theta_{\vec a,\vec b}$ for the Euclidean inner product and the identity $\imath\arccos z=\log\left(z+\sqrt{z^2-1}\right)$, however, the form \eqref{eq:dihedral_angle_definition} is more commonly used in recent literature, because it facilitates the analytic continuation needed to explore contour deformations (see \cite{Asante:2021phx}); or in the terms above, it facilitates the extension to complex coordinate variables.

\begin{figure}[h!]
    \centering
    \includegraphics[width=0.75\textwidth]{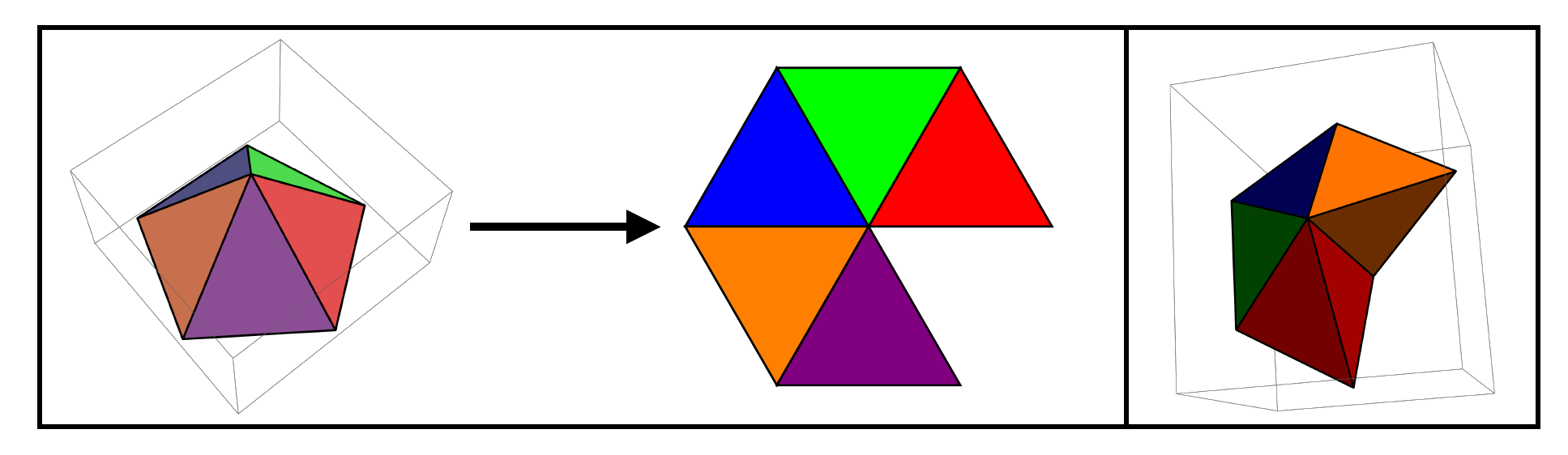}
    \caption{The deficit angle indicates whether a collection of simplices surrounding a given bone can fit together smoothly in flat spacetime. When the deficit angle non-zero, as shown in the left panel, the surrounding triangles cannot form a closed chain without distortion, requiring a break to achieve an embedding. The right panel extends this idea to three dimensions, illustrating a disrupted chain.}
    \label{fig:deficit_angle}
\end{figure}

The way $\epsilon$ captures curvature for the case of a bulk bone is that whenever it vanishes, the cycle of top-dimensional cells meeting at it can be embedded into flat spacetime without breaking. The two-dimensional representation in FIG. \ref{fig:deficit_angle} helps visualize its meaning. 

It is useful to rewrite \eqref{eq:Regge_exponent} using \eqref{eq:deficit_angle_definition} as
\begin{equation}
    W_\text{R}=\sum_b m_b\pi A(b)-\sum_c\sum_{b\subset c} A(b)\theta_{b\subset c}=:\sum_b W_b+\sum_c W_c,
    \label{eq:W_bone_cell_decomposition}
\end{equation}
where in the last expression the bone and cell contributions to $W_\text{R}$ are defined.  This decomposition will facilitate the computation of the Regge exponent for both the Hawking and wormhole topologies within the triangulation scheme described above. Additionally, for replica-symmetric spacetimes, it allows for the reproduction of an expression analogous to \eqref{eq:S_EH_additivity}.

From \eqref{eq:W_bone_cell_decomposition} it is clear that the only ingredients needed to compute the Regge exponent are the area and dihedral angle of the bones. The former are easily computed by means of Heron's formula \cite{Heron}, stating that (in both Lorentzian and Euclidean signature) the (signed)\footnote{The area is ``squared'' in the same sense in which line elements $\mathrm ds^2=\eta_{\mu\nu}\mathrm d x^\mu\mathrm d x^\nu$ are `squared'. } area squared $A^s$ of a triangle with edges of square lengths $s_a$, $s_b$ and $s_c$ is given by
\begin{equation}
    A^s=\frac1{16}(2 s_a s_b+2s_a s_c+2s_b s_c-s_a^2-s_b^2-s_c^2).
    \label{eq:Heron}
\end{equation}
Although not all bones of a `triangulation' need to be triangles —for example, in the triangulation scheme above there are the trapeziums $(v_{I_1},v_{I_2};v_{J_1},v_{J_2})$—, any polygon can be decomposed into triangles so that in general one needs to consider sums of triangle areas given by \eqref{eq:Heron} —\emph{E.g.} for the trapezium example above, the squared area can be assigned as the square of the sum of the areas of the triangles $(v_{I_1},v_{I_2};v_{J_1})$ and  $(v_{I_2};v_{J_1},v_{J_2})$.

In the particular context of the scheme above, the edge lengths squared can be computed by means of $\eta$ to then express the areas in terms of the coordinate variables,\footnote{Note that Heron's result follows from the popular `Euler's shoelace'/determinant formula for polygons on a plane, so this is also a path to find the areas in terms of coordinates.} and the areas can thus be analytically continued as the variables are complexified.

Now to the dihedral angles at a bone-cell pair $(b,c)$: These can be computed by first embedding the cell $c$ in the flat spacetime it belongs to and then constructing the relevant normal vectors. In four spacetime dimensions\footnote{The procedure described can easily be generalized to any dimension by increasing/decreasing the number of vectors to which the normals are orthogonal —and correspondingly the number of indices for the Levi-Civita symbol.} one proceeds by selecting a vertex $v_s$ in $b$ and identifies two vectors, $e_{b,1}$ and $e_{b,2}$, that extend from $v_s$ to adjacent vertices in $b$. The two codimension-one sub-cells ($f$aces) $f_1$ and $f_2$ in $c$ that share $b$ are then determined, and for each, an $e$\emph{dge vector} $e_{f_i}$ is chosen, connecting $v_s$ to a vertex in the corresponding face that is not part of $b$.

The normals of interest are normal to both $e_{b_1}$ and $e_{b_2}$, as well as one of the $e_{f_i}$'s and are therefore given by the components
\begin{equation}
    n_{b\subset f_i\subset c}^\mu=\eta^{\mu\alpha}\epsilon_{\alpha\beta\gamma\delta}e_{b,1}^\beta e_{b,2}^\gamma e_{f_i}^\delta.
    \label{eq:normals}
\end{equation}
 It is important to note that the way in which the vectors are contracted with the Levi-Civita symbol, $\epsilon_{\alpha\beta\gamma\delta}$, is chosen so that one normal points inward and the other outward.

(As an example, consider, in triangulation scheme above, $(v_{\text{L}_1},v_{\text{L}_2},v_{\text{L}_3})$ as the bone, paired with a frustum 
\begin{equation}    (v_{\text{L}_1},v_{\text{L}_2},v_{\text{L}_3};v_{\text{U}_1},v_{\text{U}_2},v_{\text{U}_3};v_{\text{U}_1'},v_{\text{U}_2'},v_{\text{U}_3'})
\end{equation}
as the cell. One can pick the $s$ource vertex $v_s$ as $v_{\text{L}_1}$, and  $e_{b_1}=v_{\text{L}_2}-v_{\text{L}_1}$ and $e_{b_2}=v_{\text{L}_3}-v_{\text{L}_1}$. Similarly, $f_1$ can be the frustum $(v_{\text{L}_1},v_{\text{L}_2},v_{\text{L}_3};v_{\text{U}_1},v_{\text{U}_2},v_{\text{U}_3})$ and therefore $f_2=(v_{\text{L}_1},v_{\text{L}_2},v_{\text{L}_3};v_{\text{U}_1'},v_{\text{U}_2'},v_{\text{U}_3'})$. Thus, $e_{f_1}=v_{\text{U}_1}-v_{\text{L}_1}$ and $e_{f_2}=v_{\text{U}_1}'-v_{\text{L}_1}$.)

Finally, it is only a matter of plugging $n_{b\subset f_1\subset c}^\mu\cdot n_{b\subset f_2\subset c}^\mu$ into \eqref{eq:deficit_angle_definition} to compute the deficit angles.

Therefore, all the necessary geometric ingredients needed to compute the Regge exponents of interest are provided by equations \eqref{eq:embedding}, \eqref{eq:Heron}, \eqref{eq:dihedral_angle_definition} and \eqref{eq:normals}. The only thing left to discuss is how to combine them to compute the exponent for the full Hawking and wormhole spacetimes. 

For the Hawking spacetimes it is important to note each $\mathcal M$ is glued to its dual ${\mathcal M^*}$ throughout $\Sigma_\text{Int}$, and that there are $n$ disconnected copies of such gluing. Consequently there are exactly $n$ copies of $\Sigma_\text{Int}$ and of $\left(\mathscr J_u^+\cup S_u^2\right)\cup\left(\mathscr I_u^+\setminus S^2_u\right)^*$. The celestial sphere at retarded time $u$, $S_u^2$, needs to be removed because, being a splitting surface, appears only once in the entire spacetime. Therefore, the corresponding Hawking-Regge exponent is
\begin{gather}
    W^{(n)}_\text{HR}(s^J_{I},z^J_{I},t^J_{I})\nonumber\\
    =\nonumber\\
    \sum_{i=1}^n\left(\sum_{c^i\subset\triangle(\mathcal M)^i}W_{c^i}+\sum_{c^{i^*}\subset\triangle(\mathcal M)^{i^*}}W_{c^{i^*}}+\sum_{b^i\subset(\triangle(\Sigma_\text{Int})^i=\triangle(\Sigma_\text{Int})^{i^*})}W_{b^i}+\sum_{b^i\subset\triangle(\mathscr J_u^+\setminus S_u^2)^i}W_{b^i}+\sum_{ b^{i^*}\subset\triangle(\mathscr J_u^+\setminus S_u^2)^{i^*}}W_{ b^{i^*}}\right)\nonumber\\
    +\nonumber\\
    \sum_{b\subset \triangle(S_u^2)}W_b,
    \label{eq:WHR}
\end{gather}
where a variable identification is required based on how (sub-)building blocks are glued together. So for example, if $c^i$ and $c^{i^*}$ are cells that are glued together along a face $f^i\subset \triangle(\Sigma_\text{Int})^i$, then one needs to identify the variables $s^J_I$, $z^J_I$ and $t^J_I$ associated to $f^i\subset \triangle(\Sigma^i_\text{Int})$ —or equivalently $f^{i^*}\subset \triangle(\Sigma_\text{Int})^{i^*}$ since they coincide. These identifications, in turn, completely determine the geometry of the bones $b^i$ within $f^i$. 

The way equation \eqref{eq:WHR} is written allows to use different discretizations for each $\mathcal M$ and also for the duals, however any undemocratic treatment requires justification, even if non-replica-symmetric saddles are expected. Thus, in the following it will be assumed that the same triangulation is used for $\mathcal M$ and $\mathcal M^*$, and their copies. In this case, if the contributions from cells amount to $N$ terms per spacetime copy, they lead to a total of $2nN$ cell contributions. In contrast, the number of bone contributions depends on their location: if there are $M_\text{Bulk}$ bones in the bulk of $\triangle(\mathcal{M})$ then there are  $2nM_{\text{Bulk}}$ contributions from bones in the bulks of copies of $\triangle(\mathcal{M})$ —or rather $n M_\text{Bulk}$ from bones in copies of $\triangle(\mathcal M)$ and $n M_\text{Bulk}$ from bones in copies of $\triangle(\mathcal M)^*$. Similarly one has $2 n M_\text{Asymptotic}$ contributions from bones in asymptotic regions (excluding the splitting surface),  $nM_{\Sigma_\text{Int}}$  contributions from bones in copies of $\triangle(\Sigma_\text{Int})$, and only one contribution per bone in the splitting surface.

Similarly, for the wormhole topology, the Wormhole-Regge exponent is
\begin{gather}
    W^{(n)}_\text{WR}(s^J_{I},z^J_{I},t^J_{I})\nonumber\\
    =\nonumber\\
    \sum_{i=1}^n\left(\sum_{c^i\subset\triangle(\mathcal M)^i}W_{c^i}+\sum_{c^{i^*}\subset\triangle({\mathcal M})^{i^*}}W_{c^{i^*}}+\sum_{b^i\subset\triangle(\Sigma_\text{Int})^i\setminus\triangle(\gamma)}W_{b^i}+\sum_{b^i\subset\triangle(\mathscr J_u^+\setminus S_u^2)^i}W_{b^i}+\sum_{b^{i^*}\subset\triangle(\mathscr J_u^+\setminus S_u^2)^{i^*}}W_{b^{i^*}}\right)\nonumber\\
    +\nonumber\\
    \sum_{b\subset \triangle(S_u^2)}W_b+\sum_{b\subset \triangle(\gamma)}W_b,
    \label{eq:WWR}
\end{gather}
where now the triangulated splitting surface $\triangle(\gamma)$ is treated as such.

Note that for configurations where the geometry of $\triangle(\gamma)$ is the same for every copy
\begin{equation}
    W^{(n)}_\text{WR}=W^{(n)}_\text{RH}-(n-1)\sum_{b\subset\triangle(\gamma)}W_b=W^{(n)}_\text{RH}-(n-1)\frac{ A(\triangle(\gamma))}{4G},
\end{equation}
where it was used that the Regge action conventions above set $\frac 1{8\pi G}=1$. If one is to take asymptotic limits, the discrete celestial sphere contribution needs to be regularized —something closely related to the ambiguity in the boundary terms of the Regge action mentioned above. Once that is addressed, for replica-symmetric geometries, $W^{(n)}_\text{RH} = nW^{(1)}_\text{RH}$, so that one recovers a discrete analogue of \eqref{eq:S_EH_additivity}.  This provides a direct means of analytically continuing the gravitational action in $n$, as $W^{(1)}_\text{RH}$ is manifestly independent of $n$.

Equations \eqref{eq:WHR} and \eqref{eq:WWR} furnish \emph{classical} gravitational dynamics to the discrete model. The elemental bone and cell terms in them can be calculated explicitly with the aforementioned geometric ingredients. The final result is notably long and not enlightening, which is why it is presented in the GitHub repository \cite{github}, together with its explicit computation.

In the continuum setup (\emph{cf.} \S\ref{sec:continuum}), the gravitational path integral is computed using the saddle point approximation once the matter has been integrated out. And therefore, in order to mimic this, it suffices to consider the gravitational dynamics encoded in the Regge exponents.

Measure factors in the path integral are expected to be sub-leading in the saddle point approximation —indeed, it would be surprising if they contributed at exponential order comparable to $e^W$. Nevertheless, the discrete framework should, in principle, allow for the (numerical) study of sub-leading corrections to the gravitational path integral. For this purpose, the measure of the geometric path integral must also be discussed.

\subsubsection{Comments on the Regge measure \label{sssec:Regge_measure}}

In continuum quantum gravity works, the gravitational measure is typically\footnote{The following summary mostly follows works done in the context of Euclidean (simplicial) quantum gravity, as summarized in \cite{Hamber:1997ut,Hamber:2009zz} —see also \cite{Loll:1998aj}. While a full real-time treatment remains open, significant deviations are unlikely, except perhaps regarding the statement of independence on $\vartheta$. In fact, the coarse graining approach discussed in the end of the section recently extended to Lorentzian signature and the findings are completely analogous to their Euclidean counterpart \cite{Borissova:2023izx}.} defined using a \emph{supermetric}, which provides a volume element on field space
\begin{equation}
    \mathbb dg^2=\int\mathrm d^D x\mathbb G^{\mu\nu,\rho\sigma}[g(x)]\mathbb dg_{\mu\nu}[x]\mathbb d g_{\rho\sigma}[x].
\end{equation}
Then the gravitational measure is defined as
\begin{equation}
    \mathrm Dg=\prod_x\sqrt{\mathbb{det}\mathbb G[g(x)]}\prod_{\mu\ge\nu}\mathrm dg_{\mu\nu}(x).
\end{equation}
However, this approach suffers from an ambiguity on how the supermetric is to be defined. For example, members in the family
\begin{equation}
    \mathbb G^{\mu\nu,\rho\sigma}[g(x)]=\frac12\sqrt{-g}^{1-\omega}\left(g^{\mu\rho}g^{\nu\sigma}+g^{\mu\rho}g^{\nu\sigma}+\lambda g^{\mu\nu}g^{\alpha\beta}\right)
\end{equation}
are in principle sensitive choices —for original arguments for the cases $\omega = 1$ and $\omega = 0$, which define the so-called Misner and DeWitt supermetrics, see \cite{Misner:1957wq} and \cite{DeWitt:1962by}, respectively.

This class of supermetrics has been widely studied, as not only is it superlocal, but can be argued to take the simple form \cite{Hamber:1997ut}
\begin{equation}
    \mathrm Dg\propto\sqrt{-g}^\vartheta \prod_{\mu\ge\nu}\mathrm dg_{\mu\nu}(x),
\end{equation}
with $\vartheta$ a dimension and $\omega$ dependent parameter.

Such considerations, have in turn motivated previous works to postulate the discretization\footnote{In cases where $\vartheta$ leads to a sufficiently singular metric, it is customary to include additional factors that regulate small-length divergences.}
\begin{equation}
    \mathrm Dg\longrightarrow\prod_v\mathcal V(\star v)^\vartheta\prod_e \mathrm dl^s_e,
\end{equation}
where $\mathcal V(\star v)$ computes a (four-)$\mathcal V$olume associated to the vertex $v$ through a \emph{dual triangulation} \cite{Voronoi1908} —next section will discuss dual volumes in more detail, albeit to edges. The right hand side differentials are of length-squared variables, reflecting the integration over metrics on the right hand side.\footnote{As noted (somewhat implicitly) in \cite{Asante:2021phx,Dittrich:2024awu} building on earlier insights from the Ponzano-Regge model for 3D quantum gravity, quantum Regge calculus can be extended to yield a formulation akin to an integration over vielbeins, by summing over simplex orientations.} Notably, for $\vartheta = -\frac{1}{2}$, this measure aligns (in the large spin limit and up to a phase) with that of the Ponzano-Regge model \cite{Ponzano:1968,Barrett:2008wh} —the first spinfoam model and a topological QFT that successfully describes three-dimensional quantum Euclidean gravity in vacuum without a cosmological constant. Moreover, it has been argued that in the continuum limit, long distance properties are independent of $\vartheta$, as long as it leads to a non-singular measure; in line with certain continuum considerations \cite{Hamber:1985qj}. However, alternative measures have also been explored—see \cite{Hamber:2009zz} for a survey. One approach, inspired by the induced metric of a simplex and the aforementioned supermetric class, involves products of (inverse) powers of length variables and has shown some success \cite{Hamber:1985qj}.

A promising alternative to the whole perspective above, tightly connected with taking the continuum limit, is to define a measure through a coarse graining procedure guaranteeing that triangulation invariance and thereof the continuum symmetries are recovered in the refinement limit \cite{Dittrich:2011vz}. This reasoning leads to results consistent with the Ponzano-Regge measure in three dimensions, but indicate the need of a non-local measure for the four dimensional theory \cite{Dittrich:2011vz,Dittrich:2014rha} (and indeed, non-local measures have also been proposed independently by studying discrete analogues of the supermetric determinant —see \cite{Hartle:1996db} and references therein– and even by considerations in the continuum \cite{Menotti:1996tm}).

In conclusion, while ambiguities remain in defining the quantum gravitational measure, resolving them can be deferred until going beyond semiclassical analysis becomes a priority in this line of research.

\subsection{(Discrete) Matter dynamics\label{ssec:matter}}
\subsubsection{Matter action(s)}

The following discussion will focus on how to minimally couple Regge gravity to a ‘free’ and massless real scalar field —although in this framework other types of fields may also be considered\footnote{See for example \cite{Sorkin:1975jz,Ren:1987is} that discuss also the cases of vectors and spinors.}.  Following \cite{Ren:1987is}, the field is discretized by assigning a degree of freedom to each vertex of the triangulation. The classical dynamics of the discrete field is governed via the Matter-Regge exponent \eqref{eq:W_Matter} —\emph{cf.} $-\int_\Sigma \sqrt{g}\frac12\left(\frac{\partial\phi}{\partial x^\mu}\right)^2$.
\begin{equation}
    W_\text{MR}=-\sum_{e\in\text{Edges}} L(e)\times V(*e)\frac12\frac{\left(\phi_{v_\mu}-\phi_{v_\nu}\right)^2}{ L(e)^2},
    \label{eq:W_Matter}
\end{equation}
where $v_\square$ are the vertices defining each edge $e$, to each of which a field degree of freedom $\phi_{v_\square}$ is attached. $L(\square)$ and $V(\square)$ compute the $L$ength and (three-)$V$olume of their arguments, resp. $*e$ is a three-dimensional (and more generally codimension-one) triangulation-dual to the edge $e$. There are several ways to define said dual such that \eqref{eq:W_Matter} approximates the scalar field action on a given background geometry when taking the continuum limit \cite{Hamber:1993gn}. The one considered here consists of a barycentric subdivision,\footnote{To give an example, a common alternative proceeds by considering intersections of multiple perpendicular bisectors, following the construction of what is known as a \emph{Voronoi dual lattice}. \cite{Voronoi1908}} defined for the case of a four-dimensional base `triangulation' $\triangle$ (but easily generalized to other dimensions) as the three-dimensional (actual) triangulation made by all possible tetrahedra with vertices $(v_{\bar e},v_{\bar b},v_{\bar f},v_{\bar c})$ that are barycenters of $e$, $b$, $f$ and $c$, with $e\subset b\subset f\subset c$ forming a proper chain of sub-blocks, in this case made of an $e$dge, a $b$one, a $f$ace and a $c$ell.%\footnote{Such a chain might very temptingly be called C6/E.}

\begin{figure}[h!]
    \centering
    \includegraphics[width=0.5\textwidth]{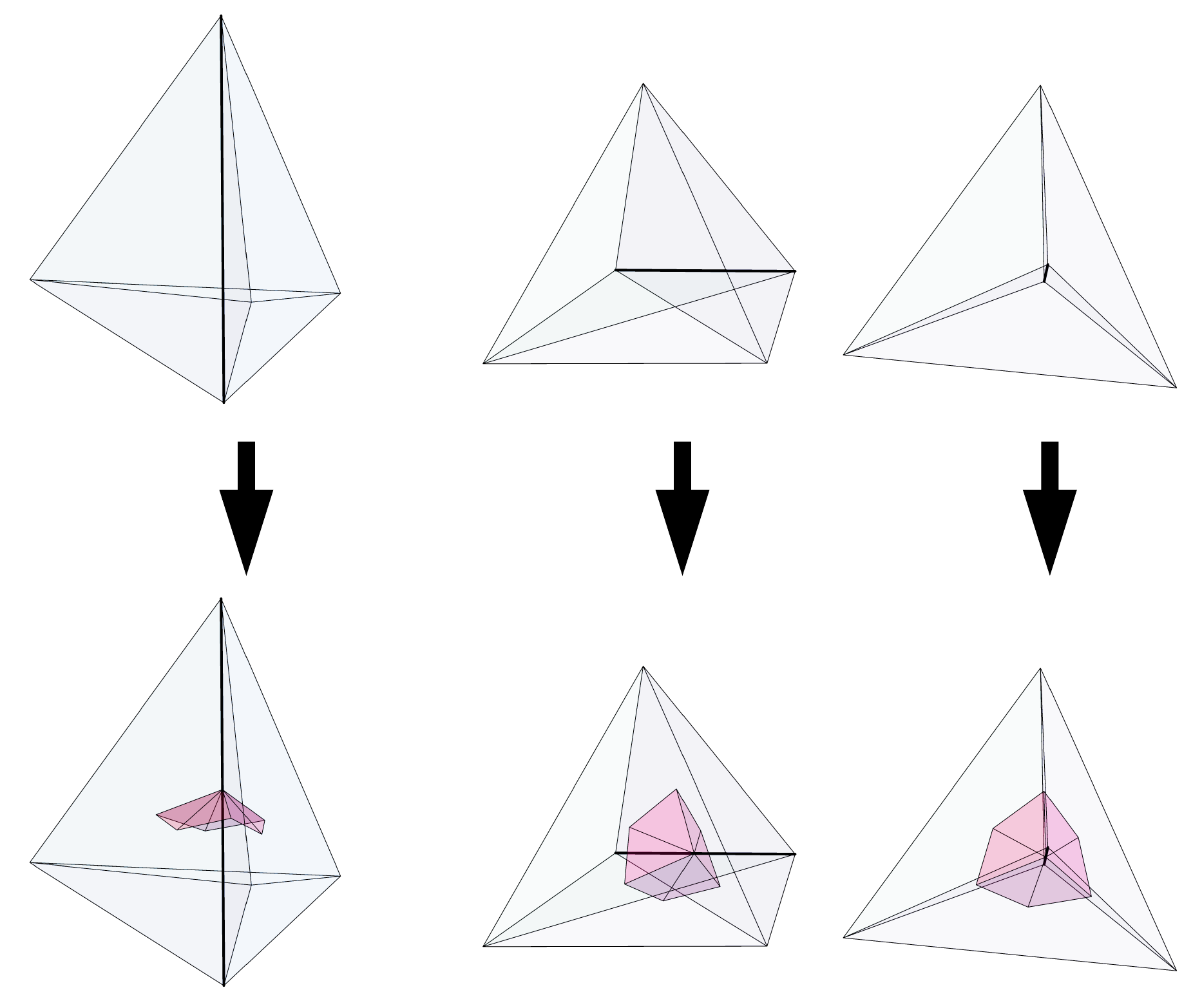}
    \caption{Visualization of the dual structure associated with an edge in a three-dimensional triangulation, shown from multiple perspectives. The base triangulation consists of three tetrahedra joined along the edge. The dual cells form another triangulation surrounding the edge.}
    \label{fig:3D_dual}
\end{figure}

To offer some intuition, the dual to an edge in a three-dimensional triangulation is presented in FIG. \ref{fig:3D_dual}. In this case such dual is two dimensional and, in contrast to what happens in four-dimensions, the chains only have one intermediate sub-cell, say $g^\sharp$, between an edge $e$ and a 3-dimensional cell $c$ —this illustrates how the D-dimensional generalization of $e^*$ is defined: by adding as many sub-cells as possible to the chains.

Therefore, computing $V(*e)$ is a simple manner of adding volumes of tetrahedra, which can in turn be calculated by a generalization of \eqref{eq:Heron} for the volume squared of a tetrahedron with sides of square lengths $s_{ij}$ (with $i$ and $j$ denoting the vertices that determine the sides) \cite{sitharam2018handbook}:
\begin{equation}
    V^s=\frac 1{2^3 (3!)^2}\det
    \begin{pmatrix}
        0 & 1 & 1 & 1 & 1\\
        1 & 0 & s_{12} & s_{13} & s_{14}\\
        1 & s_{12} & 0 & s_{23} & s_{24}\\
        1 & s_{13} & s_{23} & 0 & s_{34}\\
        1 & s_{14} & s_{24} & s_{34} & 0\\
    \end{pmatrix}.
\end{equation}
Again, for the scheme under considerations, these square edge lengths can be computed in terms of the coordinate variables using the metric $\eta$.

The points above entail all that is needed to compute the Regge exponent for this matter content. For the same reasons as before, the resulting final modular expressions for the full action are presented in \cite{github}, accompanied by a routine computing them. For these expressions the discrete version of spherical symmetry was also imposed on the scalar field degrees of freedom. It can be explicitly checked that when for Lorentzian (Euclidean) geometries $W_\text{MR}=\imath S_\text{L}$ ($W_\text{MR}=-S_\text{E}$), with $S_\square$ a real number naturally playing the role of the classical action.

\subsubsection{Matter quantum measure}

The exponent alone does not fully define the quantum dynamics of the scalar field, the path integral also includes a measure factor —which was absorbed into the symbol $\mathrm D[\phi]$ in \S\ref{sec:continuum}. These factors, when field independent, can commonly be ignored in quantum (field) theories on a fixed background, given that they always cancel when normalizing path integrals. However, they generally depend on the geometry. If gravity is dynamical, their metric dependence becomes important, as now they may not cancel out upon normalization. More concretely, consider the example of \eqref{eq:semi_effective_patt_Integral}: The metric dependence of $\mathrm D[\phi]$, which has been absorbed into the effective actions, does not necessarily factor out and cancel between the numerator and denominator, and therefore considering the right metric dependence of the measure impacts the final result, at least quantitatively. The case of the scalar field in a lattice $\Gamma$ of D-dimensional Euclidean space might help illustrate the point further. 

Consider that the lattice is such that imaginary time is discretized by steps of size $\epsilon$ and space with a regular hyper-cubical lattice of spacing $\alpha$. For such theory, the (formal) measure is \cite{Gattringer2010}
\begin{equation}
    \mathrm D_\Gamma[\phi]=\prod_{x\in\Gamma} \mathrm d\phi(x)\sqrt{\frac{\alpha^{D-1}}{\epsilon 2\pi}}.
    \label{eq:flat_matter_measure}
\end{equation}
Here $\alpha$ and $\epsilon$ can be understood as providing the metric dependent part of the measure factor $\prod_{x\in\Gamma}\sqrt{\frac{\alpha^{D-1}}{\epsilon 2\pi}}$. Note that when computing a two-point function as
\begin{equation}
    \expval{\phi(x_1)\phi(x_2)}=\frac{\int\mathrm D_{\Gamma}[\phi]\phi(x_1)\phi(x_2)e^{-S_\Gamma}}{\int\mathrm D_{\Gamma}[\phi]e^{-S_\Gamma}}
    \label{eq:two_point_function}
\end{equation}
it would cancel. However, if $\alpha$ and $\epsilon$ were site-dependent and integrated over, as would be the case when summing over metrics, the cancellation would not necessarily occur.

Equation \eqref{eq:flat_matter_measure} can be derived by mimicking the standard approach for obtaining the path integral representation of a non-relativistic particle propagator (see, \emph{e.g.}, \cite{Gattringer2010}). The strategy is to generalize\footnote{For an alternative more \emph{ab initio} reasoning, see the comments at the end of this section.} it to Regge geometries and for this some observations are in order.

For a free and massless field, $S_\Gamma$ is given by \cite{Gattringer2010}
\begin{equation}
    S_\Gamma=\frac1{2^{D-1}}\sum_{(\vec x,t)\in\Gamma}\left(\frac12 \epsilon\alpha^{D-1}\left(\frac{\phi(\vec x,t+\epsilon)-\phi(\vec x,t)}\epsilon\right)^2+\sum_{\hat{\textbf{\j}}}\frac12 \epsilon\alpha^{D-1}\left(\frac{\phi(\vec x+\hat{\textbf\j},t)-\phi(\vec x,t)}\alpha\right)^2\right),
    \label{eq:action_flat_hypercubical}
\end{equation}
with $\hat{\textbf\j}$ the vector on the j-th spatial direction of length $\alpha$.

Note that the sum in \eqref{eq:action_flat_hypercubical} can be reordered as done over edges, although there is a distinguishment between timelike and spacelike edges. By doing so, it becomes clear that up to a numerical pre-factor, which could be absorbed into a field redefinition, this action can be written in terms of dual volumes. Then, one can observe that the geometric measure factor in \eqref{eq:flat_matter_measure} matches the geometric factor in action terms associated with the timelike edges. And indeed, by following carefully the derivation of \eqref{eq:flat_matter_measure} it becomes evident that this is no coincidence. This suggests a generalization to triangulations, \emph{viz.}
\begin{equation}
    \mathrm D_\triangle[\phi]=\prod_{v_\mu}\mathrm d\phi_{v_\mu}\sqrt{\frac{\frac1{2^{D-1}n_{e_{\text T}}(v_\mu)}\sum_{e_{\text T}(v_\mu)}\frac{V(*e_{\text T}(v_\mu))}{L(e_{\text T}(v_\mu))}}{2\pi}},
    \label{eq:measure_non_cov}
\end{equation}
where the sum runs over all timelike edges incident on vertex $v_\mu$, with their total number given by $n_{e_\text T}(v_\mu)$. Note that the constant part can be ignored due to normalization as long as the same $n_{e_\text T}$ is used in numerator and denominator of expressions like \eqref{eq:two_point_function}. However, it is worth emphasizing that this may become non-trivial when taking continuum limits, which, in principle, could be taken independently for the numerator and the denominator.

Equation \eqref{eq:measure_non_cov} has the awkward property of treating spacelike and timelike edges differently. This could become a problem when doing path integrals in which the signature of edges can fluctuate and more importantly puts into question the covariance of the continuum limit. However, a more covariant choice can me made by lifting that restriction in them and in the counting factor, that is taking
\begin{equation}
    \mathrm D_\triangle[\phi]=\prod_{v_\mu}\mathrm d\phi_{v_\mu}\sqrt{\frac{\frac1{2^{D-1}n_{e}(v_\mu)}\sum_{e(v_\mu)} \frac{V(*e(v_\mu))}{L(e(v_\mu))}}{2\pi}}.
    \label{eq:measure_cov}
\end{equation}
Although heuristically motivated, this choice of measure introduces a reasonable property in quantum gravity: for a massless free field, if the volume of a region shrinks to zero, the partition function effectively behaves as if the region were absent. This property likely extends to cases with a non-trivial potential, including mass terms. The intuition behind this can be seen by considering the integration of a single field degree of freedom with this measure pre-factor and observing the limit where the surrounding region collapses to zero size.

As this research program progresses, it will be valuable to explore alternative measure choices that have been proposed in the literature and assess their impact on the conclusions and continuum limit. For instance, it would be particularly interesting to evaluate the measure introduced in \cite{Hamber:1993gn} —and explored in several subsequent works. This measure uses the induced metric of a simplex to construct a discrete analogue of a continuum measure derived from superspace invariants, analogous to those discussed in \S\ref{sssec:Regge_measure}. —In fact, the resulting geometric measure factor is also a power of $\sqrt{-g(x)}$. However, it is important to note that this alternative does not reduce to \eqref{eq:flat_matter_measure} when applied to the flat spacetime hyper-cubical lattice.

\subsubsection{Initial state preparation}

As discussed in \S\ref{sec:continuum}, the matter path integrals being discretized can be characterized by three aspects. They:
\begin{enumerate}
    \item Compute the vacuum wave-functional dependent on field configurations on each $\mathscr I^-$.
    \item Compute the time evolution propagator (propagation kernel) on every copy of the black-hole evaporation spacetime from $\mathscr I_-$ to the Cauchy slice $\Sigma_u$, and the time reversed counterpart.
    \item Glue (of `\emph{sew}') portions of copies either on $\Sigma_\text{Int}$'s or on $\mathscr I_u^+$'s.
\end{enumerate}
Each of these steps could in principle be adapted to the discrete model using the ingredients above —up to the discussion of asymptotic limits. For instance, the time evolution corresponds to a Lorentzian contour of integration in each $\Delta(\mathcal M^I)$, and the gluing just corresponds to an identification of boundary variables and integration over them. 
However, the computation of the vacuum wave-functional requires further discussion.

In the continuum model, where gravitational boundary conditions correspond to an asymptotically flat spacetime, the vacuum wave-functional is to that of Minkowski spacetime. As is well known (see \emph{e.g.} \cite{Weinberg_1995}), this can be computed via a path integral on half of the Euclidean space. A faithful adaptation to the discrete setup would thus require a method for handling asymptotic boundaries —an issue that was briefly discussed above and will be revisited in the discussion section. Ideally, such a method would allow for the direct computation of a discretized vacuum functional —or, more generally, for preparing any state.

A more practical approach (especially if one is primarily interested in qualitative features) is to \emph{discretize} directly rather than compute the vacuum wave-functional. This is the approach that will be taken in the application below.

In the continuum the vacuum wave-functional for a massless free and real scalar field is known \cite{Weinberg_1995} to take the harmonic-oscillator-vacuum-like (or Gaussian-like) form
\begin{equation}
    \bra{\phi}\ket{0}=\mathcal N\exp\left(-\frac12\int\mathrm d^{D-1}x\mathrm d^{D-1}y \phi(\vec x)\mathcal E(\vec x,\vec y)\phi(\vec y)\right),
    \label{eq:continuum_vacuum_wave_functional}
\end{equation}
where $\mathcal N$ is a normalization constant, and $\mathcal E(\vec x,\vec y)$ is the Fourier transform of the energy $\sqrt{\vec p^2}$ evaluated on $\vec x-\vec y$.

Motivated by this, it is reasonable (at least for initial applications of the framework) to explore discrete wave-functionals (\emph{i.e.}, wave-functions) with exponents of the form
\begin{equation}
    W_\text{Vacuum}=-\sum_{v\in\Delta(\mathscr I^-)}\frac12\Omega_v\phi_{v}^2,
    \label{eq:discrete_vacuum_exponent}
\end{equation}
where the sum runs over the vertices of the discretized asymptotic past null boundary. $\Omega_v$ is a matter field constant, though it may depend on the geometry of $\Delta(\mathscr J^-)$—a detail that could become relevant in asymptotic limits. Similarly, $\mathcal N$ is taken to be geometry-dependent only through $\triangle(\mathscr I^-)$, so that it cancels upon normalization.

Below, $\Omega$ will be treated as a parameter of the model. Naturally, it is important to assess whether, and to what extent, the qualitative properties found depend on its value. Within the restrictions below, they remain largely unchanged across several orders of magnitude.

\section{Application: Minimal reproduction of the Page transition \label{sec:application}}
\subsection{Triangulation}

Finally, it is time to bring all these elements together and demonstrate how the framework is to be applied. The setup will be minimal, serving as a proof of principle and an exercise toward more realistic applications. Nevertheless, it may already provide insight into how generic the existence of replica saddles is. %Particularly, the emphasis will be on the path integrals associated with gluing copies of $\mathcal M$

\begin{figure}
    \centering
    \includegraphics[width=0.4\textwidth]{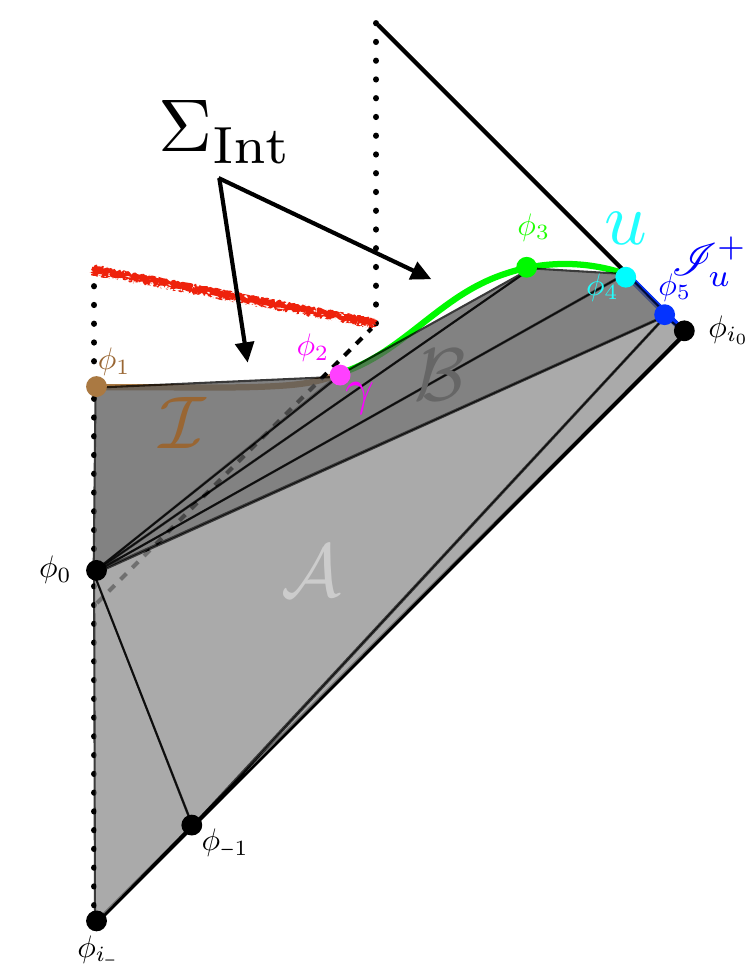}
    \caption{Initial implementation of the proposed discretization scheme. A single black hole spacetime would contain nine field degrees of freedom, the minimal number required to represent all key regions involved in swap entropy computations. The illustration is schematic, as the actual calculation has not incorporated asymptotic limits. Contrary to the previously shown Penrose diagrams, $\gamma$ was drawn to lie outside the horizon in order to emphasize that its precise location is something that should come out of (discrete) calculations.}
    \label{fig:application}
\end{figure}

The specific triangulation used is designed to provide a finitized version of (copies of) $\mathcal M$ and is represented in figure \ref{fig:application}, although note that strictly speaking the triangulation does not go to infinity. The figure also illustrates the matter field degrees of freedom being considered. Analogous to the treatment of geometry, the scalar field will be assumed to possess a discrete version of spherical symmetry, reducing the number of independent degrees of freedom to one per tetrahedral shell. Each shell will be labeled according to its corresponding field subindex.

It is useful to conceptualize the triangulation in two parts: $\triangle(\mathcal A)$ and $\triangle(\mathcal B)$, corresponding to continuum regions $\mathcal A$ and $\mathcal B$, separated by the diagram line $\overline{05}$. This division illustrates how one might localize the quantum gravitational computation to a finite region,\footnote{This is an illustrative choice, as the finite region extends to the would-be $\mathscr I^+$ but not to $\mathscr I^-$. However, the ``RG-like reduction'' exposed later provides motivation for this approach.} addressing the infinite extent of spacetime—\emph{cf.} \ref{ssec:discretization}. If (quantum) gravitational back reaction becomes significant at $\mathcal{A} \cap \mathcal{B}$, then region $\mathcal{A}$ can be treated classically with regards to the gravitational field, and Regge calculus is required only in $\mathcal{B}$. Handling calculations in the infinite region $\mathcal A$ requires further development of the exposed framework. One option is to perform them in continuum QFT on a fixed spacetime, necessitating a scheme to match the continuum and discrete data at $\mathcal{A} \cap \mathcal{B}$. Alternatively, Regge calculus could be used throughout, but this would require careful treatment of asymptotic limits. Current work primarily focuses on the latter approach, so this will be the perspective taken hereafter, and therefore Regge calculus will be applied to both regions, though without addressing the asymptotic limit —\emph{Ergo} the introduction of $\triangle(\mathcal B)$. Notice that this split is consistent with the topology change only possibly occurring in $\mathcal B$.

Some of the tetrahedral shells of the polytopes represented by the triangle diagrams must have size zero (corresponding to setting $s_I\overset{!}{=}0$), for example, those lying on the $r=0$ line. Since $i^0$ is a point in the continuum, the corresponding tetrahedral shell is also taken to have zero size. This second imposition is likely artificial, because $i^0$ being a point follows from the mathematical procedure of compactification to obtain the Penrose diagram.  However, in the absence of a defined procedure for taking asymptotic limits and uncertainty about whether (and how) this might involve blowing up spacelike infinity (more on this in \S\ref{sec:discussion}), this simplification seems practical. Nevertheless, the ``RG-like reduction'' discussed below suggests that this assumption does not qualitatively affect the results presented here.

The previous paragraph implies that some of the polytopes in question are special cases of the model in figure \ref{sfig:4D_polytope}. For example, polytopes of region $\triangle(\mathcal B)$ are glued around the \emph{point} with label $0$, so their topology can be thought of as that of four-dimensional pyramids with frustal bases. Some polytopes of $\triangle(\mathcal A)$ have yet other topologies.

This degeneracy must be handled with care when computing dual volumes. Since these involve averaging, taking the limit of barycenters as points approach coincidence does not yield the same result as computing the barycenter of the degenerate configuration directly. This discrepancy arises because the averaging procedure involves dividing by the number of points, which is discontinuous in the coincidence limit.\footnote{This can also be understood by recalling that barycenters are centers of mass, so when points coincide their combined contribution effectively outweighs the others.} Accordingly, for shells corresponding to the origin, dual volumes will be defined directly from the degenerate configuration rather than as a limiting case. The same approach will be taken for $i^0$, though this may need to be revisited as understanding on how to deal with asymptotic limits deepens.

The number of polytopes chosen is such that the gluing (or \emph{sewing}) path integrals involve field degrees of freedom on all triangulated counterparts of the key portions of (each) $\Sigma_u$ appearing in the continuum, \emph{viz.}: $\mathcal I$, $\gamma$, $\Sigma_\text{Int}$, $S_u^2$ and $\mathscr I_u^+\setminus S^2_u$. Similarly, integrating $\phi_0$ corresponds to integrating over bulk fields in (each) $\mathcal M$, while the integrals over the remaining field degrees of freedom correspond to integrating over field configurations on (each) $\mathscr J^-$ —which, as previously mentioned, will be treated as the arguments of the vacuum wave-function(s). 

\subsection{Matter effective actions}

 With these preliminaries considered, the steps of the matter path integrals are to be accounted for as follows:
 \begin{itemize}
     \item For each copy, the vacuum wave-functional is replaced by a discrete analogue and the functional integrals over its argument by integral over $\phi_{i^-}$, $\phi_{-1}$ and $\phi_{i^0}$. 
     \item For each copy, the time evolution from $\mathscr I^-$ to the Cauchy slice is discretized by the $\phi_0$ integral. 
     \item The gluing integrals are preformed over (copies of) $\phi^i$, $i=1,...,5$.
 \end{itemize}

Above the choice was made to treat the field degree of freedom at $i^0$ as part of $\mathscr I^-$ rather than $\mathscr I^+_u$. More precisely, its integral was considered part of the state preparation rather than the sewing. Alternative approaches, such as those in which $i^0$ is blown up, could allow certain portions of the blow-up to be more naturally identified exclusively with either $\mathscr I^+_u$ or $\mathscr I^-$.

In any case, in the continuum limit, the relevance of field degrees of freedom on a zero-measure set may not be significant unless distributional effects are involved, as was the case with the splitting surface and the gravitational field. However it may be, it is nevertheless unclear what can be said for the matter field degrees of freedom on the entangling surfaces, since they act as intermediaries between the field degrees of freedom on different replicas.

To sum up, in contrast to the discussion of \S\ref{sec:continuum}, the concreteness of the discrete set up forces one to consider (corner) modes and a choice was implemented above —indeed a similar discussion to the one above could be had for $\phi_{i^-}$.

Given the considerations above, the matter effective actions are computed \emph{via}
\begin{gather}
    e^{W_\text{MR,Eff}}\\
    =\nonumber\\
    \int\left(\prod_{j=1}^n\mathrm d\phi_{i^-}^j\mathrm d\phi_{-1}^j\mathrm d\phi_{i^0}^j\right)\left(\prod_{j^*=1}^n\mathrm d \phi_{i^-}^{j^*}\mathrm d \phi_{-1}^{j^*}\mathrm d\phi_{i^0}^{j^*}\right)\times\nonumber\\
    \times\left(\prod_{j=1}^n\mathrm d\phi_0^j\right)\left(\prod_{j^*=1}^n\mathrm d\phi_0^{j^*}\right)\times\left(\prod_{J=1}^n\mathrm d\phi_1^J\right)\times\left(\prod_{J=1}^n\mathrm d\phi_3^J\right)\times\left(\prod_{J=1}^n\mathrm d\phi_5^J\right)\times\left(\prod_{J=1}^m\mathrm d\phi_2^J\right)\times\mathrm d\phi_4\times\nonumber\\
    \times\mu e^{W_\text{MR}(\triangle(\cup(\mathcal A^L\cup\mathcal B^L)))}e^{ W_\text{Vacuum}},
\end{gather}
where $m=n$ ($m=1$) for the Hawking (replica) topology and likewise the way fields appear in $W_\text{MR}$ depends on which is the case in question. Correspondingly, the argument of the Regge-Matter exponents communicates that it is associated to the whole triangulation. Finally, $\mu$ is the matter measure and thus consists of products of inverse dual volumes, edge lengths and constants. Upper case indices have been used for objects that could be labeled with both starred /andor not-starred indices.

The first step will be to integrate the matter on the region whose geometry is considered to be fixed, \emph{viz.} $\cup\triangle(\mathcal A^L)$. This is straightforward, because one is only dealing with Gaussian kernels and these regions are insensitive to the gluing.

After integrating the degrees of freedom in all the $\triangle(\mathcal A^L)$'s one obtains
\begin{gather}
    e^{W^{(n)}_\text{MR,Eff}}\\
    =\nonumber\\
    \left(\prod_{j=1}^n\mathcal N_{\triangle(\mathcal A)}^{j^*}\right)\left(\prod_{j^*=1}^n\mathcal N_{\triangle(\mathcal A)}^{ j}\right)\times\nonumber\\
    \times\int\left(\prod_{j=1}^n\mathrm d\phi_0^j\right)\left(\prod_{ j^*=1}^n\mathrm d\phi_0^{j^*}\right)\times\left(\prod_{J=1}^n\mathrm d\phi_1^J\right)\times\left(\prod_{J=1}^n\mathrm d\phi_3^J\right)\times\left(\prod_{J=1}^n\mathrm d\phi_5^J\right)\times\left(\prod_{J=1}^m\mathrm d\phi_2^J\right)\times\mathrm d\phi_4\times\nonumber\\
    \times\mu e^{W_{\text{MR}}(\triangle(\cup\mathcal B^L))}\times\nonumber\\
    \times\exp\left(-\sum_j\Delta c_{05}^j(\triangle(\mathcal A^j))(\phi_0^j-\phi_5^j)^2-\sum_{j^*}\Delta c_{05}^{j^*}(\triangle(\mathcal A^{j^*}))(\phi_0^{j^*}-\phi_5^{j^*})^2\right)\times\nonumber\\
    \times\exp\sum_j\left(-\Omega_0^{j}(\triangle(\mathcal A^{j}))(\phi_0^{j})^2-\Omega_5^{j}(\triangle(\mathcal A^{j}))(\phi_5^{j})^2-\Omega_0^{j^*}(\triangle(\mathcal A^{j^*}))(\phi_0^{j^*})^2-\Omega_5^{j^*}(\triangle(\mathcal A^{j^*}))(\phi_5^{j^*})^2\right).
    \label{eq:RG_like_reduction}
\end{gather}
Several points are worth noting here.

First that the normalization factors $\mathcal N_{\triangle(\mathcal A)}^J$ arising from the Gaussian integration depend only on the vacuum wave-function $\Omega$-coefficients, and dual volumes and edge lengths in copies of $\triangle(\mathcal A)$. Then, they depend only on that geometry which may be treated as non-dynamical. Furthermore, in a replica-symmetric spacetime, these factors combine into a term that precisely cancels upon normalization against a corresponding contribution from the —discrete version of— $(\text{Tr} \rho)^n$ and can therefore be neglected.

The way the remaining integral was written in eq. \eqref{eq:RG_like_reduction} illustrates that it takes the form of a path integral computed solely over copies of $\mathcal B$, \emph{i.e.}, the region where quantum gravitational effects are expected, albeit for some modifications:
\begin{itemize}
    \item For the field degrees of freedom on $\triangle(\mathcal A^L\cap\mathcal B^L)$'s, the volume-over-length coefficients acquire additional contributions: For those in the measure $\mu$ the contribution is just the one from the region $\mathcal A$. For those in the exponent they are given by (twice) the $\Delta c$ coefficients, whose geometry and $\Omega_v$ dependence is obtained by performing the partial integration partially —see the GitHub repository \cite{github}.
    
    \item The vacuum wave-functions, now defined on the past boundaries of $\cup\triangle(\mathcal B^L)$, retain their Gaussian form, but their $\Omega$-coefficients are determined by the outer geometry. For their explicit expression also see \cite{github}. Notably, they might no longer be real. 
    
    \item Additional normalization factors $\mathcal N_{\triangle(\mathcal A)}^L$ arise, but in replica-symmetric spacetimes, they cancel upon normalization as long as the fixed-geometry region is the same in the numerator and denominator path integrals that compute the swap operator expectation value. This condition —equality of the fixed-geometry region— is well motivated in the $n\to 1^+$ limit and will be assumed in the next section.
\end{itemize}
Since this `\emph{RG-like reduction}' holds as long as the matter theory remains Gaussian, a key challenge for future work is to understand how this kind of additional information required for finite-extent calculations depends on the boundary conditions.

Having established this point, Gaussianity can once again be invoked to rewrite the last expression in a more useful form, making the connection to the continuum discussion more transparent:
\begin{gather}
     e^{W^{(n)}_\text{MR,Eff}}\\
    =\nonumber\\
   \int\left(\prod_{J=1}^n\mathrm d\phi_1^J\right)\times\left(\prod_{J=1}^n\mathrm d\phi_3^J\right)\times\left(\prod_{J=1}^n\mathrm d\phi_5^J\right)\times\left(\prod_{J=1}^m\mathrm d\phi_2^J\right)\times\mathrm d\phi_4 \mu_{\cup\triangle(\Sigma^J_\text{Int})}\prod_{L=1}^{2n} \omega^L(\phi^L),
\end{gather}
where $\mu_\square$ is the measure factor of field degrees of freedom on $\square$ and 
\begin{equation}
    \omega^L(\phi^L)=\mathcal N_{\triangle(\mathcal A)}^L\int d\phi_0^L\mu_{\mathcal \triangle(\mathcal B^L\setminus\Sigma^L_\text{Int})}e^{\widetilde W_\text{MR}(\triangle(\mathcal B^L))}e^{\widetilde W_\text{Vacuum}(\triangle(\mathcal A^L))}.
\end{equation}
where tildes indicate the modifications arising from the partial integration over $\triangle(\mathcal A)^L$. Notably, due to this partial integration, $\omega^L(\phi^L)$ can be interpreted as the discrete analogue of the transition amplitude on a single spacetime component (the $L$-th), evolving either from the vacuum state at a $\mathscr I^-$ to the field configuration on the $\Sigma_\text{Int}$ that $\phi^L$ discretized, or such evolution but backwards.

Thus, it is just a matter of sewing these amplitudes to compute the desired effective actions.

For the Hawking topology one has
\begin{gather}
     e^{W^{(n)}_\text{MHR,Eff}}\\
    =\nonumber\\
    \mu_{\cup\triangle(\Sigma^J_\text{Int})}\int\left(\prod_{j=1}^n \mathrm d\phi_5^j\right)\times\mathrm d\phi_4 \times\left(\prod_{j=1}^n\mathrm d\phi_1^j\mathrm d\phi_2^j\mathrm d\phi_3^j\right)\times\nonumber\\
    \times
    \omega^1_\text{MHR}(\phi^1_1,\phi^1_2,\phi^1_3;\phi_4;\phi^1_5) 
    \omega^{1^*}_\text{MHR}(\phi^1_1,\phi^1_2,\phi^1_3;\phi_4;\phi^2_5)
    \times
    \omega^2_\text{MHR}(\phi^2_1,\phi^2_2,\phi^2_3;\phi_4;\phi^2_5) 
    \omega^{2^*}_\text{MHR}(\phi^2_1,\phi^2_2,\phi^2_3;\phi_4;\phi^3_5)\times\nonumber\\
    \times\cdots\times\nonumber\\
    \omega^n_\text{MHR}(\phi^n_1,\phi^n_2,\phi^n_3;\phi_4;\phi^n_5)
    \omega^{n^*}_\text{MHR}(\phi^n_1,\phi^n_2,\phi^n_3;\phi_4;\phi^1_5)\nonumber\\
    =:\nonumber\\
    \mu_{\cup\triangle(\Sigma^J_\text{Int})}\int\left(\prod_{j=1}^n \mathrm d\phi_5^j\right)\times\mathrm d\phi_4\times q^1_\text{MHR}(\phi_4;\phi^1_5,\phi^2_5)\times\cdots\times q^n_\text{MHR}(\phi_4;\phi^n_5,\phi^1_5)\nonumber\\
    =:\nonumber\\
    \mu_{\cup\triangle(\Sigma^J_\text{Int})}\int\left(\prod_{j=1}^n \mathrm d\phi_5^j\right)\times r^1_\text{MHR}(\phi^1_5,\phi^2_5)\times\cdots\times r^n_\text{MHR}(\phi^n_5,\phi^1_5),
\end{gather}
where the final factorized form follows from the fact that the integral kernels are Gaussian and this class is closed under integration. In fact, the $r^J_\text{MHR}$'s are Gaussians themselves, and their amplitudes and exponent coefficients can be computed by explicitly performing of the intermediary steps above. Naturally, these parameters depend on triangulation's geometric configuration. 

Similarly, for the replica topology one has (\emph{cf.} the discussion in \S\ref{sec:continuum} about the $n$-Rényi entropy of $\Sigma_\text{Int}\setminus\mathcal I$)
\begin{equation}
    e^{W^{(n)}_\text{MWR,Eff}}=\mu_{\cup\triangle(\Sigma^J_\text{Int})}\int\left(\prod_{j=1}^n \mathrm d\phi_3^j\right)\times r^1_\text{MWR}(\phi^1_3,\phi^2_3)\times\cdots\times r^n_\text{MWR}(\phi^n_3,\phi^1_3),
\end{equation}
where the $r^J_\text{MWR}$’s, though also Gaussians, have a different geometric dependence than the $r^J_\text{MHR}$’s.

In summary, computing the matter effective actions amounts to performing a composition of $n$ Gaussian integral-kernels $r_\square$ and tracing. From the steps above it follows that these kernels are the discrete counter-parts of density matrix elements for the field degrees of freedom living either on $\cup\triangle(\mathscr I_u^+)^L$ for the Hawking topology, or on $\cup\triangle(\Sigma_\text{Int}\setminus\mathcal I)^L$ for the replica topology.

In principle such integration be done numerically for an arbitrary geometric configuration. However, assuming replica symmetry (which partly defines the minisuperspace explored in the next section when looking for saddles), it is possible obtain a closed expression in terms of $n$ that can be analytically continued.

Indeed, consider the kernel of the composition of $n$ Gaussian-functional kernels $r_{xy}=e^{\alpha x^2+\beta xy+\gamma y^2}$, \emph{i.e.}
\begin{align}
    (r^n)_{xy}&=\int\mathrm d^{n-1}z r_{xz_1}r_{z_1z_2}\cdots r_{z_{n-1}y}=\int\mathrm d^{n-1}ze^{\alpha x^2+\beta xz_1+\gamma z_1^2+\alpha z_1^2+\beta z_1 z_2+\gamma z_2^2+\dots+\alpha z_{n-1}^2+\beta z_{n-1}y+\gamma y^2}\nonumber\\
    &=:\int\mathrm d^{n-1}z\exp\left(-\frac12\vec z^{\text T}\mathbb T^{(n-1)}\vec z+\vec b^{(n-1)}(x,y)^{\text T} \vec z+c(x,y)\right)\nonumber\\
    &=\sqrt{\frac{(2\pi)^{n-1}}{\det\mathbb T^{(n-1)}}}e^{\frac12 \vec b^{(n-1)}(x,y)^{\text T}\left(\mathbb T^{(n-1)}\right)^{-1}\vec b^{(n-1)}(x,y)+c(x,y)},
    \label{eq:Gaussian_composition}
\end{align}
where $\mathbb T^{(n-1)}$ is the $(n-1)\times(n-1)$ matrix defined by
\begin{equation}
    \mathbb T^{(n-1)}:=
    \begin{pmatrix}
        \delta & \mu & 0 & \phantom{0}\cdots &  \phantom{0}0 \\
        \mu & \delta & \mu & \phantom{0}\ddots &  \phantom{0}\vdots \\
        0 & \mu & \delta & \phantom{0}\ddots &  \phantom{0}0 \\
        \vdots & \ddots & \ddots & \phantom{0}\ddots &  \phantom{0}\mu \\
        0 & \cdots & 0 & \phantom{0}\mu &  \phantom{0}\delta
    \end{pmatrix},
\end{equation}
with $\delta:=-2(\alpha+\gamma)$ and $\mu:=-\beta$; and $\vec b^{(n-1)}$ the $(n-1)$-dimensional vector
\begin{equation}
    \vec b^{(n-1)}(x,y)^{\text T}:=(\beta x\quad 0 \quad \dots \quad 0 \quad \beta y)
\end{equation}
and
\begin{equation}
    c(x,y):=\alpha x^2+\gamma y^2.
\end{equation}
One can use the particular form of $\mathbb T^{(n-1)}$, namely that it is \emph{Toeplitz tridiagonal}\footnote{It is also symmetric, but the result below has a generalization to the non-symmetric case, so such property is not really needed.} to arrive at close formulas in terms of $n$. Indeed, for such matrices the eigenvalues have known expressions as a function of the matrix's dimension $m$ \cite{kulkarni:hal-01461924}. For the symmetric matrix in question, with $m=n-1$, the eigenvalues are
\begin{equation}
    \delta-2\mu\cos\left(\frac{k\pi}{n}\right),\quad k=1,\dots,m=n-1.
\end{equation}
And therefore
\begin{equation}
    \det\mathbb T^{(n-1)}=\prod_{k=1}^{n-1}\left(\delta-2\mu\cos\left(\frac{k\pi}{n}\right)\right)=\mu^{n-1}U_{n-1}\left(\frac\delta{2\mu}\right),
\end{equation}
where $U_n$ is the $n$-Chebyshev polynomial of the second kind, one of whose representations as a product guarantees that the second identity holds \cite{Chebyshev}.\footnote{There is also a famous representation as a determinant from which the result could have been derived.} Using the fact that $U_m(\cos\theta)=\frac{\sin((m+1)\theta)}{\sin\theta}$ it is simply a matter of substituting to get a final identity.

Having an expression for the determinant,  the exponent of the last line in \eqref{eq:Gaussian_composition} can also be recast into a closed form. Indeed, thanks to the particular form of $\vec b^{(n-1)}$,
\begin{gather}
    \vec b^{(n-1)}(x,y)^{\text T}\left(\mathbb T^{(n-1)}\right)^{-1}\vec b(x,y)\nonumber\\
    =\nonumber\\
    \beta^2\left[\mathbb T^{(n-1)}\right]^{-1}_{11} x^2+\beta^2\left(\left[\mathbb T^{(n-1)}\right]^{-1}_{1n-1}+\left[\mathbb T^{(n-1)}\right]^{-1}_{n-11}\right)xy+\beta^2\left[\mathbb T^{(n-1)}\right]^{-1}_{n-1n-1}y^2.
\end{gather}
So using the relation of a matrix's inverse with its minors $M_{ij}$, namely
\begin{equation}
    \left[\mathbb A\right]^{-1}_{ij}=\frac{(-1)^{i+j}}{\det\mathbb A}M_{ji},
\end{equation}
one has 
\begin{equation}
    \left[\mathbb T^{(n-1)}\right]^{-1}_{11}=\frac{\det\mathbb T^{(n-2)}}{\det\mathbb T^{(n-1)}}=\left[\mathbb T^{(n-1)}\right]^{-1}_{n-1n-1},\quad \left[\mathbb T^{(n-1)}\right]^{-1}_{1n-1}=\frac{(-1)^n\mu^{n-2}}{\det\mathbb T^{(n-1)}}=\left[\mathbb T^{(n-1)}\right]^{-1}_{n-11},
\end{equation}
where the last two equalities follow from the fact that the minors are determinants of triangular matrices.

In sum, all of the elements in \eqref{eq:Gaussian_composition} can be expressed in an analytic form in $n$ and thereafter taking the trace results in the analytic expression in $n$
\begin{equation}
    \text{Tr}(r^n)=\int\mathrm dx (r^n)_{xx}=\sqrt{\frac{(2\pi)^n}{2\beta^n\left((-1)^n\cos\left(n\arccos\left(
    \frac{\alpha+\gamma}\beta\right)\right)-1\right)}}.
    \label{eq:closed_trace}
\end{equation}
Thus, by determining the geometric dependence for the coefficients $\alpha$, $\beta$, and $\gamma$ in the Gaussians $r_\square^J$, a closed-form expression in terms of $n$ can be obtained for the effective action for both the Hawking and replica topologies, which can then be evaluated in the $n \to 1^+$ limit —for such result, the factors produced by the partial integration ought to be considered as well.

The final expressions are shown in the GitHub repository \cite{github}. The derivation that they come with performed the matter path integrals in a different different than the one above, leaving the integrals over the degrees of freedom on the entangling surfaces for the end. The procedure only changes the discussion above slightly and the corresponding changes are the content of appendix \ref{app:entangling_integrals}

Note that an implicit assumption in the derivation above was that $n>2$. A natural question, then, is how evaluating the expression at $n=1$ or $n=2$ compares with performing the calculation directly at those values Remarkably, the results agree —and for $n = 1$, the outcome is the same for both topologies, as expected, since the replica and Hawking topologies actually coincide in that case. This confirms that the formula is also valid for computing the normalization integrals.

The remaining step in the calculation is thus to consider fluctuations in the geometry.

\subsection{Gravitational semiclassical saddles}
\subsubsection{Minisuperspace(s)}

The framework developed so far promises to be able to consider these fluctuations fully, by \emph{e.g.} numerical computations of the path integral. But for now, the geometric path integral will be computed by saddle point evaluation, effectively reducing the problem to a classical metric coupled to quantum matter. To do so, the first step is to examine the triangulation’s geometry in more detail.

\emph{A priori} it might seem that using the coordinate variables described above the geometry is parametrized by the set $\left\{s^J_I,t^J_I,z^J_I\right\}_{I\in\left\{-1,0,1,2,3,4,5,i^-,i^0\right\}}$ up to an identification upon gluing. However, not all of these are free. For example, at $i^J_-$ there is not really a (non-degenerate) tetrahedron, but a point, so accordingly $s^J_{i^-}=0$. Similarly, $s^J_1=s^J_0=s^J_{i^0}=0$. Likewise, the coordinate redundancy can be used to set $z^J_{i^-}=z^J_0=z^J_1=0$. The remaining redundancy will be used to set $t^J_{0}=0$ so that the $0$ point serves somewhat as an `origin' in the triangulation. All of these choices are only akin to gauge fixing, but there are also geometric/physical reductions that will be implemented.

The first kind defines what kind of boundary geometries are to be considered. For instance, it is of interest to make $\triangle(\mathscr I^-)^J$ and $\triangle(\mathscr I_u^+)^J$ null. This will be ensured as follows
\begin{itemize}
    \item Fixing $t^J_{-1}$ and ($t^J_{i^0}$) by the condition that the `frustum' $\overline{-1i^-}^J$ ($\overline{-1i^0}^J$) —actually a tetrahedron— would be light-like (\emph{i.e.} has vanishing volume squared) if dealing with Lorentzian geometry, \emph{i.e.} if setting all Wick-rotation angles in the polytope $\overline{-10i^-}$ ($\overline{-105}^J$) to zero (or $\pi)$\footnote{The choice of time coordinate sign corresponding to these hypothetical Wick rotation angles is irrelevant for the present discussion. However, it is emphasized here to illustrate that one would should position the diagram vertex $i^-$ below vertex 0, which is treated as an “origin” with the gauge-fixing choice described above. Failing to ensure this could introduce a branch point in the Lorentzian contour (with a transverse branch cut) due to an orientation flip of the polytopes $\overline{i^-01}$ and $\overline{012}$ \cite{DittrichConfig:2025}. And a similar reasoning may apply to the Wick rotation angle of $i^0$.\label{fnote:orientation_branch_cut}}. Then, this constraint is actually imposed on $r_{t^J_{-1}}$ ($r_{t^J_{i^0}}$).\footnote{The fact that the restriction only involves $r_\square$, the norm of a complex number, impacts the analytic structure of the exponents. This is not important for the Euclidean search below, but also means that an imposition at the level of time coordinates is what will actually be needed in future applications designed to explore the relevance of saddle points to the \emph{a priori} Lorentzian path integral, as questions of this type depend on analyticity —for such purposes, it is also important to keep footnote \ref{fnote:orientation_branch_cut} in mind.}
    \item Setting $t_5^J$ ($t_4^J$) by demanding that the frustum $\overline{i^05}^J$ ($\overline{45}^J$) is null if Lorentzian —see the previous point.
    \item Imposing $s_{-1}=s_5=s_4=:s_\odot$, corresponding to the demand that all (`finitized') celestial spheres have the same size.
\end{itemize}
It is important to remark that conditions above may have more that one root. A sensible criterion  (and the one used for the results described below) is to pick them such that the corresponding geometry resembles the Hawking-Penrose diagram in the sense that $t^J_{i^-}<t^J_{-1}<t^J_{i^0}$ or $t^J_{i^0}<t^J_5<t^J_4$, and analogues for \emph{some} $z$-variables.

The second type of geometric reduction involves bulk geometry and thus defines the minisuperspace on which saddles will be sought. The following discussion presents the minisuperspace reduction performed in the calculation below, in decreasing stages of conservativeness.

The first conditions will be replica symmetry \emph{and} CPT symmetry. That is, all copies of $\triangle(\mathcal A\cup\mathcal B)\cup\triangle(\mathcal A\cup\mathcal B)^*$ will be assumed to have the same geometry and further, the geometry of $\triangle(\mathcal A\cup\mathcal B)^*$ will be the CPT-dual to that of $\triangle(\mathcal A\cup\mathcal B)$. This CPT-reversal is straightforward to implement given the coordinate variable parametrization, and the use of real scalar fields —see \cite{Asante:2021phx,Dittrich:2024awu} for related discussions on the matter-less case.

With all these considerations, the current variable/parameter set up is as follows:
\begin{itemize}
    \item Parameters $\Omega_{i^-}$, $\Omega_{-1}$ and $\Omega_{i^0}$ associated with the vacuum wave-function.
    \item Boundary parameters and frozen geometric variables setting the geometry of (all copies of) $\triangle(\mathcal A)$ and the boundary geometry of all $\triangle(\mathcal B)$'s —and correspondingly on all CPT-duals—: $z_{-1}, z_{i^0}, z_5$ and $s_\odot$, and $t_{i^-}$. Naturally, one could also make the bulk geometry of $\triangle(\mathcal A)$ dynamical.
    \item Boundary variable $z_4$ exclusive to all $\triangle(\mathcal B)$'s.
    \item Dynamical bulk geometric variables of the $\triangle(\mathcal B)$'s: $t_1$, and $(s_I,t_I,z_I)$ with $I=1,2$ —when complexifying, the $t_J$ variables correspond to two real ones $r_{t_J}$ and $\phi_{t_J}$.
\end{itemize}

At this point, it is possible to point out that if the geometry of $\triangle(\mathcal A)$ is indeed fixed, then the roles of $t_5$ and $z_{i^0}$ can be interchanged, treating the former as a boundary variable while fixing the latter through the light-like condition above. The geometric intuition behind this is that once the relative location of $\triangle(i^0)$ is determined, the nullity condition can be used to locate shell 5, given $z_5$. Conversely, one could first fix the position of shell 5 and then, after also profiling $\triangle(\mathscr I^-)$, fully determine the relative location of $\triangle(i^0)$. This approach aligns better with the classical-‘quantum’ split of the triangulation, as it more directly defines the geometry of $\triangle(\mathcal A\cap\mathcal B)$. With that motivation, such strategy will be adopted below. 

So far these minisuperspace restrictions are quite conservative and indeed most of the previous literature on continuous models imposes them.

However, due to the large number of variables involved, the necessary infrastructure for exploring the Riemann sheets of these complex geometries in search of saddles is still under development. Attempting such an exploration without a robust framework may be premature, given that the current minimal setup might not fully capture the relevant \emph{black hole} physics. Thus, the following will focus on further refining the minisuperspace.

Motivated by the fact that $\Sigma_\text{Int}$ is a portion of a Cauchy slice, the first step will be to impose that $\overline{01}$ is spacelike if Lorentzian, which can be achieved by re-parametrizing $t_2$ \emph{via}
\begin{equation}
    t_2=m_{12}z_2+b_{12},\quad\text{with}\quad b_{12}=t_1,
    \label{eq:t2}
\end{equation}
where $m_{12}$ is now treated as variable replacing $t_2$, but is required to be real and less than unity in absolute value when corresponding to a Lorentzian geometry. In other words, in Lorentz signature, the diagram line $\overline{12}$ has slope $m_{12}$, intersects the $r=0$ line at $t=t_1$, and thus connects with the diagram vertex 1 as it should.

Similarly, $t_3$ will be re-parametrized by a slope variable subject to the same spacelike-ensuring condition when Lorentzian through
\begin{equation}
    t_3=m_{23}z_3+b_{23},\quad\text{with}\quad b_{23}=t_2-m_{23}z_2.
\end{equation}
On similar ground, for a given boundary geometry, one might further constrain the domain of bulk variables in a Lorentzian path integral by the requirement that in the integration contour $|m_{23}|:=\frac{t_4-t_3}{z_4-z_3}\le1$ is guaranteed.

Similarly, considering that for Cauchy slices the normal remains timelike at all times and is continuous, so that the surface has no kinks, one might require $z_I<z_{I+1}$. This will in particular be imposed for $z_3$ by re-parametrizing it as the convex combination
\begin{equation}
    z_3=z_2+\rho_3(z_4-z_2),
\end{equation}
with $0\le\rho_3\le1$. Now, as the shell 4 is to correspond to a celestial sphere it is desirable to ensure that the bulk shells are smaller. This will particularly be imposed on shell 3 by re-parametrizing its scale $s_3$ by means of
\begin{equation}
    s_3=\sigma_{34} s_4,
\end{equation}
and restricting to $\sigma_{34}<1$.

So far, all of these are simply re-parametrizations that do not reduce the number of variables in question, but impose some reasonable conditions.

The restrictions exemplify how flexible Regge calculus can be as well as how `irregular' its histories can become. Indeed, as described above the configurations of $\triangle(\Sigma_\text{Int})$ can become timelike, or develop kinks —when gluing polyhedra with opposite orientations. In principle any of these could contribute to the path integral.\footnote{For a more thorough discussion of the configuration space of quantum Regge calculus, considering causality as well, see \cite{DittrichConfig:2025}.} The assumptions made above just tame some of these in principle unexpected (from the continuum point of view) phenomena, but could be relaxed in future explorations. The use of coordinate variables greatly simplifies the imposition of these constraints compared to using geometric invariants. Note also that these `sensible' restrictions were only imposed partially (\emph{i.e. on some variables}); nevertheless, the saddles discussed below satisfy them in their entirety.

Now to the final and most drastic refinements of the minisuperspace, in which the number of dynamic variables will actually be reduced. Ultimately, all quantities will depend on the length scale $s_1$ of the splitting surface in the replica topology and its counterpart in the Hawking topology.

First, $z_2$ will be expressed as a function of $s_2$, with the ansatz chosen so that as $s_2 \to 0$, $z_2$ approaches a finite, positive value. This choice is motivated by the expectation that quantum extremal surfaces lie just inside the horizon, so that as the horizon shrinks, their location moves toward the latest singular point in advanced time —the intersection of the horizon with the vertical dotted line in the final black-hole-less stage in the Hawking-Penrose diagram of FIG. \ref{fig:application}. Note however, that this imposition is a weaker property. The choice for such relation will be:
\begin{equation}
    z_2 = \frac{1}{1 + s_2}.
    \label{eq:z2_reduction}
\end{equation}
Note that incidentally, this also implies that $z_2$ being close to zero corresponds to large values of $s_2$. This behavior would correspond to the interior of a would-be black hole opening up.

Equation \eqref{eq:z2_reduction} is possibly the most artificial restriction in this construction. However, it should be remarked that other choices in which $z_2$ increases as $s_2$ decreases (\emph{e.g.} $z_2=constant-s_2$, or \eqref{eq:z2_reduction} with added parameters) have been implemented and the swap entropies show the same qualitative behavior, so that the final conclusions below would likely not need to be modified.

Finally, note that with the choice \eqref{eq:z2_reduction}, the frustum $\overline{12}$ (a tetrahedron in this case) collapses to a line in the $s_2 \to 0$ limit. This degenerate configuration could be understood to include the desired \emph{spacelike} singularity, but this is being forced by construction. To avoid this imposed degeneracy and more flexibly represent the geometry in this regime, future refinements should consider introducing intermediate vertices between the origin and $\triangle(\gamma)$ —Nevertheless note that even in the continuum, this limit is expected to correspond to a partial Cauchy slice without a boundary.

Next, $m_{23}$ will be identified with $m_{24}:=\frac{t_4-t_2}{z_4-z_2}$, meaning that the diagram vertices 2, 3 and 4 are set to be co-linear. Note that $m_{24}$ depends on the location of shell 2 and thus on $s_1$. When analyzing saddles, it will be important to verify if these slopes would correspond to a spacelike edge when Wick rotating to a Lorentzian regime. And for the result presented below, they do.

Finally, $m_{12}$, $\sigma_{34}$, $\rho_3$ and $t_1$ will be frozen and treated as parameters. Further, all Wick rotation angles are set to be $\pm\pi/2$, so that the geometry is Euclidean.

In effect, this means that all bulk geometry is parametrized by $s_1$. This dependence is such that (with the exception of $\overline{-10}$ and $s_3$, and $\phi_I$), the actual edge lengths are not frozen, but dependent on $s_1$. As discussed, this dependence is such that it partially implements, by construction, some of the causal structure depicted in the Hawking-Penrose diagram, when Wick rotating. 

\subsubsection{Microsuperspace saddles}

It is at this last stage of minisuperspace refinement, which can justifiably be termed microsuperspace, that saddles were sought.

To briefly recapitulate: By assuming replica symmetry and CPT-reducing the geometric variables, the Regge exponent can be calculated, analogously to the continuum situation, by analyzing a single $\triangle(\mathcal M)$ and the result can be analytically extended in the geometric variables as well as the replica number $n$. The same situation holds for the matter effective action, where special care needs to be taken to track all normalization factors produced by partial integrations. The remaining exponents, $W^{(n)}_{\square,\text{Semiclassical}}$, which combine the classical Regge exponent and the matter effective action in their respective topology $\square$, are the quantities to be \emph{extremized} in order to compute discrete swap entropies semiclassically. (General extrema —not just maxima— are sought, in line with the rationale of the Lorentzian path integral.) 

In such calculations, the natural variable to be treated as retarded time is
\begin{equation}
    \Delta z:=z_5-z_4
\end{equation}
in the sense that increasing it from zero while keeping $z_5$ fixed (and positive), corresponds to increasing $t_4$ while keeping it in the null line representing $\triangle(\mathscr I^+)$ in Hawking-Penrose diagram.

Then, once every other boundary variable and parameter has been fixed, the final goal is to compute (discrete) saddle-point-evaluated swap entropy
\begin{equation}
    S^{(n)}_\text{Swap}=\min_\square\left\{-\frac1{n-1}\left(W^{(n)}_{\square,\text{Semiclassical}}(\text{extremum}^{(n)})-nW^{(1)}_{\square,\text{Semiclassical}}(\text{extremum}^{(1)})\right)\right\}
\end{equation}
as a function of $\Delta z$. Note that the extremas of each of these exponents are generally different because of the difference in the replica number. 

As is to be expected, at least within the current model, whether there is a Page-like-transition behavior generically depends on the choice of boundary variables and parameters. However, the following remark is in place: the fact that there are saddles for both topologies and that $s_1$ decreases as $\Delta z$ grows appears to be robust. Additionally one typically observes that the fixed-topology swap entropy decreases with $\Delta z$ for the replica case. However, the increasing behavior in the Hawking topology is not as robust, which may be unsurprising given that this minimal setup might lack key physical ingredients underlying Hawking’s result. Similarly, even in regimes where the right monotonicity is satisfied, the two curves corresponding to different topologies might not cross either within the regime or without violating the conditions described in the previous subsection that were not imposed \emph{a priori} —\emph{e.g.} $|m_{34}|\overset!<1$ when Lorentzian.

\begin{figure}
    \centering
    \includegraphics[width=0.9\textwidth]{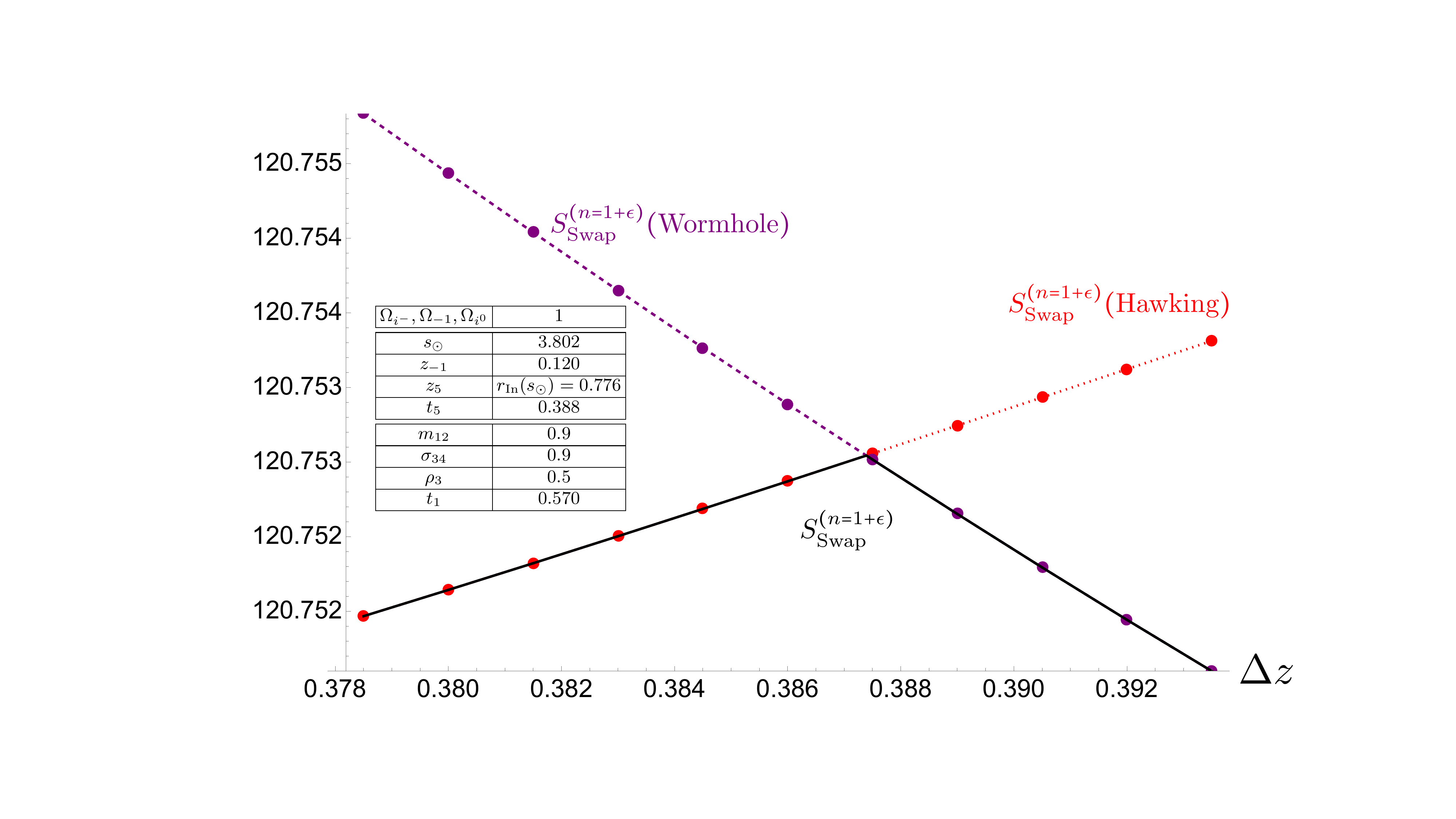}
    \caption{Semiclassical evaluation of the discrete swap entropy as $n \to 1^+$, using the parameters and boundary data from the table. As discussed in the text, $\Delta z$ serves as a discrete analogue of retarded time. The fixed-topology swap entropy for the Hawking (replica) topology, in red/dotted (purple/dashed), is seen to increase (decrease) as $\Delta z$ grows. The actual swap entropy (solid/black), computed by minimizing the two fixed-topology contributions, reveals a Page-like-transition behavior. 
    The table presents the values of the parameters and boundary values used for these calculations (and there $r_\text{In}(\square)$ represents \emph{inradius} of a regular tetrahedron of edge size $\square$, \emph{i.e.} the radius of the largest inscribed sphere that touches all four of its faces).}
    \label{fig:P-Transition}
\end{figure}

Nevertheless, it is possible to reproduce the Page transition within the setup, with saddles that satisfy all the unimplemented constraints. Indeed, in the $n\to1^+$ limit, the swap entropy behaves as shown in figure \ref{fig:P-Transition} using the parameters on its table. And, at each $\Delta z$, the constraints are satisfied.

\iffalse
\begin{table}[h]
    \centering
    \begin{tabular}{|c|c|}
        \hline
        $\Omega_{i^-},\Omega_{-1},\Omega_{i^0}$ & $1$\\
        \hline
        \hline
        $s_{\odot}$ & $3.802$\\
        \hline
        $z_{-1}$ & $0.120$\\
        \hline
        $z_{5}$ & $r_\text{In}(s_\odot)=0.776$\\
        \hline
        $t_{5}$ & $0.388$\\
        \hline
        \hline
        $m_{12}$ & $0.9$\\
        \hline
        $\sigma_{34}$ & $0.9$\\
        \hline
        $\rho_{3}$ & $0.5$\\
        \hline
        $t_{1}$ & $0.570$\\
        \hline
    \end{tabular}
\end{table}
\fi

What remains to be seen is whether the behavior shown in FIG. \ref{fig:P-Transition} persists in more detailed calculations and whether the entire Page curve can ultimately be reproduced. (The latter, in particular, may be challenging in this first implementation without violating one of the imposed constraints, as suggested by numerical explorations.) Nonetheless, given the observed robustnesses and the fact that Regge calculus approximates General Relativity in the continuum, this is an encouraging result.

Indeed, that a simple coarse-grained model within this discrete four-dimensional framework already reproduces a Page-like transition is a highly encouraging sign and suggests that the essential mechanisms behind the Page curve may be generic.  While much remains to be done (most pressingly, relaxing the minisuperspace constraints —thereby restoring Lorentzian signature— and refining the triangulation to capture black hole geometry more faithfully) the current results offer concrete evidence that these goals are within reach. Far from being a toy exercise, this first implementation provides a firm foundation on which to build. It shows that Regge calculus can connect replica-based reasoning with numerical, nonperturbative approaches to quantum gravity, opening the door to a new class of explicitly computable models of black hole evaporation.

\section{Overview, Discussion and Outlook \label{sec:discussion}}

In recent years, the replica paradigm has significantly advanced the understanding of the black hole information paradox. A growing body of evidence suggests that it can be used to reproduce the Page curve, strongly indicating that evaporating black holes evolve unitarily as quantum systems. Importantly, and in tension with old folklore, this approach does not rely on strong UV assumptions; rather it remains semiclassical, depending only on saddle point evaluations of gravitational path integrals that, however, are assumed to include topology fluctuations.

Although often treated as a “trick”, in this context the replica calculation can be operationally motivated by considering an observer measuring the entropy of radiation collected up to some retarded time $u$. Since entropy measurements inherently require replicating the system, this forces consideration of an ensemble of $n$ evaporating black holes.

Within this setup, an observable can be defined whose expectation value characterizes $n$-Rényi entropies when the spacetime topology remains trivial (disconnected in the bulk) across the replicas. The key insight is that this expectation value can be expressed as a path integral, and if the gravitational path integral sums over topologies, contributions from nontrivial topologies (such as those connecting the replicated spacetimes via an `entangling' bulk 2-surface) may appear. The presence of these contributions alters the expectation value and it no longer strictly corresponds to an $n$-Rényi entropy, but rather a more general quantity termed \emph{swap entropy}.

It has been argued that, at least in the formal limit $n\to1^+$, where Rényi entropies coincide with the von Neumann entropy, saddles with a nontrivial wormhole topology exist. Therefore, in this limit, the semiclassical path integral exhibits a competition between Hawking saddles and wormhole saddles. At early retarded times Hawking saddles dominate, leading to the expected entropy growth —consistent with the von Neumann entropy computation. However, at later times, the wormhole topology contribution takes over, causing the swap entropy to decrease and eventually vanish. This change in dominance corresponds to the Page transition and results in the emergence of the Page curve.

As remarkable as this progress is, it leaves things to be desired. First, the saddles that have been found in literature pertain to relatively simple models, such as JT gravity, and not 4-dimensional Einstein Gravity. Additionally, they have only been found in the $n\to1^+$ limit, which is not satisfactory from the operational point of view –however, see \cite{Mirbabayi:2020fyk} for numerical studies on the case $n=2$ for a two-dimensional theory. Further, with the exception of \cite{Held:2024qcl} which analyzed complex saddles in JT gravity calculations, virtually all analyses have been done in Euclidean signature and therefore if one takes the operational approach advocated here, it is important to argue that they actually contribute to the Lorentzian path integral. 

The purpose of the work presented in this paper was to begin addressing these gaps from the point of view of Regge calculus, understood as a lattice-like approach to quantum gravity. To this end, a discretization scheme has been proposed for 4-dimensional evaporating black hole spacetimes that allows one to frame the computation of swap entropies within this discrete framework.

All fundamental objects that would appear in any implementation of the framework have been computed explicitly in the spherically symmetric case, with the exception of potential dual 4-volume factors that could appear in the ambiguous gravitational measure. Although this measure is not needed for semiclassical evaluations, which are the first natural step in this line of research, it will be important to revisit its ambiguities in the future if one is to compute quantum corrections. 

The computed fundamental objects (namely, the bone and cell contributions to the Regge action, as well as the dual volumes) have been parametrized in a way that facilitates analytic continuation. This is essential for contour deformations in real-time calculations to assess whether complex saddles contribute. However, the literature indicates that the analytic structure of Regge actions is highly nontrivial, leading to intricate Riemann surface topologies \cite{Asante:2021phx}. Consequently, it is imperative to develop a technical framework capable of handling these surfaces in higher dimensions than those previously dealt with. An alternative approach is to adapt holomorphic gradient flow methods \cite{Alexandru:2020wrj} to Regge calculus, akin to the techniques used in \cite{Jia:2021xeh}; see also \cite{Han:2020npv} for related work. These advancements could, optimistically, bridge the replica paradigm with high-performance computing and foster productive new connections.

To justify commitment to these developments, it was crucial to establish a proof of principle for the proposal. Accordingly, a concrete implementation of the framework was realized. Although the chosen triangulation is coarse, it provides sufficient radiation degrees of freedom to bear swap entropies. However, this comes at the cost of introducing a significant number of gravitational degrees of freedom. The matter content was taken to be a free, massless, and minimally coupled scalar field, which allowed for its full analytical integration in the swap entropy calculations.

Given the large number of geometric variables, a minisuperspace was adopted for the discrete gravitational field to facilitate the search for semiclassical saddles. While the final minisuperspace contained only a single variable, the reduction was carried out in stages of increasing boldness, allowing for re-examination once the appropriate infrastructure is developed. Within this (in particular Euclidean) minisuperspace, saddles can be found in the $n\to 1^+$ limit (one for each topology) that reproduce the Page transition. This suggests that sufficiently refined discrete calculations could recover the Page curve and thereby address the questions outlined above. However, given the coarseness of the model, it can, at this stage, only serve as an indication of the generality of this behavior. Even more conservatively, it may simply be viewed as a useful validation of the proposal, offering several interesting questions regarding the framework.

One open issue that becomes particularly relevant here is how to determine whether a given Regge configuration corresponds to a spacetime containing a black hole. Unlike in the continuum, where event and apparent horizons, and trapped surfaces are well-defined, such features have not thoroughly been adapted to Regge calculus. A better understanding of how to diagnose black hole structures in triangulated geometries (particularly in coarse ones) would greatly enhance the interpretability of results and inform the refinement of triangulations used in swap entropy calculations.

Another key question, which has been emphasized, is how to reconcile the framework of Regge calculus with the fact that the continuum calculation being discretized extends to asymptotic infinity. Regge calculus is naturally suited for finite-region calculations, and while one might initially expect that a continuum calculation with asymptotic boundaries could be approximated by a discrete Regge calculus calculation with finite boundaries and then taking the relevant edge lengths at these boundaries to infinity, care must be taken to ensure that a subsequent continuum limit does not inadvertently truncate the number of degrees of freedom in the asymptotic region. Some considerations were briefly discussed in previous sections, but this issue remains open.

Closely related to this is the question of how to handle spacetime asymptotic corners, particularly the boundary degrees of freedom residing there. It is unclear whether these should be associated with state preparation path integrals (as with degrees of freedom in $\mathscr{I}^-$) or with sewing path integrals (as with degrees of freedom in $\mathscr{I}_u^+$). Alternatively, one might need to take a different approach, such as “blowing up” these asymptotic corners. While this issue might seem artificial for asymptotic boundaries, it is likely to resurface in finite-boundary calculations. In any case, while addressing this problem is important for future work, the results at present do not seem to be sensitive to the choices discussed above, and they should be independent of the decision made in \S\ref{sec:application} to associate the corner degrees of freedom with the preparation path integral. This expectation of independence is supported by the RG-like property observed in the matter effective actions.

One of the most promising ways to address the issue of asymptotic boundaries may be to take a different perspective: keeping the quantum Regge calculus calculation confined to a finite region where quantum gravity effects are relevant, while performing a continuum QFT calculation on a curved but fixed background where gravity does not back react. However, this requires further study of mappings between continuous and discrete boundary data and how they change with triangulation refinements —for this, considering the ``consistent boundary formulation'' of \cite{Dittrich:2012jq,Asante:2022dnj} could be useful.

The discretization scheme presented offers significant flexibility, which allows for its tailored application. For instance, different regions of spacetime can be triangulated with varying levels of coarseness. Indeed, as mentioned, the fundamental shells considered can be assembled in multiple ways, enabling arbitrarily fine triangulations of 2-spheres and an arbitrary number of cells meeting at bones. However, this flexibility comes with ambiguities that have not been thoroughly explored in the literature. Some issues related to the measure for the scalar and gravitational fields have been discussed in \S\ref{sec:discrete}, but a systematic comparison of the choices made here with alternative possibilities remains necessary —This comparison should be done both from \emph{a piori} and \emph{a posteriori} perspectives, and also to consider whether differences are relevant as one considers increasingly refined triangulation. Whatever the case, it seems unlikely that such ambiguities would lead to qualitative differences at the level of semiclassical gravitational calculations.

However, there is an ambiguity in defining the contour of the Regge path integral, which could significantly impact the semiclassical relevance of saddles. The naïve approach of summing over all discrete geometries introduces apparently anomalous histories that have no clear continuum counterpart. For instance, causal irregularities such as the ones in splitting surfaces (where an unusual number of light cones emerges) can occur even in bones that are not topologically special \cite{Asante:2021phx}. Similarly, in the replica calculation, some configurations may lack a global time foliation, conflicting with the structure suggested by Hawking-Penrose diagrams. Beyond causality, other pathologies arise, such as orientation flips in faces, which can lead to branch points in the action. The inclusion or exclusion of these configurations in the defining contour of the path integral directly affects which saddle points contribute, as dictated by the analytic structure of the integrand \emph{via} the Picard-Lefschetz theorem.

At the same time, experience from the non-relativistic path integral warns against overly restrictive choices. Wildly fluctuating histories often play an essential role in quantum theories, and excluding certain configurations without systematic justification could miss important contributions. A careful study of how these ambiguities affect the semiclassical analysis is therefore needed. While much of this remains unexplored, efforts to map the space of Regge configurations and to develop tools to impose some exclusions (if needed), particularly in relation to causal structure, have begun in \cite{DittrichConfig:2025}.

The framework herein presented has particular relevance for Loop Quantum Gravity (LQG), especially in the context of spinfoam dynamics. Spinfoams provide a covariant formulation of LQG, and several arguments \cite{Barrett:2009gg,Barrett:2009mw,Han:2021kll,Han:2023cen} suggest that, in the large quantum number limit, they reduce to a sum very closely related to a quantum Regge calculus path integral. Therefore, the type of calculation studied here can be seen as a bridge between the replica paradigm and LQG. A reasonable intermediary step for this end might be to adapt the framework to \emph{effective spin foams} \cite{Asante:2020qpa,Asante:2020iwm,Asante:2021zzh}, which have a closer relation to QRC than the “state-of-the-art” spinfoam model, the EPRL/FK model \cite{Engle:2007uq,Freidel:2007py}.

However, it is important to note that it remains unclear whether the Lorentzian EPRL/FK model admits imaginary contributions of the type \eqref{eq:imaginary_contribution}, which, as shown, play a crucial role in the replica mechanism responsible for reproducing the Page curve —see \cite{Bodendorfer:2013hla} for a related discussion.

In fact, given the close relationship between QRC and spinfoams, adapting this framework to the black-to-white-hole transition scenario \cite{Haggard:2014rza,DAmbrosio:2018wgv,DeLorenzo:2015gtx} for an evaporating black hole would be of significant interest. This transition has been a long-standing goal in the LQG community \cite{DAmbrosio:2020mut,Soltani:2021zmv}, and the modular elements computed here could provide a foundation for a QRC-based computation of this process.

Indeed, some of the (modular) results presented here apply to the Regge action of any spacetime that admits a Penrose diagram representation, as well as to the action of matter fields that could model Hawking radiation. That is, while motivated by the replica paradigm, the scheme developed here extends beyond it. This reinforces the versatility of QRC as a lattice-based approach to quantum gravity, offering a concrete and computationally accessible avenue for exploring the gravitational path integral. However, key questions remain, \emph{e.g.}, the proper definition of the discrete path integral configuration space, the gravitational measure, the treatment of asymptotic boundaries, etc. Addressing these challenges will be crucial to fully realizing the framework’s potential, both as a bridge between discrete and continuum approaches and as a tool to probe the deep structure of quantum gravity.

\centerline{\emph{Vale}}
\section*{Acknowledgments}
I thank Dongxue Qu for pointing out useful references on matter in Regge calculus, and Donald Marolf for clarifying correspondence. I am also grateful to Fedro Guillén, Leonardo A. Lessa, Harish Murali, Marc Schiffer, and Jinmin Yi for (in some cases, many) valuable discussions. Special thanks to Eugenia Colafranceschi for her helpful comments on a draft of this paper. My deepest gratitude goes to Bianca Dittrich for her long-term and persistent encouragement of this project, numerous insightful discussions, and feedback on multiple versions of this manuscript.

I am supported by a NSERC grant awarded
to Bianca Dittrich. Research at Perimeter Institute is supported in
part by the Government of Canada through the Department of Innovation, Science and Economic Development Canada and by the Province of Ontario through the Ministry of Colleges and Universities.

\appendix

\section{Performing the matter integrals of the entangling surfaces last \label{app:entangling_integrals}}

Although the order of integration presented in the main text is conceptually clearer for following the structure of the matter path integrals, it is computationally more convenient to perform the integrals over the degrees of freedom on the entangling surfaces last. Since this is the approach taken in the code presented in the GitHub repository \cite{github}, this appendix outlines how the discussion is modified accordingly.

By integrating out all degrees of freedom except those on the entangling surfaces and those involved in the sewing integrals, the computation of the effective action reduces (once again, due to the closure of Gaussian kernels under integration) to an expression of the form:
\begin{equation}
e^{W^{(n)}_\text{Eff}} =\int\mathrm d^mu_i \int \mathrm d^n z \exp(\text{quadratic terms}),
\end{equation}
where $m=1$ $(m=2)$ for the Hawking (wormhole) topology. The index $i=1, \dots, m$ labels the degrees of freedom $u_i$ on the entangling surfaces. The full expression is:
\begin{gather}
    e^{W^{(n)}_\text{Eff}}\nonumber\\
    =\nonumber\\
    \int\mathrm d^{m}u_i\mathrm d^{n}z\nonumber\\
    \times\nonumber\\
    e^{\tilde \alpha z_1^2+\tilde \beta z_1z_2+\tilde \gamma z_2^2+\dots+\tilde \alpha z_{n-1}^2+\tilde \beta z_{n-1}z_1+\tilde \gamma z_1^2+(\sum_i \kappa_i u_i)(z_1+\dots+z_{n-1})+(\sum_i\lambda_i u_i)(z_2+\dots+z_{n-1}+z_1)+\sum_i\nu_i u_i^2}
    \label{eq:corner_last_effective_action_with_composition}
\end{gather}
Tilded symbols have been used to emphasize that the coefficients \emph{do not} agree with those of \eqref{eq:Gaussian_composition}, (particularly here the integrals over $u_i$ have not been performed) —recall that they also depend on the topology in question. Two linear terms in $u_i$ have been written to communicate that there is one provided by each member of CPT spacetime-pairs. All these coefficients can be computed explicitly by performing the relevant partial integrations, and they are shown explicitly in the GitHub repository \cite{github}.

The integrals over the $z$-variables can be computed by the same arguments given in \S\ref{sec:application}, specifically
\begin{gather}
    \int\mathrm d^{n}z  e^{\tilde \alpha z_1^2+\tilde \beta z_1z_2+\tilde \gamma z_2^2+\dots+\tilde \alpha z_{n-1}^2+\tilde \beta z_{n-1}z_1+\tilde \gamma z_1^2+(\sum_i\kappa_i u_i)(z_1+\dots+z_{n-1})+(\sum_i\lambda_i u_i)(z_2+\dots+z_{n-1}+z_1)+\sum_i\nu_i u_i^2},\nonumber\\
    =:\nonumber\\
    \int\mathrm d^{n}z\exp\left(-\frac12\vec{z}^\text{T}\tilde{\mathbb T}^{(n)}\vec{z}+\vec{\tilde b}^{(n)}(u_i)^\text{T}\vec{z}+c(u_i)\right)\nonumber\\
    =\nonumber\\
    \sqrt{\frac{(2\pi)^{n}}{\det\tilde{\mathbb T}^{(n)}}}e^{\frac12\vec{\tilde b}^{(n)}(u_i)^\text{T}\left(\tilde{\mathbb T}^{(n)}\right)^{-1}\vec{\tilde b}^{(n)}(u_i)},
    \label{eq:corner_last_effective_action_with_T}
\end{gather}
where $\tilde{\mathbb T}^{(n)}\neq \mathbb T^{(n)}$ is the $n\times n$ matrix defined by
\begin{equation}
    \tilde{\mathbb T}^{(n)}:=
    \begin{pmatrix}
        \tilde\delta & \tilde\mu & 0 & \phantom{0}\cdots &  \phantom{0}\tilde\mu \\
        \tilde\mu & \tilde\delta & \tilde\mu & \phantom{0}\ddots &  \phantom{0}\vdots \\
        0 & \tilde\mu & \tilde\delta & \phantom{0}\ddots &  \phantom{0}0 \\
        \vdots & \ddots & \ddots & \phantom{0}\ddots &  \phantom{0}\tilde\mu \\
        \tilde\mu & \cdots & 0 & \phantom{0}\tilde\mu &  \phantom{0}\tilde\delta
        \label{eq:T_tilde}
    \end{pmatrix},
\end{equation}
with $\delta:=-2(\tilde\alpha+\tilde\gamma)$ and $\mu:=-\beta$; and $\vec{\tilde b}^{(n)}\neq\vec b^{(n)}$ the $n$-dimensional vector
\begin{equation}
    \vec{\tilde b}^{(n)}(u_i)^{\text T}:=\left(\sum_i(\kappa_i u_i+\lambda_i u_i)\quad\dots\quad\sum_i(\kappa_i u_i+\lambda_i u_i))\right)=\sum_i(\kappa_i+\lambda_i)u_i(1\quad\cdots\quad1)=:\sum_i(\kappa_i+\lambda_i)u_i\mathbf 1^{(n)}.
\end{equation}
The square root factor in \eqref{eq:corner_last_effective_action_with_T} can be computed with the results of \S\ref{sec:application}, because, defining $\tilde r_{xy}$ as the same gaussian kernel than in eq. \eqref{eq:Gaussian_composition} but with tilded coefficients, it is clear that
\begin{equation}
    \sqrt{\frac{(2\pi)^{n}}{\det\tilde{\mathbb T}^{(n)}}}=\text{Tr}(\tilde r^n)\overset{\eqref{eq:closed_trace}}{=}\sqrt{\frac{(2\pi)^n}{2\tilde\beta^{n}\left((-1)^n\cos\left(n\arccos\left(
    \frac{\tilde\alpha+\tilde\gamma}{\tilde\beta}\right)\right)-1\right)}}.
    \label{eq:corner_last_sqrt}
\end{equation}
Naturally, this expression could be re-written as a relation between the determinants of $\mathbb T$ and $\tilde{\mathbb T}$, so that it is also possible to find a close expression for the determinant of \eqref{eq:T_tilde}.

To compute the exponent in the last line of \eqref{eq:corner_last_effective_action_with_T} one can begin by noticing that 
\begin{equation}
    \vec{\tilde b}^{(n)}(u_i)^\text{T}\left(\tilde{\mathbb T}^{(n)}\right)^{-1}\vec{\tilde b}^{(n)}(u_i)=\left(\sum_i(\kappa_i u_i+\lambda_i u_i)\right)^2\sum_{j,k}\left[\left(\tilde{\mathbb T}^{(n)}\right)^{-1}\right]_{jk}.
\end{equation}
So the particular form of $\vec{\tilde b}^{(n)}$ implies that one only needs to perform the sum of all the entries of $\left(\tilde{\mathbb T}^{(n)}\right)^{-1}$. This can be done by hinging on the fact that the $(j+1)$-th column of $\tilde{\mathbb T}^{(n)}$ is obtained by shifting the $j$-th column downward by one entry (with wrap-around), \emph{i.e.}, that the matrix is \emph{circulant}.\footnote{The arguments that follow are just a consequence of well understood (and stronger) properties of circulant matrices.} This implies that
\begin{equation}
    \tilde{\mathbb T}^{(n)}\mathbf 1^{(n)}=\sum_{k}\left[\tilde{\mathbb T}^{(n)}\right]_{1k}\mathbf 1^{(n)}\quad\text{so that}\quad \left(\tilde{\mathbb T}^{(n)}\right)^{-1}\mathbf 1^{(n)}=\frac1{\sum_{k}\left[\tilde{\mathbb T}^{(n)}\right]_{1k}}\mathbf 1^{(n)}
\end{equation}
so that 
\begin{equation}
    \sum_{j,k}\left[\left(\tilde{\mathbb T}^{(n)}\right)^{-1}\right]_{jk}=\left(\mathbf 1^{(n)}\right)^\text{T}\left(\tilde{\mathbb T}^{(n)}\right)^{-1}\mathbf 1^{(n)}=\frac1{\sum_{k}\left[\tilde{\mathbb T}^{(n)}\right]_{1k}}\left(\mathbf 1^{(n)}\right)^\text{T}\mathbf 1^{(n)}=\frac n {\sum_{k}\left[\tilde{\mathbb T}^{(n)}\right]_{1k}}.
\end{equation}
And therefore 
\begin{equation}
    \vec{\tilde b}^{(n)}(u_i)^\text{T}\left(\tilde{\mathbb T}^{(n)}\right)^{-1}\vec{\tilde b}^{(n)}(u_i)=n\left(\sum_i(\kappa_i u_i+\lambda_i u_i)\right)^2/(2\tilde\mu+\tilde\delta).
    \label{eq:corner_last_exponent}
\end{equation}
Thus, by plugging \eqref{eq:corner_last_sqrt} and \eqref{eq:corner_last_exponent} into \eqref{eq:corner_last_effective_action_with_T}, what remains to compute the effective action in \eqref{eq:corner_last_effective_action_with_T} for any topology is to perform a gaussian integral over the entangling surface field degrees of freedom, which manifestly produces a closed expression in terms of $n$. Such expression is shown in the GitHub repository in \cite{github}.

\bibliographystyle{apsrev4-1}
\bibliography{references}

%merlin.mbs apsrev4-1.bst 2010-07-25 4.21a (PWD, AO, DPC) hacked
%Control: key (0)
%Control: author (72) initials jnrlst
%Control: editor formatted (1) identically to author
%Control: production of article title (-1) disabled
%Control: page (0) single
%Control: year (1) truncated
%Control: production of eprint (0) enabled
\begin{thebibliography}{95}%
\makeatletter
\providecommand \@ifxundefined [1]{%
 \@ifx{#1\undefined}
}%
\providecommand \@ifnum [1]{%
 \ifnum #1\expandafter \@firstoftwo
 \else \expandafter \@secondoftwo
 \fi
}%
\providecommand \@ifx [1]{%
 \ifx #1\expandafter \@firstoftwo
 \else \expandafter \@secondoftwo
 \fi
}%
\providecommand \natexlab [1]{#1}%
\providecommand \enquote  [1]{``#1''}%
\providecommand \bibnamefont  [1]{#1}%
\providecommand \bibfnamefont [1]{#1}%
\providecommand \citenamefont [1]{#1}%
\providecommand \href@noop [0]{\@secondoftwo}%
\providecommand \href [0]{\begingroup \@sanitize@url \@href}%
\providecommand \@href[1]{\@@startlink{#1}\@@href}%
\providecommand \@@href[1]{\endgroup#1\@@endlink}%
\providecommand \@sanitize@url [0]{\catcode `\\12\catcode `\$12\catcode
  `\&12\catcode `\#12\catcode `\^12\catcode `\_12\catcode `\%12\relax}%
\providecommand \@@startlink[1]{}%
\providecommand \@@endlink[0]{}%
\providecommand \url  [0]{\begingroup\@sanitize@url \@url }%
\providecommand \@url [1]{\endgroup\@href {#1}{\urlprefix }}%
\providecommand \urlprefix  [0]{URL }%
\providecommand \Eprint [0]{\href }%
\providecommand \doibase [0]{http://dx.doi.org/}%
\providecommand \selectlanguage [0]{\@gobble}%
\providecommand \bibinfo  [0]{\@secondoftwo}%
\providecommand \bibfield  [0]{\@secondoftwo}%
\providecommand \translation [1]{[#1]}%
\providecommand \BibitemOpen [0]{}%
\providecommand \bibitemStop [0]{}%
\providecommand \bibitemNoStop [0]{.\EOS\space}%
\providecommand \EOS [0]{\spacefactor3000\relax}%
\providecommand \BibitemShut  [1]{\csname bibitem#1\endcsname}%
\let\auto@bib@innerbib\@empty
%</preamble>
\bibitem [{\citenamefont {Penington}\ \emph {et~al.}(2022)\citenamefont
  {Penington}, \citenamefont {Shenker}, \citenamefont {Stanford},\ and\
  \citenamefont {Yang}}]{Penington:2019kki}%
  \BibitemOpen
  \bibfield  {author} {\bibinfo {author} {\bibfnamefont {G.}~\bibnamefont
  {Penington}}, \bibinfo {author} {\bibfnamefont {S.~H.}\ \bibnamefont
  {Shenker}}, \bibinfo {author} {\bibfnamefont {D.}~\bibnamefont {Stanford}}, \
  and\ \bibinfo {author} {\bibfnamefont {Z.}~\bibnamefont {Yang}},\ }\href
  {\doibase 10.1007/JHEP03(2022)205} {\bibfield  {journal} {\bibinfo  {journal}
  {JHEP}\ }\textbf {\bibinfo {volume} {03}},\ \bibinfo {pages} {205} (\bibinfo
  {year} {2022})},\ \Eprint {http://arxiv.org/abs/1911.11977} {arXiv:1911.11977
  [hep-th]} \BibitemShut {NoStop}%
\bibitem [{\citenamefont {Almheiri}\ \emph {et~al.}(2020)\citenamefont
  {Almheiri}, \citenamefont {Hartman}, \citenamefont {Maldacena}, \citenamefont
  {Shaghoulian},\ and\ \citenamefont {Tajdini}}]{Almheiri:2019qdq}%
  \BibitemOpen
  \bibfield  {author} {\bibinfo {author} {\bibfnamefont {A.}~\bibnamefont
  {Almheiri}}, \bibinfo {author} {\bibfnamefont {T.}~\bibnamefont {Hartman}},
  \bibinfo {author} {\bibfnamefont {J.}~\bibnamefont {Maldacena}}, \bibinfo
  {author} {\bibfnamefont {E.}~\bibnamefont {Shaghoulian}}, \ and\ \bibinfo
  {author} {\bibfnamefont {A.}~\bibnamefont {Tajdini}},\ }\href {\doibase
  10.1007/JHEP05(2020)013} {\bibfield  {journal} {\bibinfo  {journal} {JHEP}\
  }\textbf {\bibinfo {volume} {05}},\ \bibinfo {pages} {013} (\bibinfo {year}
  {2020})},\ \Eprint {http://arxiv.org/abs/1911.12333} {arXiv:1911.12333
  [hep-th]} \BibitemShut {NoStop}%
\bibitem [{\citenamefont {Marolf}\ and\ \citenamefont
  {Maxfield}(2021{\natexlab{a}})}]{Marolf:2020rpm}%
  \BibitemOpen
  \bibfield  {author} {\bibinfo {author} {\bibfnamefont {D.}~\bibnamefont
  {Marolf}}\ and\ \bibinfo {author} {\bibfnamefont {H.}~\bibnamefont
  {Maxfield}},\ }\href {\doibase 10.1007/JHEP04(2021)272} {\bibfield  {journal}
  {\bibinfo  {journal} {JHEP}\ }\textbf {\bibinfo {volume} {04}},\ \bibinfo
  {pages} {272} (\bibinfo {year} {2021}{\natexlab{a}})},\ \Eprint
  {http://arxiv.org/abs/2010.06602} {arXiv:2010.06602 [hep-th]} \BibitemShut
  {NoStop}%
\bibitem [{\citenamefont {Chandrasekaran}\ \emph {et~al.}(2022)\citenamefont
  {Chandrasekaran}, \citenamefont {Engelhardt}, \citenamefont {Fischetti},\
  and\ \citenamefont {Hern\'andez-Cuenca}}]{Chandrasekaran:2022asa}%
  \BibitemOpen
  \bibfield  {author} {\bibinfo {author} {\bibfnamefont {V.}~\bibnamefont
  {Chandrasekaran}}, \bibinfo {author} {\bibfnamefont {N.}~\bibnamefont
  {Engelhardt}}, \bibinfo {author} {\bibfnamefont {S.}~\bibnamefont
  {Fischetti}}, \ and\ \bibinfo {author} {\bibfnamefont {S.}~\bibnamefont
  {Hern\'andez-Cuenca}},\ }\href {\doibase 10.1007/JHEP11(2022)110} {\bibfield
  {journal} {\bibinfo  {journal} {JHEP}\ }\textbf {\bibinfo {volume} {11}},\
  \bibinfo {pages} {110} (\bibinfo {year} {2022})},\ \Eprint
  {http://arxiv.org/abs/2207.09472} {arXiv:2207.09472 [hep-th]} \BibitemShut
  {NoStop}%
\bibitem [{\citenamefont {Marolf}\ and\ \citenamefont
  {Maxfield}(2021{\natexlab{b}})}]{Marolf:2021ghr}%
  \BibitemOpen
  \bibfield  {author} {\bibinfo {author} {\bibfnamefont {D.}~\bibnamefont
  {Marolf}}\ and\ \bibinfo {author} {\bibfnamefont {H.}~\bibnamefont
  {Maxfield}},\ }\href {\doibase 10.1142/S021827182142027X} {\bibfield
  {journal} {\bibinfo  {journal} {Int. J. Mod. Phys. D}\ }\textbf {\bibinfo
  {volume} {30}},\ \bibinfo {pages} {2142027} (\bibinfo {year}
  {2021}{\natexlab{b}})},\ \Eprint {http://arxiv.org/abs/2105.12211}
  {arXiv:2105.12211 [hep-th]} \BibitemShut {NoStop}%
\bibitem [{\citenamefont {Rocek}\ and\ \citenamefont
  {Williams}(1981)}]{Rocek:1981ama}%
  \BibitemOpen
  \bibfield  {author} {\bibinfo {author} {\bibfnamefont {M.}~\bibnamefont
  {Rocek}}\ and\ \bibinfo {author} {\bibfnamefont {R.~M.}\ \bibnamefont
  {Williams}},\ }\href {\doibase 10.1016/0370-2693(81)90848-0} {\bibfield
  {journal} {\bibinfo  {journal} {Phys. Lett. B}\ }\textbf {\bibinfo {volume}
  {104}},\ \bibinfo {pages} {31} (\bibinfo {year} {1981})}\BibitemShut
  {NoStop}%
\bibitem [{\citenamefont {Hamber}(2009{\natexlab{a}})}]{Hamber:2009mt}%
  \BibitemOpen
  \bibfield  {author} {\bibinfo {author} {\bibfnamefont {H.~W.}\ \bibnamefont
  {Hamber}},\ }\href {\doibase 10.1007/s10714-009-0769-y} {\bibfield  {journal}
  {\bibinfo  {journal} {Gen. Rel. Grav.}\ }\textbf {\bibinfo {volume} {41}},\
  \bibinfo {pages} {817} (\bibinfo {year} {2009}{\natexlab{a}})},\ \Eprint
  {http://arxiv.org/abs/0901.0964} {arXiv:0901.0964 [gr-qc]} \BibitemShut
  {NoStop}%
\bibitem [{\citenamefont {Baez}(1998)}]{Baez:1997zt}%
  \BibitemOpen
  \bibfield  {author} {\bibinfo {author} {\bibfnamefont {J.~C.}\ \bibnamefont
  {Baez}},\ }\href {\doibase 10.1088/0264-9381/15/7/004} {\bibfield  {journal}
  {\bibinfo  {journal} {Class. Quant. Grav.}\ }\textbf {\bibinfo {volume}
  {15}},\ \bibinfo {pages} {1827} (\bibinfo {year} {1998})},\ \Eprint
  {http://arxiv.org/abs/gr-qc/9709052} {arXiv:gr-qc/9709052} \BibitemShut
  {NoStop}%
\bibitem [{\citenamefont {Perez}(2003)}]{Perez:2003vx}%
  \BibitemOpen
  \bibfield  {author} {\bibinfo {author} {\bibfnamefont {A.}~\bibnamefont
  {Perez}},\ }\href {\doibase 10.1088/0264-9381/20/6/202} {\bibfield  {journal}
  {\bibinfo  {journal} {Class. Quant. Grav.}\ }\textbf {\bibinfo {volume}
  {20}},\ \bibinfo {pages} {R43} (\bibinfo {year} {2003})},\ \Eprint
  {http://arxiv.org/abs/gr-qc/0301113} {arXiv:gr-qc/0301113} \BibitemShut
  {NoStop}%
\bibitem [{\citenamefont {Dittrich}\ \emph {et~al.}(2024)\citenamefont
  {Dittrich}, \citenamefont {Jacobson},\ and\ \citenamefont
  {Padua-Arg\"uelles}}]{Dittrich:2024awu}%
  \BibitemOpen
  \bibfield  {author} {\bibinfo {author} {\bibfnamefont {B.}~\bibnamefont
  {Dittrich}}, \bibinfo {author} {\bibfnamefont {T.}~\bibnamefont {Jacobson}},
  \ and\ \bibinfo {author} {\bibfnamefont {J.}~\bibnamefont
  {Padua-Arg\"uelles}},\ }\href {\doibase 10.1103/PhysRevD.110.046006}
  {\bibfield  {journal} {\bibinfo  {journal} {Phys. Rev. D}\ }\textbf {\bibinfo
  {volume} {110}},\ \bibinfo {pages} {046006} (\bibinfo {year} {2024})},\
  \Eprint {http://arxiv.org/abs/2403.02119} {arXiv:2403.02119 [gr-qc]}
  \BibitemShut {NoStop}%
\bibitem [{\citenamefont {Louko}\ and\ \citenamefont
  {Sorkin}(1997)}]{Louko:1995jw}%
  \BibitemOpen
  \bibfield  {author} {\bibinfo {author} {\bibfnamefont {J.}~\bibnamefont
  {Louko}}\ and\ \bibinfo {author} {\bibfnamefont {R.~D.}\ \bibnamefont
  {Sorkin}},\ }\href {\doibase 10.1088/0264-9381/14/1/018} {\bibfield
  {journal} {\bibinfo  {journal} {Class. Quant. Grav.}\ }\textbf {\bibinfo
  {volume} {14}},\ \bibinfo {pages} {179} (\bibinfo {year} {1997})},\ \Eprint
  {http://arxiv.org/abs/gr-qc/9511023} {arXiv:gr-qc/9511023} \BibitemShut
  {NoStop}%
\bibitem [{\citenamefont {Asante}\ \emph
  {et~al.}(2023{\natexlab{a}})\citenamefont {Asante}, \citenamefont
  {Dittrich},\ and\ \citenamefont {Padua-Arg\"uelles}}]{Asante:2021phx}%
  \BibitemOpen
  \bibfield  {author} {\bibinfo {author} {\bibfnamefont {S.~K.}\ \bibnamefont
  {Asante}}, \bibinfo {author} {\bibfnamefont {B.}~\bibnamefont {Dittrich}}, \
  and\ \bibinfo {author} {\bibfnamefont {J.}~\bibnamefont
  {Padua-Arg\"uelles}},\ }\href {\doibase 10.1088/1361-6382/accc01} {\bibfield
  {journal} {\bibinfo  {journal} {Class. Quant. Grav.}\ }\textbf {\bibinfo
  {volume} {40}},\ \bibinfo {pages} {105005} (\bibinfo {year}
  {2023}{\natexlab{a}})},\ \Eprint {http://arxiv.org/abs/2112.15387}
  {arXiv:2112.15387 [gr-qc]} \BibitemShut {NoStop}%
\bibitem [{\citenamefont {Jia}(2022)}]{Jia:2021xeh}%
  \BibitemOpen
  \bibfield  {author} {\bibinfo {author} {\bibfnamefont {D.}~\bibnamefont
  {Jia}},\ }\href {\doibase 10.1088/1361-6382/ac4b04} {\bibfield  {journal}
  {\bibinfo  {journal} {Class. Quant. Grav.}\ }\textbf {\bibinfo {volume}
  {39}},\ \bibinfo {pages} {065002} (\bibinfo {year} {2022})},\ \Eprint
  {http://arxiv.org/abs/2110.05953} {arXiv:2110.05953 [gr-qc]} \BibitemShut
  {NoStop}%
\bibitem [{\citenamefont {Padua-Argüelles}({\natexlab{a}})}]{github}%
  \BibitemOpen
  \bibfield  {author} {\bibinfo {author} {\bibfnamefont {J.}~\bibnamefont
  {Padua-Argüelles}},\ }\href@noop {} {} ({\natexlab{a}}),\ \bibinfo {note}
  {simplicial-replicas,
  \href{https://github.com/jospadua/Simplicial-replicas.git}{https://github.com/jospadua/Simplicial-replicas.git}}\BibitemShut
  {NoStop}%
\bibitem [{\citenamefont
  {Padua-Argüelles}({\natexlab{b}})}]{PaduaArguellesQRC:2025}%
  \BibitemOpen
  \bibfield  {author} {\bibinfo {author} {\bibfnamefont {J.}~\bibnamefont
  {Padua-Argüelles}},\ }\href@noop {} {\enquote {\bibinfo {title} {To
  appear},}\ } ({\natexlab{b}}),\ \bibinfo {note} {2025}\BibitemShut {NoStop}%
\bibitem [{\citenamefont {Hawking}(1975)}]{Hawking:1975vcx}%
  \BibitemOpen
  \bibfield  {author} {\bibinfo {author} {\bibfnamefont {S.~W.}\ \bibnamefont
  {Hawking}},\ }\href {\doibase 10.1007/BF02345020} {\bibfield  {journal}
  {\bibinfo  {journal} {Commun. Math. Phys.}\ }\textbf {\bibinfo {volume}
  {43}},\ \bibinfo {pages} {199} (\bibinfo {year} {1975})},\ \bibinfo {note}
  {[Erratum: Commun.Math.Phys. 46, 206 (1976)]}\BibitemShut {NoStop}%
\bibitem [{\citenamefont {Hawking}(1976)}]{Hawking:1976ra}%
  \BibitemOpen
  \bibfield  {author} {\bibinfo {author} {\bibfnamefont {S.~W.}\ \bibnamefont
  {Hawking}},\ }\href {\doibase 10.1103/PhysRevD.14.2460} {\bibfield  {journal}
  {\bibinfo  {journal} {Phys. Rev. D}\ }\textbf {\bibinfo {volume} {14}},\
  \bibinfo {pages} {2460} (\bibinfo {year} {1976})}\BibitemShut {NoStop}%
\bibitem [{\citenamefont {Susskind}(1995)}]{Susskind:1995da}%
  \BibitemOpen
  \bibfield  {author} {\bibinfo {author} {\bibfnamefont {L.}~\bibnamefont
  {Susskind}},\ }\href@noop {} {\  (\bibinfo {year} {1995})},\ \Eprint
  {http://arxiv.org/abs/hep-th/9501106} {arXiv:hep-th/9501106} \BibitemShut
  {NoStop}%
\bibitem [{\citenamefont {Banks}\ and\ \citenamefont
  {Seiberg}(2011)}]{Banks:2010zn}%
  \BibitemOpen
  \bibfield  {author} {\bibinfo {author} {\bibfnamefont {T.}~\bibnamefont
  {Banks}}\ and\ \bibinfo {author} {\bibfnamefont {N.}~\bibnamefont
  {Seiberg}},\ }\href {\doibase 10.1103/PhysRevD.83.084019} {\bibfield
  {journal} {\bibinfo  {journal} {Phys. Rev. D}\ }\textbf {\bibinfo {volume}
  {83}},\ \bibinfo {pages} {084019} (\bibinfo {year} {2011})},\ \Eprint
  {http://arxiv.org/abs/1011.5120} {arXiv:1011.5120 [hep-th]} \BibitemShut
  {NoStop}%
\bibitem [{\citenamefont {Page}(1993{\natexlab{a}})}]{Page:1993up}%
  \BibitemOpen
  \bibfield  {author} {\bibinfo {author} {\bibfnamefont {D.~N.}\ \bibnamefont
  {Page}},\ }in\ \href@noop {} {\emph {\bibinfo {booktitle} {{5th Canadian
  Conference on General Relativity and Relativistic Astrophysics (5CCGRRA)}}}}\
  (\bibinfo {year} {1993})\ \Eprint {http://arxiv.org/abs/hep-th/9305040}
  {arXiv:hep-th/9305040} \BibitemShut {NoStop}%
\bibitem [{\citenamefont {Mathur}(2009)}]{Mathur:2009hf}%
  \BibitemOpen
  \bibfield  {author} {\bibinfo {author} {\bibfnamefont {S.~D.}\ \bibnamefont
  {Mathur}},\ }\href {\doibase 10.1088/0264-9381/26/22/224001} {\bibfield
  {journal} {\bibinfo  {journal} {Class. Quant. Grav.}\ }\textbf {\bibinfo
  {volume} {26}},\ \bibinfo {pages} {224001} (\bibinfo {year} {2009})},\
  \Eprint {http://arxiv.org/abs/0909.1038} {arXiv:0909.1038 [hep-th]}
  \BibitemShut {NoStop}%
\bibitem [{\citenamefont {Page}(1993{\natexlab{b}})}]{Page:1993df}%
  \BibitemOpen
  \bibfield  {author} {\bibinfo {author} {\bibfnamefont {D.~N.}\ \bibnamefont
  {Page}},\ }\href {\doibase 10.1103/PhysRevLett.71.1291} {\bibfield  {journal}
  {\bibinfo  {journal} {Phys. Rev. Lett.}\ }\textbf {\bibinfo {volume} {71}},\
  \bibinfo {pages} {1291} (\bibinfo {year} {1993}{\natexlab{b}})},\ \Eprint
  {http://arxiv.org/abs/gr-qc/9305007} {arXiv:gr-qc/9305007} \BibitemShut
  {NoStop}%
\bibitem [{\citenamefont {Araki}(1964)}]{10.1143/PTP.32.956}%
  \BibitemOpen
  \bibfield  {author} {\bibinfo {author} {\bibfnamefont {H.}~\bibnamefont
  {Araki}},\ }\href {\doibase 10.1143/PTP.32.956} {\bibfield  {journal}
  {\bibinfo  {journal} {Progress of Theoretical Physics}\ }\textbf {\bibinfo
  {volume} {32}},\ \bibinfo {pages} {956} (\bibinfo {year} {1964})},\ \Eprint
  {http://arxiv.org/abs/https://academic.oup.com/ptp/article-pdf/32/6/956/5311286/32-6-956.pdf}
  {https://academic.oup.com/ptp/article-pdf/32/6/956/5311286/32-6-956.pdf}
  \BibitemShut {NoStop}%
\bibitem [{\citenamefont {Witten}(2018)}]{Witten:2018zxz}%
  \BibitemOpen
  \bibfield  {author} {\bibinfo {author} {\bibfnamefont {E.}~\bibnamefont
  {Witten}},\ }\href {\doibase 10.1103/RevModPhys.90.045003} {\bibfield
  {journal} {\bibinfo  {journal} {Rev. Mod. Phys.}\ }\textbf {\bibinfo {volume}
  {90}},\ \bibinfo {pages} {045003} (\bibinfo {year} {2018})},\ \Eprint
  {http://arxiv.org/abs/1803.04993} {arXiv:1803.04993 [hep-th]} \BibitemShut
  {NoStop}%
\bibitem [{\citenamefont {Weinberg}(1995)}]{Weinberg_1995}%
  \BibitemOpen
  \bibfield  {author} {\bibinfo {author} {\bibfnamefont {S.}~\bibnamefont
  {Weinberg}},\ }\href@noop {} {\emph {\bibinfo {title} {The Quantum Theory of
  Fields}}}\ (\bibinfo  {publisher} {Cambridge University Press},\ \bibinfo
  {year} {1995})\BibitemShut {NoStop}%
\bibitem [{\citenamefont {Ambj\o{}rn}\ and\ \citenamefont
  {Loll}(2024)}]{Ambjorn:2024pyv}%
  \BibitemOpen
  \bibfield  {author} {\bibinfo {author} {\bibfnamefont {J.}~\bibnamefont
  {Ambj\o{}rn}}\ and\ \bibinfo {author} {\bibfnamefont {R.}~\bibnamefont
  {Loll}}\ }(\bibinfo {year} {2024})\ \Eprint {http://arxiv.org/abs/2401.09399}
  {arXiv:2401.09399 [hep-th]} \BibitemShut {NoStop}%
\bibitem [{\citenamefont {Lewkowycz}\ and\ \citenamefont
  {Maldacena}(2013)}]{Lewkowycz:2013nqa}%
  \BibitemOpen
  \bibfield  {author} {\bibinfo {author} {\bibfnamefont {A.}~\bibnamefont
  {Lewkowycz}}\ and\ \bibinfo {author} {\bibfnamefont {J.}~\bibnamefont
  {Maldacena}},\ }\href {\doibase 10.1007/JHEP08(2013)090} {\bibfield
  {journal} {\bibinfo  {journal} {JHEP}\ }\textbf {\bibinfo {volume} {08}},\
  \bibinfo {pages} {090} (\bibinfo {year} {2013})},\ \Eprint
  {http://arxiv.org/abs/1304.4926} {arXiv:1304.4926 [hep-th]} \BibitemShut
  {NoStop}%
\bibitem [{\citenamefont {Sorkin}(2019)}]{Sorkin:2019llw}%
  \BibitemOpen
  \bibfield  {author} {\bibinfo {author} {\bibfnamefont {R.~D.}\ \bibnamefont
  {Sorkin}},\ }\href@noop {} {\  (\bibinfo {year} {2019})},\ \Eprint
  {http://arxiv.org/abs/1908.10022} {arXiv:1908.10022 [gr-qc]} \BibitemShut
  {NoStop}%
\bibitem [{\citenamefont {Neiman}(2013{\natexlab{a}})}]{Neiman:2013ap}%
  \BibitemOpen
  \bibfield  {author} {\bibinfo {author} {\bibfnamefont {Y.}~\bibnamefont
  {Neiman}},\ }\href {\doibase 10.1007/JHEP04(2013)071} {\bibfield  {journal}
  {\bibinfo  {journal} {JHEP}\ }\textbf {\bibinfo {volume} {04}},\ \bibinfo
  {pages} {071} (\bibinfo {year} {2013}{\natexlab{a}})},\ \Eprint
  {http://arxiv.org/abs/1301.7041} {arXiv:1301.7041 [gr-qc]} \BibitemShut
  {NoStop}%
\bibitem [{\citenamefont {Bodendorfer}\ and\ \citenamefont
  {Neiman}(2013)}]{Bodendorfer:2013hla}%
  \BibitemOpen
  \bibfield  {author} {\bibinfo {author} {\bibfnamefont {N.}~\bibnamefont
  {Bodendorfer}}\ and\ \bibinfo {author} {\bibfnamefont {Y.}~\bibnamefont
  {Neiman}},\ }\href {\doibase 10.1088/0264-9381/30/19/195018} {\bibfield
  {journal} {\bibinfo  {journal} {Class. Quant. Grav.}\ }\textbf {\bibinfo
  {volume} {30}},\ \bibinfo {pages} {195018} (\bibinfo {year} {2013})},\
  \Eprint {http://arxiv.org/abs/1303.4752} {arXiv:1303.4752 [gr-qc]}
  \BibitemShut {NoStop}%
\bibitem [{\citenamefont {Neiman}(2013{\natexlab{b}})}]{Neiman:2013lxa}%
  \BibitemOpen
  \bibfield  {author} {\bibinfo {author} {\bibfnamefont {Y.}~\bibnamefont
  {Neiman}},\ }\href {\doibase 10.1103/PhysRevD.88.024037} {\bibfield
  {journal} {\bibinfo  {journal} {Phys. Rev. D}\ }\textbf {\bibinfo {volume}
  {88}},\ \bibinfo {pages} {024037} (\bibinfo {year} {2013}{\natexlab{b}})},\
  \Eprint {http://arxiv.org/abs/1305.2207} {arXiv:1305.2207 [gr-qc]}
  \BibitemShut {NoStop}%
\bibitem [{\citenamefont {Colin-Ellerin}\ \emph {et~al.}(2021)\citenamefont
  {Colin-Ellerin}, \citenamefont {Dong}, \citenamefont {Marolf}, \citenamefont
  {Rangamani},\ and\ \citenamefont {Wang}}]{Colin-Ellerin:2020mva}%
  \BibitemOpen
  \bibfield  {author} {\bibinfo {author} {\bibfnamefont {S.}~\bibnamefont
  {Colin-Ellerin}}, \bibinfo {author} {\bibfnamefont {X.}~\bibnamefont {Dong}},
  \bibinfo {author} {\bibfnamefont {D.}~\bibnamefont {Marolf}}, \bibinfo
  {author} {\bibfnamefont {M.}~\bibnamefont {Rangamani}}, \ and\ \bibinfo
  {author} {\bibfnamefont {Z.}~\bibnamefont {Wang}},\ }\href {\doibase
  10.1007/JHEP05(2021)117} {\bibfield  {journal} {\bibinfo  {journal} {JHEP}\
  }\textbf {\bibinfo {volume} {05}},\ \bibinfo {pages} {117} (\bibinfo {year}
  {2021})},\ \Eprint {http://arxiv.org/abs/2012.00828} {arXiv:2012.00828
  [hep-th]} \BibitemShut {NoStop}%
\bibitem [{\citenamefont {Halliwell}\ and\ \citenamefont
  {Hartle}(1990)}]{HalliwellContours}%
  \BibitemOpen
  \bibfield  {author} {\bibinfo {author} {\bibfnamefont {J.~J.}\ \bibnamefont
  {Halliwell}}\ and\ \bibinfo {author} {\bibfnamefont {J.~B.}\ \bibnamefont
  {Hartle}},\ }\href {\doibase 10.1103/PhysRevD.41.1815} {\bibfield  {journal}
  {\bibinfo  {journal} {Phys. Rev. D}\ }\textbf {\bibinfo {volume} {41}},\
  \bibinfo {pages} {1815} (\bibinfo {year} {1990})}\BibitemShut {NoStop}%
\bibitem [{\citenamefont {Kontsevich}\ and\ \citenamefont
  {Segal}(2021)}]{Kontsevich:2021dmb}%
  \BibitemOpen
  \bibfield  {author} {\bibinfo {author} {\bibfnamefont {M.}~\bibnamefont
  {Kontsevich}}\ and\ \bibinfo {author} {\bibfnamefont {G.}~\bibnamefont
  {Segal}},\ }\href {\doibase 10.1093/qmath/haab027} {\bibfield  {journal}
  {\bibinfo  {journal} {Quart. J. Math. Oxford Ser.}\ }\textbf {\bibinfo
  {volume} {72}},\ \bibinfo {pages} {673} (\bibinfo {year} {2021})},\ \Eprint
  {http://arxiv.org/abs/2105.10161} {arXiv:2105.10161 [hep-th]} \BibitemShut
  {NoStop}%
\bibitem [{\citenamefont {Witten}(2021)}]{Witten:2021nzp}%
  \BibitemOpen
  \bibfield  {author} {\bibinfo {author} {\bibfnamefont {E.}~\bibnamefont
  {Witten}},\ }\href@noop {} {\  (\bibinfo {year} {2021})},\ \Eprint
  {http://arxiv.org/abs/2111.06514} {arXiv:2111.06514 [hep-th]} \BibitemShut
  {NoStop}%
\bibitem [{\citenamefont {Lehners}(2022)}]{Lehners:2021mah}%
  \BibitemOpen
  \bibfield  {author} {\bibinfo {author} {\bibfnamefont {J.-L.}\ \bibnamefont
  {Lehners}},\ }\href {\doibase 10.1103/PhysRevD.105.026022} {\bibfield
  {journal} {\bibinfo  {journal} {Phys. Rev. D}\ }\textbf {\bibinfo {volume}
  {105}},\ \bibinfo {pages} {026022} (\bibinfo {year} {2022})},\ \Eprint
  {http://arxiv.org/abs/2111.07816} {arXiv:2111.07816 [hep-th]} \BibitemShut
  {NoStop}%
\bibitem [{\citenamefont {Jonas}\ \emph {et~al.}(2022)\citenamefont {Jonas},
  \citenamefont {Lehners},\ and\ \citenamefont {Quintin}}]{Jonas:2022uqb}%
  \BibitemOpen
  \bibfield  {author} {\bibinfo {author} {\bibfnamefont {C.}~\bibnamefont
  {Jonas}}, \bibinfo {author} {\bibfnamefont {J.-L.}\ \bibnamefont {Lehners}},
  \ and\ \bibinfo {author} {\bibfnamefont {J.}~\bibnamefont {Quintin}},\ }\href
  {\doibase 10.1007/JHEP08(2022)284} {\bibfield  {journal} {\bibinfo  {journal}
  {JHEP}\ }\textbf {\bibinfo {volume} {08}},\ \bibinfo {pages} {284} (\bibinfo
  {year} {2022})},\ \Eprint {http://arxiv.org/abs/2205.15332} {arXiv:2205.15332
  [hep-th]} \BibitemShut {NoStop}%
\bibitem [{\citenamefont {Lehners}(2023)}]{Lehners:2022xds}%
  \BibitemOpen
  \bibfield  {author} {\bibinfo {author} {\bibfnamefont {J.-L.}\ \bibnamefont
  {Lehners}},\ }\href {\doibase 10.1103/PhysRevD.107.046004} {\bibfield
  {journal} {\bibinfo  {journal} {Phys. Rev. D}\ }\textbf {\bibinfo {volume}
  {107}},\ \bibinfo {pages} {046004} (\bibinfo {year} {2023})},\ \Eprint
  {http://arxiv.org/abs/2209.14669} {arXiv:2209.14669 [hep-th]} \BibitemShut
  {NoStop}%
\bibitem [{\citenamefont {Marolf}(2022)}]{Marolf:2022ybi}%
  \BibitemOpen
  \bibfield  {author} {\bibinfo {author} {\bibfnamefont {D.}~\bibnamefont
  {Marolf}},\ }\href {\doibase 10.1007/JHEP07(2022)108} {\bibfield  {journal}
  {\bibinfo  {journal} {JHEP}\ }\textbf {\bibinfo {volume} {07}},\ \bibinfo
  {pages} {108} (\bibinfo {year} {2022})},\ \Eprint
  {http://arxiv.org/abs/2203.07421} {arXiv:2203.07421 [hep-th]} \BibitemShut
  {NoStop}%
\bibitem [{\citenamefont {Dittrich}\ and\ \citenamefont
  {Padua-Argüelles}()}]{DittrichConfig:2025}%
  \BibitemOpen
  \bibfield  {author} {\bibinfo {author} {\bibfnamefont {B.}~\bibnamefont
  {Dittrich}}\ and\ \bibinfo {author} {\bibfnamefont {J.}~\bibnamefont
  {Padua-Argüelles}},\ }\href@noop {} {\enquote {\bibinfo {title} {To
  appear},}\ }\bibinfo {note} {2025}\BibitemShut {NoStop}%
\bibitem [{\citenamefont {Johnson}(2001)}]{987181}%
  \BibitemOpen
  \bibfield  {author} {\bibinfo {author} {\bibfnamefont {P.}~\bibnamefont
  {Johnson}},\ }in\ \href {\doibase 10.1109/PAC.2001.987181} {\emph {\bibinfo
  {booktitle} {PACS2001. Proceedings of the 2001 Particle Accelerator
  Conference (Cat. No.01CH37268)}}},\ Vol.~\bibinfo {volume} {3}\ (\bibinfo
  {year} {2001})\ pp.\ \bibinfo {pages} {1781--1783 vol.3}\BibitemShut
  {NoStop}%
\bibitem [{\citenamefont {Gall}(2016)}]{gall2016brownian}%
  \BibitemOpen
  \bibfield  {author} {\bibinfo {author} {\bibfnamefont {J.}~\bibnamefont
  {Gall}},\ }\href {https://books.google.ca/books?id=G00WDAAAQBAJ} {\emph
  {\bibinfo {title} {Brownian Motion, Martingales, and Stochastic Calculus}}},\
  Graduate Texts in Mathematics\ (\bibinfo  {publisher} {Springer International
  Publishing},\ \bibinfo {year} {2016})\BibitemShut {NoStop}%
\bibitem [{\citenamefont {Penington}(2020)}]{Penington:2019npb}%
  \BibitemOpen
  \bibfield  {author} {\bibinfo {author} {\bibfnamefont {G.}~\bibnamefont
  {Penington}},\ }\href {\doibase 10.1007/JHEP09(2020)002} {\bibfield
  {journal} {\bibinfo  {journal} {JHEP}\ }\textbf {\bibinfo {volume} {09}},\
  \bibinfo {pages} {002} (\bibinfo {year} {2020})},\ \Eprint
  {http://arxiv.org/abs/1905.08255} {arXiv:1905.08255 [hep-th]} \BibitemShut
  {NoStop}%
\bibitem [{\citenamefont {Almheiri}\ \emph {et~al.}(2019)\citenamefont
  {Almheiri}, \citenamefont {Engelhardt}, \citenamefont {Marolf},\ and\
  \citenamefont {Maxfield}}]{Almheiri:2019psf}%
  \BibitemOpen
  \bibfield  {author} {\bibinfo {author} {\bibfnamefont {A.}~\bibnamefont
  {Almheiri}}, \bibinfo {author} {\bibfnamefont {N.}~\bibnamefont
  {Engelhardt}}, \bibinfo {author} {\bibfnamefont {D.}~\bibnamefont {Marolf}},
  \ and\ \bibinfo {author} {\bibfnamefont {H.}~\bibnamefont {Maxfield}},\
  }\href {\doibase 10.1007/JHEP12(2019)063} {\bibfield  {journal} {\bibinfo
  {journal} {JHEP}\ }\textbf {\bibinfo {volume} {12}},\ \bibinfo {pages} {063}
  (\bibinfo {year} {2019})},\ \Eprint {http://arxiv.org/abs/1905.08762}
  {arXiv:1905.08762 [hep-th]} \BibitemShut {NoStop}%
\bibitem [{\citenamefont {Polchinski}\ and\ \citenamefont
  {Strominger}(1994)}]{Polchinski:1994zs}%
  \BibitemOpen
  \bibfield  {author} {\bibinfo {author} {\bibfnamefont {J.}~\bibnamefont
  {Polchinski}}\ and\ \bibinfo {author} {\bibfnamefont {A.}~\bibnamefont
  {Strominger}},\ }\href {\doibase 10.1103/PhysRevD.50.7403} {\bibfield
  {journal} {\bibinfo  {journal} {Phys. Rev. D}\ }\textbf {\bibinfo {volume}
  {50}},\ \bibinfo {pages} {7403} (\bibinfo {year} {1994})},\ \Eprint
  {http://arxiv.org/abs/hep-th/9407008} {arXiv:hep-th/9407008} \BibitemShut
  {NoStop}%
\bibitem [{\citenamefont {Mirbabayi}(2020)}]{Mirbabayi:2020fyk}%
  \BibitemOpen
  \bibfield  {author} {\bibinfo {author} {\bibfnamefont {M.}~\bibnamefont
  {Mirbabayi}},\ }\href@noop {} {\  (\bibinfo {year} {2020})},\ \Eprint
  {http://arxiv.org/abs/2008.09626} {arXiv:2008.09626 [hep-th]} \BibitemShut
  {NoStop}%
\bibitem [{\citenamefont {Held}\ \emph {et~al.}(2024)\citenamefont {Held},
  \citenamefont {Liu}, \citenamefont {Marolf},\ and\ \citenamefont
  {Wang}}]{Held:2024qcl}%
  \BibitemOpen
  \bibfield  {author} {\bibinfo {author} {\bibfnamefont {J.}~\bibnamefont
  {Held}}, \bibinfo {author} {\bibfnamefont {X.}~\bibnamefont {Liu}}, \bibinfo
  {author} {\bibfnamefont {D.}~\bibnamefont {Marolf}}, \ and\ \bibinfo {author}
  {\bibfnamefont {Z.}~\bibnamefont {Wang}},\ }\href@noop {} {\  (\bibinfo
  {year} {2024})},\ \Eprint {http://arxiv.org/abs/2409.17428} {arXiv:2409.17428
  [hep-th]} \BibitemShut {NoStop}%
\bibitem [{\citenamefont {Dittrich}(2012)}]{Dittrich:2012jq}%
  \BibitemOpen
  \bibfield  {author} {\bibinfo {author} {\bibfnamefont {B.}~\bibnamefont
  {Dittrich}},\ }\href {\doibase 10.1088/1367-2630/14/12/123004} {\bibfield
  {journal} {\bibinfo  {journal} {New J. Phys.}\ }\textbf {\bibinfo {volume}
  {14}},\ \bibinfo {pages} {123004} (\bibinfo {year} {2012})},\ \Eprint
  {http://arxiv.org/abs/1205.6127} {arXiv:1205.6127 [gr-qc]} \BibitemShut
  {NoStop}%
\bibitem [{\citenamefont {Dittrich}\ and\ \citenamefont
  {Padua-Arg\"uelles}(2024)}]{Dittrich:2023ava}%
  \BibitemOpen
  \bibfield  {author} {\bibinfo {author} {\bibfnamefont {B.}~\bibnamefont
  {Dittrich}}\ and\ \bibinfo {author} {\bibfnamefont {J.}~\bibnamefont
  {Padua-Arg\"uelles}},\ }\href {\doibase 10.1103/PhysRevD.109.026002}
  {\bibfield  {journal} {\bibinfo  {journal} {Phys. Rev. D}\ }\textbf {\bibinfo
  {volume} {109}},\ \bibinfo {pages} {026002} (\bibinfo {year} {2024})},\
  \Eprint {http://arxiv.org/abs/2302.11586} {arXiv:2302.11586 [gr-qc]}
  \BibitemShut {NoStop}%
\bibitem [{\citenamefont {Cheeger}\ \emph {et~al.}(1984)\citenamefont
  {Cheeger}, \citenamefont {Muller},\ and\ \citenamefont
  {Schrader}}]{Cheeger:1983vq}%
  \BibitemOpen
  \bibfield  {author} {\bibinfo {author} {\bibfnamefont {J.}~\bibnamefont
  {Cheeger}}, \bibinfo {author} {\bibfnamefont {W.}~\bibnamefont {Muller}}, \
  and\ \bibinfo {author} {\bibfnamefont {R.}~\bibnamefont {Schrader}},\ }\href
  {\doibase 10.1007/BF01210729} {\bibfield  {journal} {\bibinfo  {journal}
  {Commun. Math. Phys.}\ }\textbf {\bibinfo {volume} {92}},\ \bibinfo {pages}
  {405} (\bibinfo {year} {1984})}\BibitemShut {NoStop}%
\bibitem [{\citenamefont {Feinberg}\ \emph {et~al.}(1984)\citenamefont
  {Feinberg}, \citenamefont {Friedberg}, \citenamefont {Lee},\ and\
  \citenamefont {Ren}}]{Feinberg:1984he}%
  \BibitemOpen
  \bibfield  {author} {\bibinfo {author} {\bibfnamefont {G.}~\bibnamefont
  {Feinberg}}, \bibinfo {author} {\bibfnamefont {R.}~\bibnamefont {Friedberg}},
  \bibinfo {author} {\bibfnamefont {T.~D.}\ \bibnamefont {Lee}}, \ and\
  \bibinfo {author} {\bibfnamefont {H.~C.}\ \bibnamefont {Ren}},\ }\href
  {\doibase 10.1016/0550-3213(84)90436-X} {\bibfield  {journal} {\bibinfo
  {journal} {Nucl. Phys. B}\ }\textbf {\bibinfo {volume} {245}},\ \bibinfo
  {pages} {343} (\bibinfo {year} {1984})}\BibitemShut {NoStop}%
\bibitem [{\citenamefont {Rocek}\ and\ \citenamefont
  {Williams}(1984)}]{Rocek:1982tj}%
  \BibitemOpen
  \bibfield  {author} {\bibinfo {author} {\bibfnamefont {M.}~\bibnamefont
  {Rocek}}\ and\ \bibinfo {author} {\bibfnamefont {R.~M.}\ \bibnamefont
  {Williams}},\ }\href {\doibase 10.1007/BF01581603} {\bibfield  {journal}
  {\bibinfo  {journal} {Z. Phys. C}\ }\textbf {\bibinfo {volume} {21}},\
  \bibinfo {pages} {371} (\bibinfo {year} {1984})}\BibitemShut {NoStop}%
\bibitem [{\citenamefont {Barrett}(1988)}]{barrett1988convergence}%
  \BibitemOpen
  \bibfield  {author} {\bibinfo {author} {\bibfnamefont {J.~W.}\ \bibnamefont
  {Barrett}},\ }\href@noop {} {\bibfield  {journal} {\bibinfo  {journal}
  {Classical and Quantum Gravity}\ }\textbf {\bibinfo {volume} {5}},\ \bibinfo
  {pages} {1187} (\bibinfo {year} {1988})}\BibitemShut {NoStop}%
\bibitem [{\citenamefont {Barrett}\ and\ \citenamefont
  {Williams}(1988)}]{Barrett:1988wd}%
  \BibitemOpen
  \bibfield  {author} {\bibinfo {author} {\bibfnamefont {J.~W.}\ \bibnamefont
  {Barrett}}\ and\ \bibinfo {author} {\bibfnamefont {R.~M.}\ \bibnamefont
  {Williams}},\ }\href {\doibase 10.1088/0264-9381/5/12/007} {\bibfield
  {journal} {\bibinfo  {journal} {Class. Quant. Grav.}\ }\textbf {\bibinfo
  {volume} {5}},\ \bibinfo {pages} {1543} (\bibinfo {year} {1988})}\BibitemShut
  {NoStop}%
\bibitem [{\citenamefont {{Misner}}\ \emph {et~al.}(1973)\citenamefont
  {{Misner}}, \citenamefont {{Thorne}},\ and\ \citenamefont
  {{Wheeler}}}]{Misner1973}%
  \BibitemOpen
  \bibfield  {author} {\bibinfo {author} {\bibfnamefont {C.~W.}\ \bibnamefont
  {{Misner}}}, \bibinfo {author} {\bibfnamefont {K.~S.}\ \bibnamefont
  {{Thorne}}}, \ and\ \bibinfo {author} {\bibfnamefont {J.~A.}\ \bibnamefont
  {{Wheeler}}},\ }\href@noop {} {\emph {\bibinfo {title} {San Francisco:
  W.H.~Freeman and Co., 1973}}},\ edited by\ \bibinfo {editor} {\bibnamefont
  {{Misner, C.~W., Thorne, K.~S., \& Wheeler, J.~A.}}}\ (\bibinfo {year}
  {1973})\BibitemShut {NoStop}%
\bibitem [{\citenamefont {Williams}(1992)}]{Williams:1991pj}%
  \BibitemOpen
  \bibfield  {author} {\bibinfo {author} {\bibfnamefont {R.~M.}\ \bibnamefont
  {Williams}},\ }\href {\doibase 10.1142/S0217979292001043} {\bibfield
  {journal} {\bibinfo  {journal} {Int. J. Mod. Phys. B}\ }\textbf {\bibinfo
  {volume} {6}},\ \bibinfo {pages} {2097} (\bibinfo {year} {1992})}\BibitemShut
  {NoStop}%
\bibitem [{\citenamefont {of~Alexandria}(85AD)}]{Heron}%
  \BibitemOpen
  \bibfield  {author} {\bibinfo {author} {\bibfnamefont {H.}~\bibnamefont
  {of~Alexandria}},\ }\href@noop {} {\emph {\bibinfo {title} {Metrica}}}\
  (\bibinfo {year} {10-85AD})\BibitemShut {NoStop}%
\bibitem [{\citenamefont {Hamber}\ and\ \citenamefont
  {Williams}(1999)}]{Hamber:1997ut}%
  \BibitemOpen
  \bibfield  {author} {\bibinfo {author} {\bibfnamefont {H.~W.}\ \bibnamefont
  {Hamber}}\ and\ \bibinfo {author} {\bibfnamefont {R.~M.}\ \bibnamefont
  {Williams}},\ }\href {\doibase 10.1103/PhysRevD.59.064014} {\bibfield
  {journal} {\bibinfo  {journal} {Phys. Rev. D}\ }\textbf {\bibinfo {volume}
  {59}},\ \bibinfo {pages} {064014} (\bibinfo {year} {1999})},\ \Eprint
  {http://arxiv.org/abs/hep-th/9708019} {arXiv:hep-th/9708019} \BibitemShut
  {NoStop}%
\bibitem [{\citenamefont {Hamber}(2009{\natexlab{b}})}]{Hamber:2009zz}%
  \BibitemOpen
  \bibfield  {author} {\bibinfo {author} {\bibfnamefont {H.~W.}\ \bibnamefont
  {Hamber}},\ }\href {\doibase 10.1007/978-3-540-85293-3} {\emph {\bibinfo
  {title} {{Quantum gravitation: The Feynman path integral approach}}}}\
  (\bibinfo  {publisher} {Springer},\ \bibinfo {address} {Berlin},\ \bibinfo
  {year} {2009})\BibitemShut {NoStop}%
\bibitem [{\citenamefont {Loll}(1998)}]{Loll:1998aj}%
  \BibitemOpen
  \bibfield  {author} {\bibinfo {author} {\bibfnamefont {R.}~\bibnamefont
  {Loll}},\ }\href {\doibase 10.12942/lrr-1998-13} {\bibfield  {journal}
  {\bibinfo  {journal} {Living Rev. Rel.}\ }\textbf {\bibinfo {volume} {1}},\
  \bibinfo {pages} {13} (\bibinfo {year} {1998})},\ \Eprint
  {http://arxiv.org/abs/gr-qc/9805049} {arXiv:gr-qc/9805049} \BibitemShut
  {NoStop}%
\bibitem [{\citenamefont {Borissova}\ and\ \citenamefont
  {Dittrich}(2023)}]{Borissova:2023izx}%
  \BibitemOpen
  \bibfield  {author} {\bibinfo {author} {\bibfnamefont {J.~N.}\ \bibnamefont
  {Borissova}}\ and\ \bibinfo {author} {\bibfnamefont {B.}~\bibnamefont
  {Dittrich}},\ }\href {\doibase 10.1007/JHEP09(2023)069} {\bibfield  {journal}
  {\bibinfo  {journal} {JHEP}\ }\textbf {\bibinfo {volume} {09}},\ \bibinfo
  {pages} {069} (\bibinfo {year} {2023})},\ \Eprint
  {http://arxiv.org/abs/2303.07367} {arXiv:2303.07367 [hep-th]} \BibitemShut
  {NoStop}%
\bibitem [{\citenamefont {Misner}(1957)}]{Misner:1957wq}%
  \BibitemOpen
  \bibfield  {author} {\bibinfo {author} {\bibfnamefont {C.~W.}\ \bibnamefont
  {Misner}},\ }\href {\doibase 10.1103/RevModPhys.29.497} {\bibfield  {journal}
  {\bibinfo  {journal} {Rev. Mod. Phys.}\ }\textbf {\bibinfo {volume} {29}},\
  \bibinfo {pages} {497} (\bibinfo {year} {1957})}\BibitemShut {NoStop}%
\bibitem [{\citenamefont {DeWitt}(1962)}]{DeWitt:1962by}%
  \BibitemOpen
  \bibfield  {author} {\bibinfo {author} {\bibfnamefont {B.~S.}\ \bibnamefont
  {DeWitt}},\ }\href {\doibase 10.1063/1.1724266} {\bibfield  {journal}
  {\bibinfo  {journal} {J. Math. Phys.}\ }\textbf {\bibinfo {volume} {3}},\
  \bibinfo {pages} {625} (\bibinfo {year} {1962})}\BibitemShut {NoStop}%
\bibitem [{\citenamefont {Voronoi}(1908)}]{Voronoi1908}%
  \BibitemOpen
  \bibfield  {author} {\bibinfo {author} {\bibfnamefont {G.}~\bibnamefont
  {Voronoi}},\ }\href {http://eudml.org/doc/149291} {\bibfield  {journal}
  {\bibinfo  {journal} {Journal für die reine und angewandte Mathematik}\
  }\textbf {\bibinfo {volume} {134}},\ \bibinfo {pages} {198} (\bibinfo {year}
  {1908})}\BibitemShut {NoStop}%
\bibitem [{\citenamefont {Ponzano}\ and\ \citenamefont
  {Regge}(1968)}]{Ponzano:1968}%
  \BibitemOpen
  \bibfield  {author} {\bibinfo {author} {\bibfnamefont {G.}~\bibnamefont
  {Ponzano}}\ and\ \bibinfo {author} {\bibfnamefont {T.}~\bibnamefont
  {Regge}},\ }\href@noop {} {\bibfield  {journal} {\bibinfo  {journal}
  {Spectroscopic and Group Theoretical Methods in Physics}\ }\textbf {\bibinfo
  {volume} {edited by F. Bloch}} (\bibinfo {year} {1968})}\BibitemShut
  {NoStop}%
\bibitem [{\citenamefont {Barrett}\ and\ \citenamefont
  {Naish-Guzman}(2009)}]{Barrett:2008wh}%
  \BibitemOpen
  \bibfield  {author} {\bibinfo {author} {\bibfnamefont {J.~W.}\ \bibnamefont
  {Barrett}}\ and\ \bibinfo {author} {\bibfnamefont {I.}~\bibnamefont
  {Naish-Guzman}},\ }\href {\doibase 10.1088/0264-9381/26/15/155014} {\bibfield
   {journal} {\bibinfo  {journal} {Class. Quant. Grav.}\ }\textbf {\bibinfo
  {volume} {26}},\ \bibinfo {pages} {155014} (\bibinfo {year} {2009})},\
  \Eprint {http://arxiv.org/abs/0803.3319} {arXiv:0803.3319 [gr-qc]}
  \BibitemShut {NoStop}%
\bibitem [{\citenamefont {Hamber}\ and\ \citenamefont
  {Williams}(1985)}]{Hamber:1985qj}%
  \BibitemOpen
  \bibfield  {author} {\bibinfo {author} {\bibfnamefont {H.~W.}\ \bibnamefont
  {Hamber}}\ and\ \bibinfo {author} {\bibfnamefont {R.~M.}\ \bibnamefont
  {Williams}},\ }\href {\doibase 10.1016/0370-2693(85)90382-X} {\bibfield
  {journal} {\bibinfo  {journal} {Phys. Lett. B}\ }\textbf {\bibinfo {volume}
  {157}},\ \bibinfo {pages} {368} (\bibinfo {year} {1985})}\BibitemShut
  {NoStop}%
\bibitem [{\citenamefont {Dittrich}\ and\ \citenamefont
  {Steinhaus}(2012)}]{Dittrich:2011vz}%
  \BibitemOpen
  \bibfield  {author} {\bibinfo {author} {\bibfnamefont {B.}~\bibnamefont
  {Dittrich}}\ and\ \bibinfo {author} {\bibfnamefont {S.}~\bibnamefont
  {Steinhaus}},\ }\href {\doibase 10.1103/PhysRevD.85.044032} {\bibfield
  {journal} {\bibinfo  {journal} {Phys. Rev. D}\ }\textbf {\bibinfo {volume}
  {85}},\ \bibinfo {pages} {044032} (\bibinfo {year} {2012})},\ \Eprint
  {http://arxiv.org/abs/1110.6866} {arXiv:1110.6866 [gr-qc]} \BibitemShut
  {NoStop}%
\bibitem [{\citenamefont {Dittrich}\ \emph {et~al.}(2014)\citenamefont
  {Dittrich}, \citenamefont {Kami\'nski},\ and\ \citenamefont
  {Steinhaus}}]{Dittrich:2014rha}%
  \BibitemOpen
  \bibfield  {author} {\bibinfo {author} {\bibfnamefont {B.}~\bibnamefont
  {Dittrich}}, \bibinfo {author} {\bibfnamefont {W.}~\bibnamefont
  {Kami\'nski}}, \ and\ \bibinfo {author} {\bibfnamefont {S.}~\bibnamefont
  {Steinhaus}},\ }\href {\doibase 10.1088/0264-9381/31/24/245009} {\bibfield
  {journal} {\bibinfo  {journal} {Class. Quant. Grav.}\ }\textbf {\bibinfo
  {volume} {31}},\ \bibinfo {pages} {245009} (\bibinfo {year} {2014})},\
  \Eprint {http://arxiv.org/abs/1404.5288} {arXiv:1404.5288 [gr-qc]}
  \BibitemShut {NoStop}%
\bibitem [{\citenamefont {Hartle}\ \emph {et~al.}(1997)\citenamefont {Hartle},
  \citenamefont {Miller},\ and\ \citenamefont {Williams}}]{Hartle:1996db}%
  \BibitemOpen
  \bibfield  {author} {\bibinfo {author} {\bibfnamefont {J.~B.}\ \bibnamefont
  {Hartle}}, \bibinfo {author} {\bibfnamefont {W.~A.}\ \bibnamefont {Miller}},
  \ and\ \bibinfo {author} {\bibfnamefont {R.~M.}\ \bibnamefont {Williams}},\
  }\href {\doibase 10.1088/0264-9381/14/8/013} {\bibfield  {journal} {\bibinfo
  {journal} {Class. Quant. Grav.}\ }\textbf {\bibinfo {volume} {14}},\ \bibinfo
  {pages} {2137} (\bibinfo {year} {1997})},\ \Eprint
  {http://arxiv.org/abs/gr-qc/9609028} {arXiv:gr-qc/9609028} \BibitemShut
  {NoStop}%
\bibitem [{\citenamefont {Menotti}\ and\ \citenamefont
  {Peirano}(1997)}]{Menotti:1996tm}%
  \BibitemOpen
  \bibfield  {author} {\bibinfo {author} {\bibfnamefont {P.}~\bibnamefont
  {Menotti}}\ and\ \bibinfo {author} {\bibfnamefont {P.~P.}\ \bibnamefont
  {Peirano}},\ }\href {\doibase 10.1016/S0550-3213(97)00017-5} {\bibfield
  {journal} {\bibinfo  {journal} {Nucl. Phys. B}\ }\textbf {\bibinfo {volume}
  {488}},\ \bibinfo {pages} {719} (\bibinfo {year} {1997})},\ \Eprint
  {http://arxiv.org/abs/hep-th/9607071} {arXiv:hep-th/9607071} \BibitemShut
  {NoStop}%
\bibitem [{\citenamefont {Sorkin}(1975)}]{Sorkin:1975jz}%
  \BibitemOpen
  \bibfield  {author} {\bibinfo {author} {\bibfnamefont {R.}~\bibnamefont
  {Sorkin}},\ }\href {\doibase 10.1063/1.522483} {\bibfield  {journal}
  {\bibinfo  {journal} {J. Math. Phys.}\ }\textbf {\bibinfo {volume} {16}},\
  \bibinfo {pages} {2432} (\bibinfo {year} {1975})},\ \bibinfo {note}
  {[Erratum: J.Math.Phys. 19, 1800 (1978)]}\BibitemShut {NoStop}%
\bibitem [{\citenamefont {Ren}(1988)}]{Ren:1987is}%
  \BibitemOpen
  \bibfield  {author} {\bibinfo {author} {\bibfnamefont {H.-c.}\ \bibnamefont
  {Ren}},\ }\href {\doibase 10.1016/0550-3213(88)90281-7} {\bibfield  {journal}
  {\bibinfo  {journal} {Nucl. Phys. B}\ }\textbf {\bibinfo {volume} {301}},\
  \bibinfo {pages} {661} (\bibinfo {year} {1988})}\BibitemShut {NoStop}%
\bibitem [{\citenamefont {Hamber}\ and\ \citenamefont
  {Williams}(1994)}]{Hamber:1993gn}%
  \BibitemOpen
  \bibfield  {author} {\bibinfo {author} {\bibfnamefont {H.~W.}\ \bibnamefont
  {Hamber}}\ and\ \bibinfo {author} {\bibfnamefont {R.~M.}\ \bibnamefont
  {Williams}},\ }\href {\doibase 10.1016/0550-3213(94)90119-8} {\bibfield
  {journal} {\bibinfo  {journal} {Nucl. Phys. B}\ }\textbf {\bibinfo {volume}
  {415}},\ \bibinfo {pages} {463} (\bibinfo {year} {1994})},\ \Eprint
  {http://arxiv.org/abs/hep-th/9308099} {arXiv:hep-th/9308099} \BibitemShut
  {NoStop}%
\bibitem [{\citenamefont {Sitharam}\ \emph {et~al.}(2018)\citenamefont
  {Sitharam}, \citenamefont {John},\ and\ \citenamefont
  {Sidman}}]{sitharam2018handbook}%
  \BibitemOpen
  \bibfield  {author} {\bibinfo {author} {\bibfnamefont {M.}~\bibnamefont
  {Sitharam}}, \bibinfo {author} {\bibfnamefont {A.}~\bibnamefont {John}}, \
  and\ \bibinfo {author} {\bibfnamefont {J.}~\bibnamefont {Sidman}},\ }\href
  {https://books.google.ca/books?id=gglYvgAACAAJ} {\emph {\bibinfo {title}
  {Handbook of Geometric Constraint Systems Principles}}},\ Discrete
  Mathematics and Its Applications Series\ (\bibinfo  {publisher} {CRC Press,
  Taylor \& Francis Group},\ \bibinfo {year} {2018})\BibitemShut {NoStop}%
\bibitem [{\citenamefont {Gattringer}\ and\ \citenamefont
  {Lang}(2010)}]{Gattringer2010}%
  \BibitemOpen
  \bibfield  {author} {\bibinfo {author} {\bibfnamefont {C.}~\bibnamefont
  {Gattringer}}\ and\ \bibinfo {author} {\bibfnamefont {C.~B.}\ \bibnamefont
  {Lang}},\ }\enquote {\bibinfo {title} {The path integral on the lattice},}\
  in\ \href {\doibase 10.1007/978-3-642-01850-3_1} {\emph {\bibinfo {booktitle}
  {Quantum Chromodynamics on the Lattice: An Introductory Presentation}}}\
  (\bibinfo  {publisher} {Springer Berlin Heidelberg},\ \bibinfo {address}
  {Berlin, Heidelberg},\ \bibinfo {year} {2010})\ pp.\ \bibinfo {pages}
  {1--23}\BibitemShut {NoStop}%
\bibitem [{\citenamefont {Kulkarni}\ \emph {et~al.}(1999)\citenamefont
  {Kulkarni}, \citenamefont {Schmidt},\ and\ \citenamefont
  {Tsui}}]{kulkarni:hal-01461924}%
  \BibitemOpen
  \bibfield  {author} {\bibinfo {author} {\bibfnamefont {D.}~\bibnamefont
  {Kulkarni}}, \bibinfo {author} {\bibfnamefont {D.}~\bibnamefont {Schmidt}}, \
  and\ \bibinfo {author} {\bibfnamefont {S.-K.}\ \bibnamefont {Tsui}},\ }\href
  {\doibase 10.1016/S0024-3795(99)00114-7} {\bibfield  {journal} {\bibinfo
  {journal} {{Linear Algebra and its Applications}}\ }\textbf {\bibinfo
  {volume} {297}},\ \bibinfo {pages} {63} (\bibinfo {year} {1999})}\BibitemShut
  {NoStop}%
\bibitem [{Che()}]{Chebyshev}%
  \BibitemOpen
  \href@noop {} {}\bibinfo {note} {Chebyshev Polynomial of the Second Kind.
  From MathWorld--A Wolfram Web Resource.
  \href{https://mathworld.wolfram.com/ChebyshevPolynomialoftheSecondKind.html}{https://mathworld.wolfram.com/ChebyshevPolynomialoftheSecondKind.html}}\BibitemShut
  {NoStop}%
\bibitem [{\citenamefont {Alexandru}\ \emph {et~al.}(2022)\citenamefont
  {Alexandru}, \citenamefont {Basar}, \citenamefont {Bedaque},\ and\
  \citenamefont {Warrington}}]{Alexandru:2020wrj}%
  \BibitemOpen
  \bibfield  {author} {\bibinfo {author} {\bibfnamefont {A.}~\bibnamefont
  {Alexandru}}, \bibinfo {author} {\bibfnamefont {G.}~\bibnamefont {Basar}},
  \bibinfo {author} {\bibfnamefont {P.~F.}\ \bibnamefont {Bedaque}}, \ and\
  \bibinfo {author} {\bibfnamefont {N.~C.}\ \bibnamefont {Warrington}},\ }\href
  {\doibase 10.1103/RevModPhys.94.015006} {\bibfield  {journal} {\bibinfo
  {journal} {Rev. Mod. Phys.}\ }\textbf {\bibinfo {volume} {94}},\ \bibinfo
  {pages} {015006} (\bibinfo {year} {2022})},\ \Eprint
  {http://arxiv.org/abs/2007.05436} {arXiv:2007.05436 [hep-lat]} \BibitemShut
  {NoStop}%
\bibitem [{\citenamefont {Han}\ \emph {et~al.}(2021)\citenamefont {Han},
  \citenamefont {Huang}, \citenamefont {Liu}, \citenamefont {Qu},\ and\
  \citenamefont {Wan}}]{Han:2020npv}%
  \BibitemOpen
  \bibfield  {author} {\bibinfo {author} {\bibfnamefont {M.}~\bibnamefont
  {Han}}, \bibinfo {author} {\bibfnamefont {Z.}~\bibnamefont {Huang}}, \bibinfo
  {author} {\bibfnamefont {H.}~\bibnamefont {Liu}}, \bibinfo {author}
  {\bibfnamefont {D.}~\bibnamefont {Qu}}, \ and\ \bibinfo {author}
  {\bibfnamefont {Y.}~\bibnamefont {Wan}},\ }\href {\doibase
  10.1103/PhysRevD.103.084026} {\bibfield  {journal} {\bibinfo  {journal}
  {Phys. Rev. D}\ }\textbf {\bibinfo {volume} {103}},\ \bibinfo {pages}
  {084026} (\bibinfo {year} {2021})},\ \Eprint
  {http://arxiv.org/abs/2012.11515} {arXiv:2012.11515 [gr-qc]} \BibitemShut
  {NoStop}%
\bibitem [{\citenamefont {Asante}\ \emph
  {et~al.}(2023{\natexlab{b}})\citenamefont {Asante}, \citenamefont
  {Dittrich},\ and\ \citenamefont {Steinhaus}}]{Asante:2022dnj}%
  \BibitemOpen
  \bibfield  {author} {\bibinfo {author} {\bibfnamefont {S.~K.}\ \bibnamefont
  {Asante}}, \bibinfo {author} {\bibfnamefont {B.}~\bibnamefont {Dittrich}}, \
  and\ \bibinfo {author} {\bibfnamefont {S.}~\bibnamefont {Steinhaus}},\
  }\enquote {\bibinfo {title} {{Spin Foams, Refinement Limit, and
  Renormalization}},}\ \ (\bibinfo {year} {2023})\ \Eprint
  {http://arxiv.org/abs/2211.09578} {arXiv:2211.09578 [gr-qc]} \BibitemShut
  {NoStop}%
\bibitem [{\citenamefont {Barrett}\ \emph {et~al.}(2009)\citenamefont
  {Barrett}, \citenamefont {Dowdall}, \citenamefont {Fairbairn}, \citenamefont
  {Gomes},\ and\ \citenamefont {Hellmann}}]{Barrett:2009gg}%
  \BibitemOpen
  \bibfield  {author} {\bibinfo {author} {\bibfnamefont {J.~W.}\ \bibnamefont
  {Barrett}}, \bibinfo {author} {\bibfnamefont {R.~J.}\ \bibnamefont
  {Dowdall}}, \bibinfo {author} {\bibfnamefont {W.~J.}\ \bibnamefont
  {Fairbairn}}, \bibinfo {author} {\bibfnamefont {H.}~\bibnamefont {Gomes}}, \
  and\ \bibinfo {author} {\bibfnamefont {F.}~\bibnamefont {Hellmann}},\ }\href
  {\doibase 10.1063/1.3244218} {\bibfield  {journal} {\bibinfo  {journal} {J.
  Math. Phys.}\ }\textbf {\bibinfo {volume} {50}},\ \bibinfo {pages} {112504}
  (\bibinfo {year} {2009})},\ \Eprint {http://arxiv.org/abs/0902.1170}
  {arXiv:0902.1170 [gr-qc]} \BibitemShut {NoStop}%
\bibitem [{\citenamefont {Barrett}\ \emph {et~al.}(2010)\citenamefont
  {Barrett}, \citenamefont {Dowdall}, \citenamefont {Fairbairn}, \citenamefont
  {Hellmann},\ and\ \citenamefont {Pereira}}]{Barrett:2009mw}%
  \BibitemOpen
  \bibfield  {author} {\bibinfo {author} {\bibfnamefont {J.~W.}\ \bibnamefont
  {Barrett}}, \bibinfo {author} {\bibfnamefont {R.~J.}\ \bibnamefont
  {Dowdall}}, \bibinfo {author} {\bibfnamefont {W.~J.}\ \bibnamefont
  {Fairbairn}}, \bibinfo {author} {\bibfnamefont {F.}~\bibnamefont {Hellmann}},
  \ and\ \bibinfo {author} {\bibfnamefont {R.}~\bibnamefont {Pereira}},\ }\href
  {\doibase 10.1088/0264-9381/27/16/165009} {\bibfield  {journal} {\bibinfo
  {journal} {Class. Quant. Grav.}\ }\textbf {\bibinfo {volume} {27}},\ \bibinfo
  {pages} {165009} (\bibinfo {year} {2010})},\ \Eprint
  {http://arxiv.org/abs/0907.2440} {arXiv:0907.2440 [gr-qc]} \BibitemShut
  {NoStop}%
\bibitem [{\citenamefont {Han}\ \emph {et~al.}(2022)\citenamefont {Han},
  \citenamefont {Huang}, \citenamefont {Liu},\ and\ \citenamefont
  {Qu}}]{Han:2021kll}%
  \BibitemOpen
  \bibfield  {author} {\bibinfo {author} {\bibfnamefont {M.}~\bibnamefont
  {Han}}, \bibinfo {author} {\bibfnamefont {Z.}~\bibnamefont {Huang}}, \bibinfo
  {author} {\bibfnamefont {H.}~\bibnamefont {Liu}}, \ and\ \bibinfo {author}
  {\bibfnamefont {D.}~\bibnamefont {Qu}},\ }\href {\doibase
  10.1103/PhysRevD.106.044005} {\bibfield  {journal} {\bibinfo  {journal}
  {Phys. Rev. D}\ }\textbf {\bibinfo {volume} {106}},\ \bibinfo {pages}
  {044005} (\bibinfo {year} {2022})},\ \Eprint
  {http://arxiv.org/abs/2110.10670} {arXiv:2110.10670 [gr-qc]} \BibitemShut
  {NoStop}%
\bibitem [{\citenamefont {Han}\ \emph {et~al.}(2023)\citenamefont {Han},
  \citenamefont {Liu},\ and\ \citenamefont {Qu}}]{Han:2023cen}%
  \BibitemOpen
  \bibfield  {author} {\bibinfo {author} {\bibfnamefont {M.}~\bibnamefont
  {Han}}, \bibinfo {author} {\bibfnamefont {H.}~\bibnamefont {Liu}}, \ and\
  \bibinfo {author} {\bibfnamefont {D.}~\bibnamefont {Qu}},\ }\href {\doibase
  10.1103/PhysRevD.108.026010} {\bibfield  {journal} {\bibinfo  {journal}
  {Phys. Rev. D}\ }\textbf {\bibinfo {volume} {108}},\ \bibinfo {pages}
  {026010} (\bibinfo {year} {2023})},\ \Eprint
  {http://arxiv.org/abs/2301.02930} {arXiv:2301.02930 [gr-qc]} \BibitemShut
  {NoStop}%
\bibitem [{\citenamefont {Asante}\ \emph {et~al.}(2020)\citenamefont {Asante},
  \citenamefont {Dittrich},\ and\ \citenamefont {Haggard}}]{Asante:2020qpa}%
  \BibitemOpen
  \bibfield  {author} {\bibinfo {author} {\bibfnamefont {S.~K.}\ \bibnamefont
  {Asante}}, \bibinfo {author} {\bibfnamefont {B.}~\bibnamefont {Dittrich}}, \
  and\ \bibinfo {author} {\bibfnamefont {H.~M.}\ \bibnamefont {Haggard}},\
  }\href {\doibase 10.1103/PhysRevLett.125.231301} {\bibfield  {journal}
  {\bibinfo  {journal} {Phys. Rev. Lett.}\ }\textbf {\bibinfo {volume} {125}},\
  \bibinfo {pages} {231301} (\bibinfo {year} {2020})},\ \Eprint
  {http://arxiv.org/abs/2004.07013} {arXiv:2004.07013 [gr-qc]} \BibitemShut
  {NoStop}%
\bibitem [{\citenamefont {Asante}\ \emph
  {et~al.}(2021{\natexlab{a}})\citenamefont {Asante}, \citenamefont
  {Dittrich},\ and\ \citenamefont {Haggard}}]{Asante:2020iwm}%
  \BibitemOpen
  \bibfield  {author} {\bibinfo {author} {\bibfnamefont {S.~K.}\ \bibnamefont
  {Asante}}, \bibinfo {author} {\bibfnamefont {B.}~\bibnamefont {Dittrich}}, \
  and\ \bibinfo {author} {\bibfnamefont {H.~M.}\ \bibnamefont {Haggard}},\
  }\href {\doibase 10.1088/1361-6382/ac011b} {\bibfield  {journal} {\bibinfo
  {journal} {Class. Quant. Grav.}\ }\textbf {\bibinfo {volume} {38}},\ \bibinfo
  {pages} {145023} (\bibinfo {year} {2021}{\natexlab{a}})},\ \Eprint
  {http://arxiv.org/abs/2011.14468} {arXiv:2011.14468 [gr-qc]} \BibitemShut
  {NoStop}%
\bibitem [{\citenamefont {Asante}\ \emph
  {et~al.}(2021{\natexlab{b}})\citenamefont {Asante}, \citenamefont
  {Dittrich},\ and\ \citenamefont {Padua-Arguelles}}]{Asante:2021zzh}%
  \BibitemOpen
  \bibfield  {author} {\bibinfo {author} {\bibfnamefont {S.~K.}\ \bibnamefont
  {Asante}}, \bibinfo {author} {\bibfnamefont {B.}~\bibnamefont {Dittrich}}, \
  and\ \bibinfo {author} {\bibfnamefont {J.}~\bibnamefont {Padua-Arguelles}},\
  }\href {\doibase 10.1088/1361-6382/ac1b44} {\bibfield  {journal} {\bibinfo
  {journal} {Class. Quant. Grav.}\ }\textbf {\bibinfo {volume} {38}},\ \bibinfo
  {pages} {195002} (\bibinfo {year} {2021}{\natexlab{b}})},\ \Eprint
  {http://arxiv.org/abs/2104.00485} {arXiv:2104.00485 [gr-qc]} \BibitemShut
  {NoStop}%
\bibitem [{\citenamefont {Engle}\ \emph {et~al.}(2007)\citenamefont {Engle},
  \citenamefont {Pereira},\ and\ \citenamefont {Rovelli}}]{Engle:2007uq}%
  \BibitemOpen
  \bibfield  {author} {\bibinfo {author} {\bibfnamefont {J.}~\bibnamefont
  {Engle}}, \bibinfo {author} {\bibfnamefont {R.}~\bibnamefont {Pereira}}, \
  and\ \bibinfo {author} {\bibfnamefont {C.}~\bibnamefont {Rovelli}},\ }\href
  {\doibase 10.1103/PhysRevLett.99.161301} {\bibfield  {journal} {\bibinfo
  {journal} {Phys. Rev. Lett.}\ }\textbf {\bibinfo {volume} {99}},\ \bibinfo
  {pages} {161301} (\bibinfo {year} {2007})},\ \Eprint
  {http://arxiv.org/abs/0705.2388} {arXiv:0705.2388 [gr-qc]} \BibitemShut
  {NoStop}%
\bibitem [{\citenamefont {Freidel}\ and\ \citenamefont
  {Krasnov}(2008)}]{Freidel:2007py}%
  \BibitemOpen
  \bibfield  {author} {\bibinfo {author} {\bibfnamefont {L.}~\bibnamefont
  {Freidel}}\ and\ \bibinfo {author} {\bibfnamefont {K.}~\bibnamefont
  {Krasnov}},\ }\href {\doibase 10.1088/0264-9381/25/12/125018} {\bibfield
  {journal} {\bibinfo  {journal} {Class. Quant. Grav.}\ }\textbf {\bibinfo
  {volume} {25}},\ \bibinfo {pages} {125018} (\bibinfo {year} {2008})},\
  \Eprint {http://arxiv.org/abs/0708.1595} {arXiv:0708.1595 [gr-qc]}
  \BibitemShut {NoStop}%
\bibitem [{\citenamefont {Haggard}\ and\ \citenamefont
  {Rovelli}(2015)}]{Haggard:2014rza}%
  \BibitemOpen
  \bibfield  {author} {\bibinfo {author} {\bibfnamefont {H.~M.}\ \bibnamefont
  {Haggard}}\ and\ \bibinfo {author} {\bibfnamefont {C.}~\bibnamefont
  {Rovelli}},\ }\href {\doibase 10.1103/PhysRevD.92.104020} {\bibfield
  {journal} {\bibinfo  {journal} {Phys. Rev. D}\ }\textbf {\bibinfo {volume}
  {92}},\ \bibinfo {pages} {104020} (\bibinfo {year} {2015})},\ \Eprint
  {http://arxiv.org/abs/1407.0989} {arXiv:1407.0989 [gr-qc]} \BibitemShut
  {NoStop}%
\bibitem [{\citenamefont {D'Ambrosio}\ and\ \citenamefont
  {Rovelli}(2018)}]{DAmbrosio:2018wgv}%
  \BibitemOpen
  \bibfield  {author} {\bibinfo {author} {\bibfnamefont {F.}~\bibnamefont
  {D'Ambrosio}}\ and\ \bibinfo {author} {\bibfnamefont {C.}~\bibnamefont
  {Rovelli}},\ }\href {\doibase 10.1088/1361-6382/aae499} {\bibfield  {journal}
  {\bibinfo  {journal} {Class. Quant. Grav.}\ }\textbf {\bibinfo {volume}
  {35}},\ \bibinfo {pages} {215010} (\bibinfo {year} {2018})},\ \Eprint
  {http://arxiv.org/abs/1803.05015} {arXiv:1803.05015 [gr-qc]} \BibitemShut
  {NoStop}%
\bibitem [{\citenamefont {De~Lorenzo}\ and\ \citenamefont
  {Perez}(2016)}]{DeLorenzo:2015gtx}%
  \BibitemOpen
  \bibfield  {author} {\bibinfo {author} {\bibfnamefont {T.}~\bibnamefont
  {De~Lorenzo}}\ and\ \bibinfo {author} {\bibfnamefont {A.}~\bibnamefont
  {Perez}},\ }\href {\doibase 10.1103/PhysRevD.93.124018} {\bibfield  {journal}
  {\bibinfo  {journal} {Phys. Rev. D}\ }\textbf {\bibinfo {volume} {93}},\
  \bibinfo {pages} {124018} (\bibinfo {year} {2016})},\ \Eprint
  {http://arxiv.org/abs/1512.04566} {arXiv:1512.04566 [gr-qc]} \BibitemShut
  {NoStop}%
\bibitem [{\citenamefont {D'Ambrosio}\ \emph {et~al.}(2021)\citenamefont
  {D'Ambrosio}, \citenamefont {Christodoulou}, \citenamefont {Martin-Dussaud},
  \citenamefont {Rovelli},\ and\ \citenamefont {Soltani}}]{DAmbrosio:2020mut}%
  \BibitemOpen
  \bibfield  {author} {\bibinfo {author} {\bibfnamefont {F.}~\bibnamefont
  {D'Ambrosio}}, \bibinfo {author} {\bibfnamefont {M.}~\bibnamefont
  {Christodoulou}}, \bibinfo {author} {\bibfnamefont {P.}~\bibnamefont
  {Martin-Dussaud}}, \bibinfo {author} {\bibfnamefont {C.}~\bibnamefont
  {Rovelli}}, \ and\ \bibinfo {author} {\bibfnamefont {F.}~\bibnamefont
  {Soltani}},\ }\href {\doibase 10.1103/PhysRevD.103.106014} {\bibfield
  {journal} {\bibinfo  {journal} {Phys. Rev. D}\ }\textbf {\bibinfo {volume}
  {103}},\ \bibinfo {pages} {106014} (\bibinfo {year} {2021})},\ \Eprint
  {http://arxiv.org/abs/2009.05016} {arXiv:2009.05016 [gr-qc]} \BibitemShut
  {NoStop}%
\bibitem [{\citenamefont {Soltani}\ \emph {et~al.}(2021)\citenamefont
  {Soltani}, \citenamefont {Rovelli},\ and\ \citenamefont
  {Martin-Dussaud}}]{Soltani:2021zmv}%
  \BibitemOpen
  \bibfield  {author} {\bibinfo {author} {\bibfnamefont {F.}~\bibnamefont
  {Soltani}}, \bibinfo {author} {\bibfnamefont {C.}~\bibnamefont {Rovelli}}, \
  and\ \bibinfo {author} {\bibfnamefont {P.}~\bibnamefont {Martin-Dussaud}},\
  }\href {\doibase 10.1103/PhysRevD.104.066015} {\bibfield  {journal} {\bibinfo
   {journal} {Phys. Rev. D}\ }\textbf {\bibinfo {volume} {104}},\ \bibinfo
  {pages} {066015} (\bibinfo {year} {2021})},\ \Eprint
  {http://arxiv.org/abs/2105.06876} {arXiv:2105.06876 [gr-qc]} \BibitemShut
  {NoStop}%
\end{thebibliography}%

\end{document}